# Social Media and Political Contributions: The Impact of New Technology on Political Competition

Maria Petrova[1], Ananya Sen[2], and Pinar Yildirim[*,3]

[1]Universitat Pompeu Fabra, ICREA, Barcelona IPEG, Barcelona GSE, New Economic School
[2]Carnegie Mellon University
[3]Wharton School, University of Pennsylvania

[*]We thank Yann Algan, Pablo Barbera, Daniel Chen, Sylvain Ferret, Jacques Cremer, Stefano DellaVigna, Ruben Enikolopov, Yosh Halberstam, Pooyan Khashabi, Brian Knight, Ting Li, Jennifer Pan, Paul Seabright, Jas Sekhon, James Snyder, David Strömberg, Alessandro Tarozzi, Joel Waldfogel, Christophe Van den Bulte, Karine Van der Straeten, Yanhao Wei, and seminar participants at TSE, UPF, Stanford Digital Marketing Conference, 19th ISNIE Conference (Harvard), SMaPP Conference (NYU Florence), 9th IP, Software and Internet Conference (Toulouse), Economics of Network Industries Workshop (Paris), ZEW ICT conference (Mannheim), Munich Summer Institute, Media Economics Workshop (Zurich), University of Virginia Marketing Seminar and CODE (MIT) for helpful comments. We are grateful to Yosh Halberstam and Brian Knight for kindly sharing the data on followers for 2012 candidates, and to Andrei Simonov for kindly sharing his data on political advertising. We thank Jack Beckwith, Renata Gaineddenova, Erqi Ge, Fernando Fernandez, Maria Ptashkina, and Shawn Zamechek for excellent research assistance. Maria Petrova thanks Spanish Ministry of Economy and Competitiveness (Grant ECO2014-55555-P) and Ministry of Education and Science of the Russian Federation (Grant No. 14.U04.31.0002) for financial support. Ananya Sen thanks Jean Jacques Laffont Digital Chair for financial support. Pinar Yildirim thanks Wharton Dean's Research Fund and Mack Institute for financial support.

**Abstract**

Political campaigns are among the most sophisticated marketing exercises in the United States. As part of their marketing communication strategy, an increasing number of politicians adopt social media to inform their constituencies. This study documents the returns from adopting a new technology, namely Twitter, for politicians running for Congress by focusing on the change in campaign contributions received. We compare weekly donations received just before and just after a politician opens a Twitter account in regions with high and low levels of Twitter penetration, controlling for politician-month fixed effects. Specifically, over the course of a political campaign, we estimate that the differential effect of opening a Twitter account in regions with high vs low levels of Twitter penetration amounts to an increase of 0.7-2% in donations for all politicians and 1-3.1% for new politicians, who were never elected to the Congress before. In contrast, the effect of joining Twitter for experienced politicians remains negligibly small. We find some evidence consistent with the explanation that the effect is driven by new information about the candidates, e.g., the effect is primarily driven by new donors rather than past donors, by candidates without Facebook accounts and tweeting more informatively. Overall, our findings imply that social media can intensify political competition by lowering costs of disseminating information for new entrants to their constituents and thus may reduce the barriers to enter politics.

# 1   Introduction

Political campaigns are among the most expensive and sophisticated marketing efforts in the United States. In the 2018 electoral cycle, campaign spending by House and Senate candidates reached a total of more than $1 billion for each of the two houses.[1] To inform and persuade voters, candidates rely on a variety of communication strategies, including advertising, print and broadcast media, and speeches along the campaign trail. Moreover, during the past few years we observe politicians adopting new communication technologies such as social media. So far, the literature has produced mixed evidence that adopting social media as a communication technology helps brands and consumers (e.g., Gong et al., 2017; Gans et al., 2016; Hoffman and Fodor, 2010; Naylor et al., 2012; Laroche et al., 2013; Kumar et al., 2013). In a similar vain, it is not clear if this adoption is useful for politicians and alters any aspects of political competition. This question is particularly timely as we head into the 2020 Presidential Election, due to increased interest in social media's influence on electoral races.

We study the effect of new communication technology on campaign contributions. More specifically, we focus on politicians' adoption of Twitter and its effect on political donations received while they are running for the U.S. Congress. Throughout the analysis, we focus on how Twitter adoption may lead to differential benefits for new politicians relative to the more experienced ones. Documenting such heterogeneity in benefits from adopting social media is particularly important, because there is well-documented evidence that incumbents in the United States hold significant financial and informational advantages over newcomers (Ansolabehere and Snyder, 2000; Prat, 2002; Prior, 2006). We ask whether adopting social media can mitigate this incumbency advantage by giving new politicians access to an alternate, relatively cost-effective technology to communicate with their constituency about their candidacy and raise awareness.

Identifying the causal impact of Twitter on political donations is not trivial, mainly because of correlated unobservables, which could influence both a politician's decision to join Twitter and the amount of political donations she raised. To study how adopting Twitter influences the amount of political donations received more rigorously, we use a difference-in-differences approach. We combine data on 1,834 politicians who opened a personal Twitter account between 2009 and 2014, campaign contributions received, expenditures incurred by candidates, and Twitter's penetration in the politician's region, which we use as a proxy for users' exposure to information disseminated from Twitter relative to other information. We ask whether joining Twitter had a differential effect on donations in high- and low- Twitter penetration states. We control for the politician-month fixed effects to account for politician-specific unobserved time-varying factors, such as being more progressive, more tech-savvy, or being at a different stage of campaigning. Our identifying assumption is that unobserved determinants of donations did not differentially affect donations in high- vs low- Twitter penetration states. Said differently, the differences between contribution flows, unexplained by politician-month fixed effects, would remain the same in regions with different levels of Twitter penetration in the absence of politicians' entry to Twitter, also known as parallel trends assumption.[2]

---

[1] Center for Responsive Politics, https://www.opensecrets.org/overview/index.php?display=T&type=A&cycle=2018

We find that right after a candidate started to post on Twitter, his/her weekly aggregate donations went up, more so in states where Twitter penetration is high. We focus on donations below $1,000, since smaller donations are more likely to respond to information disseminated via social media. We find that the differential increase in donations between high- and low- Twitter penetration states ranges from 2.9% for 2009 to 22.8% in 2014. However, this gain is significantly positive only for new politicians who have never been elected to the Congress before, i.e., not for the experienced candidates. The persuasion rate associated with donations after opening a Twitter channel is approximately 1% (DellaVigna and Gentzkow, 2010). This rate is lower than the average media persuasion rates reported in the literature but is comparable to the rates reported for direct mailing (Gerber and Green, 2000) and political advertising (Spenkuch and Toniatti, 2016). Overall, our results are consistent with new technology being helpful to promote competition, specifically among politicians.

To strengthen our identification claims, we use a series of placebo tests. Our identifying assumption is that joining Twitter is not related to within-politician-month unobserved determinants of donations, in high- vs. low- Twitter penetration states. This assumption could be violated in case some other fund-raising or marketing activity is happening at the same time, even though, a priori, there are no particular reasons to think that these activities should affect donations differentially depending on the level of Twitter penetration. To cope with this possibility, we investigate how politicians' key marketing activities change around the time of adopting Twitter. First, using Federal Election Commission's (FEC) data on campaign spending data, we study if there are any contemporaneous increases in 12 different expense categories, including general advertising, fundraising events, campaign materials and travel. We find that there is no discontinuous increase in various types of campaign spending around the time of Twitter entry across high- and low- Twitter penetration areas, even though political contributions are strongly correlated with campaign spending in a given week. Second, we find that media and blog coverage of the politicians do not change significantly around the time of Twitter entry. Third, we find that there is no increase in advertising spending on TV around the time of Twitter entry, even though political ads seem to be a significant driver of political donations (see, e.g., Urban and Niebler (2014)). Finally, we test if Twitter penetration is just a proxy for other characteristics such as income, education, political preferences, internet access, party affiliation, and racial composition, which may influence donations in the same way, and we find that it is unlikely to be the case. Overall, the results in all of our placebo specifications are consistent with our identifying assumptions, i.e., unobserved heterogeneity is unlikely to explain our results. To alleviate concerns that there could be some residual bias coming from unobservables, in the online appendix, we conduct Altonji-Elder-Taber tests (Altonji et al., 2005). We show that in our particular setting, to attribute our entire difference-in-differences estimates to selection effects, selection on unobservables would have to be over 10.4 times greater than the selection on observables.

---

[2] Because all states had nonzero Twitter penetration in our sample, we assume that the results from nonzero Twitter penetration states can be linearly extrapolated to calculate the counterfactual for a state with no Twitter usage.

To explain the mechanism through which donations increase, we borrow from the literature on advertising (Nelson, 1974). Adopting a new communication technology, similar to advertising, helps politicians to gain support through an information and/or persuasion channel. The information mechanism suggests that voters, who were not previously familiar with a candidate, become informed about the candidate and their policy proposals. The persuasion mechanism suggests that voters who are already informed about the candidates and their policy stances would be further persuaded or mobilized to provide support. We provide some evidence consistent with the information mechanism in our setting. First, we find that entry to Twitter increases political donations from the donors who did not support the politicians before, but not from repeat donors. Second, we find that the effect is driven by the politicians, who did not have a Facebook account before, in contrast to politicians, who joined Facebook before Twitter. Third, the analysis of tweet content and frequency suggests that more frequent and more informative tweets (e.g., including links to websites, responding to news fast, or more anti-establishment Tweets) are associated with receiving higher contributions after adopting Twitter. Fourth, we find that the donations to a candidate come predominantly from the same-state donors as opposed to donors from other states. Finally, political candidates for the House of Representatives in high Twitter penetration regions see a quantitatively higher jump in donations after adopting Twitter, as compared to Senate candidates whose name recognition is arguably higher. Overall, these findings are consistent with new information driving political contributions as a mechanism by increasing potential donors' awareness about the politicians and their proposed policies.

A broader implication of our study is that the adoption of Twitter can reduce the gap in fundraising opportunities between new and experienced politicians, which, in turn, can lower barriers to entry to national politics and increase political competition. Furthermore, we find qualitatively similar results when we analyze the impact of Facebook adoption on political donations, within a similar empirical framework. Thus, new technologies indeed have a potential to change the incentives of prospective entrants and, ultimately, make American politics more competitive. Since social media provides access to a relatively cheap advertising platform, our results have more general implications for platforms such as Twitter and Facebook allowing new brands and products to enter some markets and inform consumers, making these markets more competitive.

Our paper contributes to several areas in the literature. First, we contribute to studies of the impact of media and information and communication technologies (ICTs) on competition (Fudenberg et al., 1983; Fudenberg and Tirole, 1985; Aghion et al., 2005), consumer demand and financial gain (Bollinger et al., 2013; Fang and Peress, 2009; Goh et al., 2011; Engelberg and Parsons, 2011), voting behavior (DellaVigna and Kaplan, 2007; Enikolopov et al., 2011; Gentzkow et al., 2011; Chiang and Knight, 2011; Gentzkow et al., 2014; Rotesi, 2019), social outcomes (Jensen and Oster, 2009; La Ferrara et al., 2012; Enikolopov et al., 2016), and policy decisions (Strömberg, 2004; Eisensee and Strömberg, 2007; Snyder Jr and Strömberg, 2010). We complement this literature by highlighting a mechanism through which media could influence political competition. We demonstrate how the advent of a new communication technology, social media in general and Twitter in particular, can alter political competition by improving opportunities for new candidates to raise funds and inform voters in a cost-effective fashion.

Second, our paper is also related to a stream of studies in marketing, management, and economics that discuss the outcomes and returns from adoption of social media (e.g., Culotta and Cutler, 2016; Hoffman and Fodor, 2010; Kumar et al., 2013). Gong et al. (2017) and Seiler et al. (2016) study the impact of advertising of TV content in Chinese micro blogs on subsequent TV series viewership. Enikolopov et al. (2016) study the impact of social media on corporate accountability. Gans et al. (2016) and Ma et al. (2015) focus on social media as a tool for consumers to exhibit voice and for firms to respond to consumer complaints. Acemoglu et al. (2017) and Enikolopov et al. (2016) analyze the effects of social media content and penetration on subsequent protest participation. Qin et al. (2016) study the content and the impact of social media in China for collective action outcomes, while Qin (2013) look at the relationship between a Chinese microblog penetration on drug quality. In contrast with these studies, we focus our investigation on the strategic benefit of entry into an online social network for public personalities, specifically for politicians, by quantifying their financial gain and investigating the mechanisms behind it in detail.

Finally, we also contribute to the studies of political campaigns and political communication in marketing. Marketing scholars have long been interested in the determinants and design of political races (Rosenthal and Sen, 1969, 1973, 1977), with the interest resurfacing more recently (Gordon et al., 2012). Soberman and Sadoulet (2007) studied how campaign limits influence political spending and found that tighter limits stimulate more aggressive advertising of competing parties. Gordon and Hartmann (2013, 2016) estimate the effectiveness of advertising on political races and also focus on how the structure of political competition in the U.S. shapes advertising spending. Xiang and Sarvary (2007); Gal-Or et al. (2012); Yildirim et al. (2013), and Zhu and Dukes (2015) focus on how media outlets strategically adopt political biases to maximize their revenue. This paper contributes to these earlier studies by demonstrating the effect of communication on social media on politicians' fundraising ability and electoral competition.

The rest of the paper is organized as follows. Section 2 provides a review of the literature and institutional details on the use of social media by politicians and Section 3 provides a summary of the data we use. Section 4 provides the framework for the empirical analysis, Section 5 details the results, and Section 6 discusses the mechanism. Finally, Section 7 concludes.

## 2  Background

**Use of Social Media by Politicians**

Until recently, traditional media held the role of being the primary information channel for politicians, so obtaining coverage in newspapers and TV outlets has been crucial for electoral success. Candidates further disseminate information about their candidacy and policy goals through speeches they give along the campaign trail and through public appearances (Garcia-Jimeno and Yildirim, 2015). Today, a reported 80% of heads of states around the world use Twitter to communicate with their constituencies.[3] Compared to campaign messages, the content of this communication is more personal and includes information about politicians' lives and activities outside of politics. While politicians who are well-known and hold high-level positions typically reach out to several million followers on Twitter, lesser-known politicians communicate with several hundred to several thousand individuals. In 2018, Barack Obama had more than 100 million Twitter followers, Senator Orin

---
[3]http://www.adweek.com/socialtimes/world-leaders-twitter/495103

Hatch had more than 100 thousand, and Representative Paul Cook had more than 14,000. In our data, the total number of candidates who had been using Twitter increases from 741 in 2009 to 1,024 in 2010, to 1,488 in 2012, and to 1,834 in 2014.

After the 2008 presidential election, scholars predicted increased and targeted web use by political campaigns at the federal and local level (Towner and Dulio, 2012). This included use of Social Networking Services (SNSs), which allow candidates to build profiles and showcase connections within a delimited system (Boyd and Ellison, 2010; Boyd and Marwick, 2011). Among these sites, Twitter is unique due to its confinement to 140-characters (which was extended to 280 characters in November 2017) and the lack of restrictions on viewing messages from those with whom one is not directly connected to. Connections on Twitter are created based on the content of messages rather than real-life relationships, resulting in ties that span physical and social disparities (Virk et al., 2011). As of today, Twitter and other online SNSs are seen as complements to traditional outreach mediums (Towner and Dulio, 2012; Campante et al., 2013).

The primary benefits of the SNS as a campaign tool include low cost, ability to recruit volunteers and receive contributions, and its accessibility to all candidates, whether well-known or lesser-known (Gueorguieva, 2008). In addition to these benefits, Twitter brings with it new possibilities for candidate-voter interaction as the "@username" function allows candidates to reply directly to other users and promote a dialogue, allowing a candidate to bypass traditional media outlets (Lassen and Brown, 2010).

Given these benefits and the ubiquity of Twitter in campaigns, scholars and pundits have begun studying whether the overall use of social media by politicians actually matters for political outcomes (Kushin and Yamamoto, 2010; Baumgartner and Morris, 2010; Zhang et al., 2010). A number of studies provide correlational evidence on how social media influences campaigns. Metaxas and Eni (2012) comment on the relationship between social media use and electoral outcomes, while Hong and Nadler (2011) demonstrate how the use of Twitter correlates with shifts in polling outcomes during election periods. There are also reported challenges of managing a Twitter account, such as the need for constant monitoring and responding to audience interests (Boyd and Marwick, 2011), absence of authoritative hierarchies (Metzgar and Maruggi, 2009), and possible loss of control over a message (Gueorguieva, 2008; Johnson and Perlmutter, 2010). As the number of Twitter users continues to increase, so does the fraction of them who report using the site to gather political information (Smith and Rainie, 2008; Smith, 2011). According to a University of Oxford and Reuters joint report, in 2017, 56% of the surveyed individuals in the U.S. followed at least one politician on Twitter (Kalogeropoulos, 2017). For politicians, policy makers, and consumers of social media, documenting the causal impact of Twitter with mechanisms at play is essential.

**Media and Incumbency Advantage**

Incumbency advantage is among the best-documented electoral patterns in the United States (Ansolabehere et al., 2006a). Incumbents reportedly enjoyed increasing levels of electoral wins, starting with a 1-2% point advantage in the 1940s and rising to an 8-10% advantage in the 2000s. Explanations for why known or incumbent politicians enjoy an advantage include the incumbents actually being higher-quality candidates (Jacobson and Kernell, 1982), the access to the resources of the office they held (including the staff and committee positions to raise campaign funds)(Cox and Katz, 1996), and the extensive media attention they receive, compared with inexperienced politicians.

Low political competition and incumbency advantage emerge when challengers do not have enough opportunities to inform voters about their candidacy and policy positions (Ansolabehere and Snyder, 2000; Prat, 2002; Strömberg, 2004; Prior, 2006). The persistent advantage enjoyed by experienced politicians over challengers is well-documented. Incumbents are reported to achieve re-election rates at around 90% (Levitt and Wolfram, 1997). They also receive higher levels of media coverage and endorsements, creating additional barriers to entry for new politicians. By documenting how social media can benefit new candidates relative to experienced ones, we complement the literature that documents the positive impact of political competition and lowering barriers to entering politics (Myerson, 1993; Persson et al., 2003; Besley et al., 2010; Ferraz and Finan, 2011; Galasso and Nannicini, 2011). While we recognize that there may be downsides to more competitive political races, we posit that more information about candidates and their policy positions stands to benefit voters.

Traditional media can influence voter decisions through its coverage and candidate endorsements. Voters also favor candidates whom they can recognize (Jacobsen, 1987). Survey-based findings suggest that incumbents enjoy higher media coverage and more frequent endorsements (Goldenberg and Traugott, 1980; Clarke and Evans, 1983; Ansolabehere et al., 2006a), as compared with their opponents. Ansolabehere et al. (2006b) find that endorsements influence the outcome of an election by about 1-5% points. These findings suggest that the experience of a candidate in politics - both due to her public recognition and due to holding a public office - can put new politicians at a disadvantage (Cox and Katz, 1996). Lower incentives for running for an office by entrants translate into less competitive races, which is correlated with lower responsibility and accountability towards constituents by politicians (Carson et al., 2007). These concerns together suggest that new technologies, which can reduce the incumbency advantage, can result in elections to be more competitive. While there are benefits to more competitive elections, they may also bring along less desirable outcomes such as more polarized and negative tones in campaigns and excessive spending for political marketing. Complementing these earlier studies, our study finds that new rather than the experienced politicians have an advantage in opening an account on social media promising to mitigate the incumbency advantage.

## 3 Data

Our study uses data from a variety of sources, each of which we discuss below.

**List of Politicians:** First, using the Federal Election Commission (FEC) database, we compiled the list of all politicians who either registered with the FEC or whose name is mentioned on the state ballot for an election to the U.S. Senate or House of Representatives in the election cycles of 2010, 2012, and 2014.[4] We refined this list based on the Twitter accounts, detailed below.

**Contributions to Politicians:** The data source for the political donations is the FEC database which makes data on campaign contributions for each candidate publicly available.[5] We use data on the contributions to candidates, rather than to PACs or other organizations. In most of our analysis, we limit our attention to donations under $1,000,[6] as larger donations may be motivated by other concerns than supporting a politician (e.g., lobbying efforts). The database details the amount of each contribution, its date, and the name and occupation of the donor as well as her location. We provide detailed summary statistics for the donations and politician characteristics in the online appendix Tables A8 and A9. In our analysis, we aggregate donations at politician-week level.

**Campaign Expenditures:** The source of our data for campaign expenses of a politician is the FEC. The FEC database on disbursements lists the exact amount, date, purpose, payee of each expenditure item by each candidate. Note that by law, for any politician to accept donations they need to be registered by the FEC and have to report all of their expenses with the above details if the payment exceeds $200. Note that recorded expenditures can be negative due to refunds given back to contributors.

---

[4]Elections are held every two years in even-numbered years.

[5]The FEC requires candidates to identify individuals who give them more than $200 in an election cycle. Additionally, they must disclose expenditures exceeding $200 per election cycle to any individual or vendor.

[6]We also study and report the results for donations between $1,000 and $3,000, and over $3,000 in the paper and in Section A.5.

In addition, we record the purpose of each expense labeled based on the 12 categories specified by the FEC: administrative, travel, solicitation and fundraising, advertising, polling, campaign materials, campaign events, transfers, loan repayments, refunds, contributions, donations to charitable or civic organizations. Detailed descriptions of each category can be found in the data appendix, on page A29.

**Twitter Account Data:** For each politician in our initial list from the FEC, we searched for their Twitter handle.[7] We combine an automated script with manual check to gather information about whether a politician has a Twitter account or not, and collect data from her account. We identify the date that the account was first activated and supplement it with data on the number of tweets and retweets, text of the tweets and the number of followers.

Figure 1 demonstrates the Twitter account opening dates for politicians in our initial list between 2007 and 2014. The distribution shows that politicians opened accounts on Twitter continuously between 2009 and 2014 and there was little entry prior to 2009. Variation in entry dates reduces the concern that politicians' entry time may be strategic, coinciding with the timing of specific events such as election years. Moreover, the majority of politicians adopt Twitter in a non-election year, again indicating that their account opening date is not always strategic.

The increase in the donations received by a politician may partially be an artifact of the calls for donations and donation links posted on Twitter. If this is the case, Twitter adoption would increase donations not only due to information about politicians, but because these links act as reminders to donate. Moreover, campaign accounts are typically temporary, as they are opened strategically before an election and closed after the election. To reduce these concerns, we remove these accounts dedicated to campaigns in our benchmark analysis. We drop Twitter accounts with "2010", "2012", "2014" or "4" (e.g., "@chip4congress", "@MCarey2012") in the handle string because use of these numbers tends to indicate that the account was started for a particular upcoming election campaign.[8]

**Twitter Penetration:** We need a measure of Twitter penetration, which would essentially capture the probability with which an average person would see a Twitter message from a political candidate. This probability depends on both the extensive margin (how many people are on Twitter) and the intensive margin (how much time each person spends on Twitter), as it is easier to notice content from a candidate for somebody who uses Twitter more often. One problem is that neither the total number of Twitter users by geography nor the average intensity of using Twitter are observable, as

---

[7] A detailed description of the data collection process is given in the online data appendix.

[8] The results including the campaign accounts can be found in the online appendix Table A11. Inclusion of these accounts does not seem to change our qualitative results.

Twitter does not share this type of data with researchers. Some scholars use geotagged tweets to infer penetration by geography, but given that only 2% of all tweets are geotagged, this approach is likely to lead to skewed estimates (Kinder-Kurlanda et al., 2017). [9]

Our approach to overcome this problem is to use comScore's representative panel of Internet users. Our preferred measure is based on counting the number of visits to Twitter and to all other sites by comScore respondents. To create this measure, we aggregate the total number of Twitter visits by comScore respondents in a state and divide it by the total number of all site visits by comScore respondents in that state, thus obtaining our proxy for a probability that an average Internet user in a state would get exposed to some content on Twitter.[10] More formally, We calculate the average pageviews for each state $s$ and year $y$ as:

$$\text{Twitter Penetration}_{sy} = \left( \frac{\text{Number pageviews on Twitter}_{sy}}{\text{All pageviews}_{sy}} \right)$$

We use 11 other measures of Twitter penetration, 9 of which use the comScore panel (total number of comScore households on Twitter, total comScore duration spent on Twitter, average and median weekly share of time spent on Twitter; average and median weekly share of sessions on Twitter; median share of Twitter pageviews; and average and median weekly proportion of households visiting Twitter) and two measures (proportion of households using Twitter in the last 7 days, proportion of households using Twitter in the last 30 days) using Simmons Oneview, an alternative panel to comScore's. In the appendix, we show that our results are robust to using any of those alternative measures. We also run a population weighted regression with our baseline penetration measure to find similar results. Details on these measures and summary statistics are provided in the online appendix page A27 and the benchmark specification with these alternate measures are provided in Tables A14 and A15.

**Advertising Spending Data:** We collect additional advertising spending data from Kantar Media's Adspender database, dating back to as early as 2009. This dataset covers dollar value of broadcast TV advertising for each week. Notice that the advertising spending values here may differ from advertising campaign expenses for two reasons. First, because the FEC expenses may cover other formats of advertising, second because Adspender reports the spending for the ads broadcast in that week, whereas campaign expenses may reflect a purchase for a future date.

---

[9]Note that the decision to put geotagging to a tweet is not random and can be endogenous to local characteristics.

[10]Note that our measure is highly correlated with the share of Twitter users among comScore respondents in a state (0.70 correlation coefficient), but this correlation is weaker for the average share of Twitter visits among Twitter users (0.59 correlation coefficient). This implies that the variation in our measure comes primarily from the extensive rather than the intensive margin.

**News and Blogging Data:** For each politician in our list, we collect information on the number of media mentions for a window of ten weeks before and after opening an account on Twitter. We run a search for the number of times the politician's name appeared in Google News and Google Blogs. We use this information to check whether there are systematically more media mentions of a politician around the time her Twitter account is started. If there are other events related to a politician's campaign around the time of opening a Twitter account which affect donations, media may also cover them, resulting in higher number of mentions. So using media mentions we can also test for the presence of other contemporaneous events.

**Politician's Personal Data:** We collect data on the personal characteristics of the politicians using two different data sources. The first source is the FEC, and the second is the VoteSmart database, which provides information about their age, education, and voting history. More than 50% of the data is missing on votesmart.org, and when possible, we support the missing information through manual data collection.

In our empirical analysis, we extensively use the classification of politicians into two groups: new and experienced. A politician is classified as a new politician if at the time of opening a Twitter account she had never been elected to the Congress before. If she already won an election in the past, then she is classified as an experienced politician. We present summary statistics separately for the experienced and the new politicians in the online appendix Table A9. Throughout the paper, we check for heterogeneous effects of Twitter adoption by classifying the politicians as new vs. experienced.[11]

**Facebook Adoption:** We also collect data on politicians' adoption of the most prominent competing social network, Facebook. We collected data on the dates of the first public post on Facebook for all the politicians in our list and use them as the date of adopting Facebook. Less than 1% of politicians open an account in the same week on Twitter and on Facebook. We use a dummy variable equal to one if a politician adopted Facebook before joining Twitter, and zero otherwise (for the politicians with a Twitter account).

**State Demographic and Voting Characteristics:** In our analyses, we control for state characteristics including demographics such as household income and population obtained from the Cen-sus. In extensions of the analysis, we also consider correlations with the share of rich (i.e., share of households with over 250K income), share with college education, and share of African-American population using data from the Census. We use the data on the Republican vote share (received by George W. Bush in 2004 Presidential elections) obtained from uselectionatlas.org.

---

[11]We prefer the new vs. experienced classification to the incumbent vs. challenger classification because an experienced politician may end up being a challenger in a future election while still benefiting from having been in the Congress before (e.g, greater name recognition, well-known policy stance, higher coverage by media, etc.) Our classification captures the short as well as the long term incumbency advantage an experienced politician holds.

## 4 Empirical Framework

### 4.1 Identification

Our key empirical hypothesis is that politicians who adopt Twitter see gains in campaign contributions. In the online appendix Section A.3, we formally model how politicians gain from adopting Twitter.[12] Figure 3 demonstrates how political donations evolve in high- and low- Twitter penetration states, controlling only for politician and week fixed effects, before and after Twitter entry.[13] There are two takeaway points from this figure. First, donations indeed seem to increase after joining Twitter, but not before, and, furthermore, this effect is stronger in places with high Twitter penetration. Second, there seem to be no significant differences in donations to politicians between high and low Twitter penetration states before they join Twitter, but there is a visible difference after they join. Overall, Figure 3 illustrates our main point: entry to Twitter seems to help politicians to increase their political donations, and the support is higher in high Twitter penetration places.

Technically, we aim to estimate the following equation:

$$Outcome_{it} = \beta_0 + \beta_1 onTwitter_{it} + \beta_2 onTwitter_{it} \times Penetration_{sy} + \gamma_{pm} + X_{it} + \varepsilon_{it}, \quad (1)$$

Here $Penetration_{sy}$ is Twitter penetration at state $s$ in year $y$, $\gamma_{pm}$ is a politician-month fixed effect, while $X_{it}$ is a vector of controls[14]. To identify $\beta_2$, we need an assumption that the error term is not *differentially* correlated with unobserved factors. For example, even if some unobserved campaign activity is present and is changing sharply at the time of the Twitter entry, it should not be a problem

---
[12]We analyze two different channels through which Twitter could affect the behavior of donors. An information channel implies that opening a Twitter account allows the politicians to access a new and relatively inexpensive channel of communication with their constituency. For donors who do not know about a candidate or are uninformed of her policies, this creates awareness. A persuasion channel could allow potential donors who already know the candidate to get repeated exposure to information via Twitter and persuade them to contribute more.

[13]Note that our baseline empirical specification is more saturated. To construct this figure, we use practically raw data, for the purpose of illustration.

[14]In the paper, we present the baseline results for both a specification without and with controls in which state-level controls interacted with being on Twitter are included.

for our estimation, as long as this activity does not differentially affect political donations in high- and low- Twitter penetration states. Formally, the following assumption should hold:

**Assumption 1.** $corr\left(onTwitter_{it} \times Penetration_{sy}, \varepsilon_{it} | onTwitter_{it}, X_{it}, \gamma_{pm}\right) = 0.$

Under this assumption, we can correctly estimate the differential impact of joining Twitter in high- vs low- Twitter penetration states, even though the estimate of the direct effect of Twitter, $\beta_1$, or of the full effect of Twitter, $\beta_1 + \beta_2 \times Penetration_{sy}$ could be biased. It is, in fact, a parallel trend assumption, applied to our specific empirical framework.[15]

In more detail, the expected bias of the OLS estimate in equation (1) is

$$E\begin{pmatrix}\hat{\beta}_0\\\hat{\beta}_1\\\hat{\beta}_2\\...\end{pmatrix} = \begin{pmatrix}\beta_0\\\beta_1\\\beta_2\\...\end{pmatrix} + (X'X)^{-1}\begin{pmatrix}E\left(\varepsilon_{it}|X_{it},\gamma_{pm}\right)\\cov\left(onTwitter_{it},\varepsilon_{it}|X_{it},\gamma_{pm}\right)\\cov\left(onTwitter_{it} \times Penetration_{st},\varepsilon_{it}|X_{it},\gamma_{pm}\right)\\...\end{pmatrix},$$

Assumption 1 is needed to consistently identify $\beta_2$ without bias in the expectation. At the same time, to identify both $\beta_1$ and $\beta_2$ we need a stronger assumption that the error term needs to be uncorrelated with unobserved determinants of decisions to join Twitter, conditional on fixed effects and other observables. [16] [17]

Note that, while we have regions with varying levels of Twitter penetration, we do not have

---

[15] Note that the difference-in-difference analysis with staggered adoption was recently used in several papers, including Athey and Imbens (2018) and Xiong et al. (2019) who provide a more general discussions related to this approach.

[16] Technically, it is

**Assumption 2.** $\begin{cases} corr\left(onTwitter_{it},\varepsilon_{it}|X_{it},\gamma_{pm}\right) = 0 \\ corr\left(onTwitter_{it} \times Penetration_{st},\varepsilon_{it}|X_{it},\gamma_{pm}\right) = 0 \end{cases}$

This is a strong condition, which holds if the so-called "precise timing" assumption is satisfied, i.e., if other potential determinants of the outcome of interest move smoothly around the time of a discrete change of the Twitter entry (see also Gentzkow et al. (2011) for this argument applied to newspapers entering the market). However, some unobserved high frequency campaign activity could violate this assumption.

[17] More formally, let's assume that there is some other unmeasurable activity that happens during the same time as joining Twitter, $a_{it}$, such that $corr\left(a_{it},\varepsilon_{it}\right) \neq 0$, but $corr\left(a_{it} \times Penetration_{it},\varepsilon_{it}\right) = 0$ (i.e., assumption 1 still holds). As before, we estimate:
$Outcome_{it} = \beta_0 + \beta_1\left(TrueTwitter_{it} + a_{it}\right) + \beta_2\left(TrueTwitter_{it} + a_{it}\right) \times Penetration_{it} + \gamma_{pm} + \varepsilon_{it}$ where $TrueTwitter_{it}$ is a quasi-random component of joining Twitter. Then the estimated coefficients are given by

$$E\begin{pmatrix}\hat{\beta}_0\\\hat{\beta}_1\\\hat{\beta}_2\\...\end{pmatrix} = \begin{pmatrix}\beta_0\\\beta_1\\\beta_2\\...\end{pmatrix} + (X'X)^{-1}\begin{pmatrix}E(\varepsilon_{it}|\gamma_{pm})\\cov\left(onTwitter_{it}+a_{it},\varepsilon_{it}|\gamma_{pm}\right)\\cov\left([onTwitter_{it}+a_{it}] \times Penetration_{st},\varepsilon_{it}|\gamma_{pm}\right)\\...\end{pmatrix}$$

thus $\beta_1$ would be biased, while $\beta_2$ would not.

a proper control market where Twitter penetration is exactly zero. Here we implicitly make the assumption that, using the interaction term, we can linearly extrapolate to markets with no penetration. This assumption is not far from reality as the Twitter penetration scores are indeed very close to zero in some markets (e.g. Wyoming or Montana). Under this assumption, $OnTwitter$ dummy captures any activity that happens when politicians join Twitter, but that is uncorrelated with penetration (see the discussion in footnote 16 above).[18]

Note also that in principle the decision to join Twitter, $onTwitter_{it}$, could be a function of a (slowly changing) Twitter penetration rate without violation of Assumptions 1 and 2, as long as Twitter penetration rate, $Penetration_{sy}$, available in our data at state-year level, is perfectly collinear with politician-month fixed effects.

In what follows, we are going to use the most conservative approach to estimate 1, thus we rely on Assumption 1. Using this, we focus on estimating the differential impact of joining Twitter in high- vs low- Twitter penetration states. We will talk more about the value of Twitter for politicians in Section 5.3.

### 4.2 Identification checks

Because specification in (1) controls flexibly for time-invariant and time-variant characteristics of the politicians and their states, the main threat to identification is contemporaneous unobserved marketing activity of these politicians, which, deliberately or by chance, could lead to the jump in donations at the same week when these politicians joined Twitter, and more so in the states with higher Twitter penetration.

We unfortunately cannot directly test the identifying assumption and check whether joining Twitter coincides with some unobservable factor. Instead, we conduct two exercises with the aim of strengthening our identification argument. First, we check if the observable potential drivers of donations are systematically related to joining Twitter, differentially in high- vs low- Twitter penetration states by providing placebo or 'balance' tests (see Pei et al. (2019)). Second, in the online appendix, we consider the potential bias due to unobservables, that is, if they are positively correlated with observables, using the Altonji-Elder-Taber framework (Altonji et al., 2005). We estimate how large the selection on unobservables has to be in order to explain our findings.

We report the results of the placebo tests which test if joining Twitter indeed coincides with some other measures of activity in Section 5.4. These tests check if when politicians adopt Twitter there is a differential activity of higher news and blog coverage, higher spending on TV advertising or other types of campaign activities in high- and low- Twitter penetration regions. Figures 4, 5, and 6 illustrate how the news coverage, blog mentions and advertising spending change around the week of Twitter adoption in high- and low- Twitter penetration regions. Figures do not show significant changes in these activities before and after Twitter entry for different penetration regions, and this finding is consistent with our identifying assumptions.

---
[18] We thank the Associate Editor for suggesting this point.

## 5 Empirical results

### 5.1 Baseline results

We present the results from the specification given in (1) for aggregate donations in Table 1, for donating at least once in Table 2, and for the number of donations in Table 3. Table 1 demonstrates a positive and significant impact of the interaction of Twitter and Twitter penetration on aggregate political donations for all politicians. In the tables, we also report the implied effect of joining Twitter by comparing the gains in higher Twitter-penetration states relative to the lower-penetration states, where high- and low- are defined as the 75th and the 25th percentile in state penetration levels, respectively. This difference is given in the last two rows of all the tables and separately for the beginning (2009) and end year (2014) in our data. We do not report the aggregate effects of Twitter at any particular level of Twitter penetration (e.g. mean or median), as under Assumption 1 we cannot guarantee that those numbers are estimated consistently.[19]

The results in these tables imply that for a simple specification without additional controls, the differential impact of joining Twitter ranges from 2.9% weekly increase for 2009 to 22.8% weekly increase for 2014 (column 1, Table 1). These numbers describe the average increase in donations during the month of joining Twitter; and we discuss potential ways to aggregate effects over the course of the campaign in Section 5.3. Even-numbered columns in Tables 1, 2 and 3 report the results conditional on two additional controls; joining Twitter interacted with population and median household income. We add these controls since the impact of joining Twitter could be higher in larger markets, where potential donors are richer, and we do not want our coefficient of interest to pick up those relationships. Nevertheless, one can see that in all three tables and specifications, adding these controls does not change the coefficient of interest largely, and increases it rather than decreasing it, which is consistent with no unobserved heterogeneity explaining away our results (Altonji et al. (2005)). The corresponding results with controls for these interactions are indeed slightly more precise in most cases.[20] The direct coefficient for joining Twitter is mostly insignificant, consistent with the prediction that joining Twitter should not affect markets with very small penetration.[21]

---

[19] Another potential problem with the estimates of the aggregate effect is that those are based on linear interpolation to the markets with zero penetration. While doing this seems reasonable for the earlier years, it could be problematic for the later years in our time period.

Tables 1, 2, and 3 also report an important dimension of heterogeneity of the results, as joining Twitter seems to help new politicians, who were never elected before (columns (3) and (4) in Tables 1, 2, and 3), without much of the effect for the experienced politicians, who were elected at least once before deciding to open a Twitter account. More specifically, for the new politicians the differential impact of joining Twitter on weekly donations during the first month after entry ranges from 4.7% for 2009 to 41.4% for 2014 (column 3 of Table 1). In contrast, the interaction coefficients for experienced politicians are negative, though not statistically significant (columns (5)-(6) of Table 1), with the absolute value of the coefficients being at least 4 times smaller than that for the new politicians.[22] Overall, these results are consistent with the finding that social media is helpful for the candidates who are lesser-known, but not for the more experienced politicians. The results of the estimation for the probability of receiving at least one donation (Table 2) and for the number of donations (Table 3) are largely consistent with the results for the aggregate donations, with the effects for the new politicians, again, being the largest and most precisely estimated. To derive the implications of these coefficients in a detailed manner taking the observed decline of the Twitter effect over time in Section 5.3.

## 5.2 Baseline Results with Varying Window Specifications

The results we discussed so far do not allow us to interpret the magnitudes beyond the month when politicians open their Twitter account. Politician-month fixed effects in equation (1) absorb any effect after the new month starts. Interpreting the magnitudes given in Tables 1, 2 and 3 for the months in a campaign following the month of opening a Twitter account thus requires making strong assumptions. One way to deal with this problem is to use less restrictive fixed effects such as politician fixed effects. The primary reason of this section is to provide some alternative estimates to understand the magnitudes.

---

[20]In online appendix Table A16, as an illustration, we provide the coefficients of the interactions with controls and add the controls one by one for log(amount of donations).

[21]Note that these results should be interpreted with caution as, under Assumption 1, we cannot consistently estimate the direct effect. Moreover, as we do not observe markets with zero penetration in the data, interpreting the direct effect as one in a market with zero penetration is based on interpolation.

[22]Note that the direct coefficient for joining Twitter is marginally significant for experienced politicians, though it loses the significance and changes its sign once controls are added.

In this subsection, we offer an alternate way to estimate the differential impact of Twitter using politician fixed effects, in a way that is qualitatively similar to the results in Figure 3. Specifically, we report what happens before and after a politician opens a Twitter account for windows of 5,6,7,8,9,10, and 20 weeks after opening and 8 weeks before the opening. We estimate the following window specification:

$$Outcome_{it} = \beta_0 + \beta_1 onTwitter_{it} + \beta_2 onTwitter_{it} \times Penetration_{sy} + \delta_i + X_{it} + \varepsilon_{it}$$

for $t_0 - 8 < t < t_0 + k$ time window, where $\delta_i$ is a politician fixed effect.[23]

These results, presented in Table 4, show that the coefficients for joining Twitter interacted with Twitter penetration are consistent with our main results. The impact of Twitter on aggregate political donations is particularly strong for new politicians, but much smaller in magnitude and not statistically significant for the experienced ones. The magnitudes in columns 1 and 2 imply that the differential impact of Twitter is 2.4% in 2009 and 20.9% in 2014 for the sample of all politicians (column 1 of Table 4) and from 3.5% to 28.7% for new politicians (column 2 of Table 4) for the same years.[24] These magnitudes are slightly smaller than the ones in Table 1. To the extent that we trust both estimates, we can use the comparison of coefficients in Table 1 and Table 4 to understand the decline of the Twitter effect over the course of the campaign.

### 5.3 Discussion of Magnitudes

The results from Table 1 imply that the differential impact of Twitter on weekly donations ranges from a 2.9% increase for 2009 to a 22.8% increase for 2014, when using a specification with politician-month fixed effects, all based on the donations below $1,000. In this section, we discuss what magnitudes these numbers could imply, referring to the differential Twitter effect in high- vs low-Twitter penetration states (75th vs. 25th percentile). Note that under Assumption 1 we are only able to consistently estimate the differential effect of onTwitter, in contrast to full aggregate effect of joining Twitter at some particular level of Twitter penetration.[25]

---

[23]We find that our placebo tests related to campaign expenditures for new politicians are still valid for an 8-week window specification after which they fail consistently. Hence, the assumption of quasi-random entry on Twitter appears to be still plausible in this window.

[24]Note that this approach delivers results quantitatively similar to our baseline results (Table 3) if we focus on 2 weeks window instead of 8 week window, so our two approaches are consistent with each other. We provide the results with various window lengths in Tables 4 – 7 using a pre-period of 8 weeks and then varying the window length after.

[25]Comparing 25 percentile with 75 percentile corresponds to the comparison of median penetration below the median vs median penetration above the median, which is in line with the descriptive evidence in Figure 3. Using the same approach, one can compute the differential effects for other pairs of percentiles.

Note that the average weekly donation amount in the months the politicians join Twitter is $1,534 (Table A8). On average, politicians join Twitter in the second week of the month and are active for 2.79 weeks. Based on Table 1, the average differential effect of Twitter ranges from $1,534 \times 0.029 = \$44$ per week in 2009 to $1,534 \times 0.228 = \$350$ per week in 2014, amounting to $\$44 \times 2.79 = \$124$ gain for the month of opening an account in 2009 and $350 \times 2.79 = \$976$ for 2014.

A lower bound of the estimated impact of Twitter can be calculated by assuming that the effect of Twitter disappears at the end of a short period. The results in Table 4 provide us with the opportunity to understand how the implied effect of Twitter changes across the 8 weeks after opening the account. Using the implied Twitter effect for 2009 and 2014 from column (1) of this table indicates a gain of $\$1,534 \times 0.024 \times 8 = \$294$ in 2009 and $\$1,534 \times 0.209 \times 8 = \$2,565$ in 2014. These numbers are the lower bound of the Twitter effect under the assumption of zero effect after these 8 weeks. The latter number suggests that the implied effect of Twitter corresponds to at least 0.7% of all donations received by an average House candidate over the course of the campaign and 1.7% of all donations of $1,000 and below.[26] These numbers constitute, on average, 1% of all donations to new House candidates and 2.7% of all donations below $1,000 over the course of the campaign. [27]

Note that these 8 weeks consist of, on average, 2.79 weeks from the first month, which our fixed effects estimation is based upon, and the remaining 5.21 weeks. Thus, from these numbers, we can deduce that the average differential weekly increase associated with Twitter after the first month of data was ($294-$124)/5.21= $33 per week for 2009 and ($2,565-$976)/5.21=$304 per week for 2014. The weekly magnitudes become smaller after the first month of donations.

An upper bound of the effect can be calculated assuming that the effect of Twitter remains at the same level till the end of the campaign period as it did in the second month. On average, politicians open Twitter accounts about 25 weeks before the elections. For this period, the implied gain translates to $124 (the effect during the first month) + $33×(25 − 2.79) =$857 (the effect during the remaining campaign period) in 2009 and $976 + $304×(25 − 2.79) = $7,728 in 2014. Based on the latter number, the upper bound on the implied effect of Twitter is 2% for all donations received by an average House candidate over the course of the campaign, and 5.4% for all donations under $1,000. Similarly, for a new politician, the upper bound calculated corresponds to 3.1% of all campaign donations to new House candidates and 8.2% of all donations below $1,000.

As an additional exercise to put our estimates in perspective, we compute persuasion rates

---

[26] The average total donation received by a House candidate over a two-year campaign is $386,877 and $143,611 for donations under $1000.

[27] The average total donations to a new House candidate over a two-year campaign is $249,810, and $94,849 for donations under $1,000.

(DellaVigna and Kaplan, 2007; DellaVigna and Gentzkow, 2010) (see online appendix A.1). The main takeaway from the exercise is that the persuasion rates implied by our estimates are remarkably similar to those in political advertising (Spenkuch and Toniatti, 2016).

## 5.4 Placebo Tests

Our identifying assumption is that, conditional on politician-month fixed effects and other controls, entry on Twitter across states of high- and low- Twitter penetration is not related to unobserved heterogeneity in donations. In other words, we assume that within a given month for a particular politician the exact timing of joining Twitter is good as random across areas of high- and low- Twitter penetration. This assumption is violated if some other fundraising or marketing activity coincides in time with Twitter adoption differentially across areas with different levels of Twitter penetration. While we cannot test this assumption directly, we conduct a number of tests to ensure that at least the observable potential determinants of donations cannot explain away our results. In this subsection, following the theoretical arguments by Pei et al. (2019), we carry out 'placebo' or 'balance' tests to analyze based on observables whether there are any obvious violations of our identifying assumption. Pei et al. (2019) argue that a 'balance test' approach, which puts the observables on the 'left hand side' as the dependent variable is a helpful exercise similar to those used in randomized trials comparing baseline or pre-treatment characteristics.[28]

**Campaign Expenditures**

A potential threat to our casual claim is the possibility of a correlation between the timing of Twitter entry and other unobserved marketing activities. While unobserved marketing activities are, by definition, the activities for which we do not have data, the campaign expenditures of politicians can be a reasonable proxy for them. When and how campaign funds are spent are reported to FEC in detail (please see Section 3 and the data appendix), and the FEC classifies campaign disbursements into 12 categories. The categories most relevant to marketing activities include advertising expenses, campaign material expenses, fundraising expenses, and travel expenses (which may include town visits on the campaign trail). We estimate a placebo specification using the weekly expenditure in each category (as well as the total expenditure) as the dependent variable. One concern with the expenditure data is how noisy it is, however, we find that the weekly total campaign expenditure is highly correlated with the aggregate campaign contributions received in the same week, as demonstrated in Table A17 for each expense category. This reduces the concern that the expenditure data contains high levels of noise.

---

[28] In the online appendix Section A.2, we compute tests based on methods proposed by Altonji et al. (2005) to quantify how large the impact of unobservables would have to be relative to selection on observables in order to fully explain our results. Using this approach, we find that to attribute our entire difference-in-differences estimates to selection effects, selection on unobservables would have to be at least 10.4 times greater than the selection on observables and, on average, over forty-seven times greater. This strengthens confidence in our estimates. The coefficient ratio tests are provided in Table A1 in the online appendix for a variety of specifications.

Table 8 provides the placebos with campaign expenditures in different categories. If there are other unobserved marketing activities that coincide with opening a Twitter account and these activities vary across the high- and low- Twitter penetration regions, then such activities may pose an alternate explanation to the effect we attribute to Twitter. Reducing this concern, we find that the coefficient of entry on Twitter interacted with Twitter penetration is statistically insignificant for all listed expense categories as well as the total expenses. The differential effects of Twitter across high- and low- penetration areas are small in magnitude and never statistically significant. We also report the baseline results with inclusion of contemporaneous and lagged campaign expenditures in Table A18. To the extent that campaign expenditures capture other activities of the politicians the same week they open a Twitter account, these findings are consistent with the causal interpretation of results in Tables 1-3.

**Political Advertising**

We next test for the potential simultaneous increase in advertising activity by the candidates using a second dataset from Kantar Media's AdSpender database. We check if there was a differential increase in the political ad spending around the time a candidate joined Twitter, for high- vs low- Twitter penetration states. Columns (1)-(3) of Table 9 presents these results. The results are consistent with joining Twitter not being associated with an increase in political ad spending after controlling for politician-month fixed effects. The interaction term and the direct effect are insignificant for all and new politicians. For the experienced politicians (column 3), the interaction term is negative, though statistically insignificant. Thus, in all samples joining Twitter is not associated with an increase in political advertising spending, and, therefore, a spurious relationship between opening a Twitter account and political advertising cannot explain our results.

Note that we do not observe digital advertising which may include social media advertising, email marketing, search advertising, display advertising, and our advertising spending placebo does not address these activities. We do not, however, see a significant increase in the disaggregated campaign expenditures (which include categories such as advertising and campaign materials) in the week a politician joins Twitter. If the campaigns allocated extra funds on digital marketing at this time, as payments for digital ads are typically billed within days, these efforts would likely be captured in the weekly campaign expenses.

**News and Blogs Coverage**

Next, we test if the timing of adopting Twitter may coincide with other external events reported on media, possibly as part of a larger PR campaign. Media mentions of a politician capture both additional information shocks voters receive and the events a politician is involved in, which may drive donations independently of Twitter. To address this concern, we collect data on the mentions of a politician in traditional and social media. We run a search for each politician's name in Google News and Google Blogs for a 10-week window around the time of opening a Twitter account.[29]

Columns (4)-(6) of Table 9 reports a placebo test which uses the logged weekly news reports of a politician as the dependent variable. Overall, the estimates suggest that opening a Twitter account is not correlated with the differential number of news articles about a politician in the overall sample (columns (4)) as well as new (columns (5)) and experienced (columns (6)) politicians across areas with high- and low- Twitter penetration.

Columns (7)-(9) of Table 9 reports a placebo specification with the logged number of blog mentions. The effect of adopting Twitter is not significantly correlated with the number of blog posts for the overall sample. For new politicians (columns (8)), the interaction term is negative, rather than positive, and marginally significant. This negative sign may suggest, for instance, a displacement of politician-related blog content from the more traditional blogging platforms to Twitter after a politician herself joins Twitter. Nevertheless, as the relationship is negative, the coverage on other blogs cannot explain away the effect of Twitter adoption. Note that Google Blogs data used does not include Twitter and Facebook.[30] [31]

---

[29]We search for the full name of the politician and record the number of hits we find on Google News and Google Blogs.

[30]We also checked for politicians' Facebook account opening dates. We find that only 3 politicians in our sample opened a Facebook account in the same week of and only 9 opened a Facebook account within 4 weeks of opening a Twitter account. Therefore, opening of a Facebook account does not seem to be coordinated with opening a Twitter account in time. Moreover, we find no robust relationship between having a Facebook account before opening a Twitter account, reported in Table A13.

[31]We also checked that our baseline results are robust to the inclusion of any of the placebo variables as controls, instead of running placebos. Results are available upon request.

# 6 Mechanisms

Our main findings suggest that a politician's adoption of Twitter causes an increase in the aggregate donations s/he receives. We consider two potential mechanisms driving donations: information and persuasion. Adoption and activity on Twitter may help a politician to increase awareness about her candidacy and policies, which in turn can increase her support from the electorate. Intuitively, we expect the gains to be higher for the new politicians, compared with the experienced ones, since experienced politicians' policy positions and candidacy are often better known. Alternatively, adopting Twitter and communicating through it may mainly raise donations by persuading donors who are already aware of the name and policy positions of a politician by encouraging them to donate more. Through either mechanism, the donations raised by a politician can be expected to increase after Twitter adoption. If information is the main channel, however, we expect the effect to be more pronounced for the new politicians and for the donations from first-time donors. Similarly, gains from being on Twitter are expected to be higher for the in-state, compared to the out-of-state donations, and for the House vs. the Senate candidates.

Our baseline findings demonstrate that social media raises donations only for the new politicians and not for the experienced ones. Our theoretical framework given in the online appendix and our main results are consistent with the information mechanism, proposed above. With this information mechanism, the marginal return to information provision through Twitter is likely lower for the experienced candidates since their quality, experience, and policy positions are better known. For a political newcomer, there is more new information to share.

In this section, we present a number of additional tests that allow us to provide some evidence in line with the mechanisms that our data is consistent with. First, we check whether our estimates are stronger for new vs. repeat donors. We classify each donor as new if no donor with the same first and last name has contributed to a particular Congressional candidate before. Second, we test if the effect is driven by the politicians, who did not have Facebook account before, or by politicians, who joined Facebook before Twitter and are thus more experienced with the social media. Third, we check if the candidates for the U.S. Senate, who run at state-level elections and have better name recognition, gain more than the candidates for the U.S. House of Representatives, who are elected from smaller districts. Finally, we also analyze tweeting activity by politicians to document how differences in tweeting activity and content of tweets affect donations.

**New vs. Repeat Donors**

A politician's presence on social media has two possible ways of influencing donors. First, it is possible that a politician's presence simply changes the amount individuals contribute, without altering the actual donor population. Second, in line with the information channel, it is plausible that Twitter helps politicians to expand their donor base, with some new donors hearing about and contributing to the campaign for the first time. In this section, we investigate if the behavior of new and repeat donors respond differentially to opening a Twitter account, estimating Equation 1 separately for these groups of donors.

These results in Table 10 show that opening a Twitter account increases the amount and the number of donations received by politicians, more so in high-Twitter penetration states. This effect is especially pronounced for new politicians and is negative but not significant for experienced politicians. Numerically, the magnitude of the effect is a 2.8% increase in 2009 and a 22.2% increase in 2014 for all politicians when considering aggregate donations from the new donors (column 1). For the new politicians, the implied effect for donations from new donors is 5% and 44.8% for the same years (column (2)). In contrast, for the donations from repeat donors, we do not see any significant increase in donations, and the corresponding coefficients for the full sample (column(4) in panels A and B) are small and negative.[32]

**Politicians with Facebook Accounts Prior to Joining Twitter**

The information mechanism implies that the impact of Twitter might be smaller for politicians who already use some other social media, such as Facebook. In this section, we test whether our results are substantially different for politicians who joined Facebook before Twitter, as compared with those for whom their Twitter account is their first social media account. These results are summarized in Table 12. As one can see, there was no significant impact of opening a Twitter account for those who had a Facebook account before. The coefficients are negative, but not statistically significant.[33] In contrast, in Table A12, we show that excluding politicians, who had a Facebook account before,

---

[32] The difference between the aggregate donations from new and repeat donors is statistically significant in seemingly unrelated framework for all and new politicians but not for the experienced ones. We report the associated p-values in Table 10 notes.

[33] In these estimations, the sample of politicians who joined Facebook before Twitter is relatively small. Thus, in these results we cluster only at state level, because in this restricted sample we do not have enough clusters to compute standard errors. We control for week fixed effects instead of week of month fixed effects, to ensure that standard errors could be computed. Due to power issues, we focus on the full sample.

leads to the results qualitatively similar to the baseline. [34] [35]

**House and Senate Candidates**

In this section, we compare the gains for the candidates running for the Senate and the House of Representatives. The name recognition of candidates running for the Senate is generally higher, as compared to the candidates running for the House of Representatives, as all states are represented by only two senators, but often by a higher number of representatives. Moreover, senators are appointed for a six year term, compared to the two year term of a candidate elected to the House. Thus, we expect the average candidate for the House to obtain higher gains from communicating via Twitter compared to the average candidate for the Senate.

Focusing on the sign and magnitude of the interaction effect, we find the results, reported in Table 11, to be in line with this prediction. More specifically, the interaction effect of Twitter is positive and significant for the full sample and new House candidates for aggregate donations (columns 4 and 5 of panel A) and for the number of donations (column 5 of Panel B) in Table 11. The effect for the average House candidate ranges from 3% for 2009 to 24.8% for 2014. For the experienced politicians, who are part of the House, we see a negative, but statistically insignificant result, consistent with our previous findings. In contrast, all the estimates for Senate candidates (columns 1-3 of panels A and B) are small in magnitude and are far from being statistically significant.[36] [37] [38]

**Tweeting Activity and Tweet Content**

In this section, we study whether the gain of new politicians is correlated with their tweeting activity and specific content. To this end, we focus on the coefficient of the triple interaction term between joining Twitter,[39] Twitter penetration, and (various measures of) tweeting activity. We consider a 25-week window around politicians' adoption of Twitter.

---

[34] The results reported in Table A23 of the online appendix demonstrate that our Twitter findings hold more generally. Looking at the interaction effect between Facebook penetration and Facebook adoption shows a statistically significant increase in the amount and number of weekly donations and this effect holds at the 1% level of statistical significance, though only for new politicians and not for experienced ones. Overall, these results suggest that our Twitter estimates have external validity and can be viewed more generally as representative of the impact of social media adoption on political donations.

[35] If we run a triple interaction estimation, we find the differences in the amount and number of donations to be statistically significant for new politicians.

[36] These estimates are very similar without controlling for interactions of being on Twitter with market size and income.

[37] The difference between the aggregate and number of donations to House vs. Senate candidates is statistically significant for new politicians but not for the experienced ones. We report the associated p-values in Table 11 notes.

[38] It is computationally challenging to estimate triple interaction specification with standard errors clustered two ways at state and week levels, so despite our relatively large dataset, we feel that we might face power issues here.

[39] Note that it is computationally challenging to estimate triple interaction specification with standard errors clustered two ways both at state and week levels. Thus despite the large dataset that we have, the confidence intervals can be quite large.

The results from the analysis of tweeting activity are given in online appendix Tables A21 and A22. In sum, we find that donations go up for politicians who post more original tweets, rather than retweets, use more hyperlinks, use more anti-establishment related words or appear to be more "plugged in." We use a psychological approach to text analysis using the Linguistic Inquiry and Word Count methods (Pennebaker et al., 2015) which analyze the use of adjectives and pronouns to assess personality traits of individuals using these words. The scale that we use is developed by James Pennebaker and is intended to measure personal characteristics of an individual. We find no differential impact for politicians with various psychological traits, but we find some marginal increase for those politicians, who are "plugged in," i.e., those who have a social style related to staying informed about recent news and developments. However, these results should be interpreted with caution, as they use all subsequent Twitter posting behavior of a politician to assess the impact of joining Twitter during the first weeks after opening an account.

**Within and Out-of-State Donations**

An implication of using a state-level Twitter penetration measure is that we expect the residents of a state to be the ones predominantly donating to the candidates running for an office from the same state. To validate the use of the within state Twitter penetration measure, we compare if donations coming from residents within and out of state respond differently to a politician's Twitter entry.

The results in Table 13 demonstrate that this intuition is supported by the data. The table shows that the within state estimates are statistically significant and positive across all (columns (1)-(2)) with the strongest effect for those who are new (column (3)-(4)). Quantitatively, the magnitudes are comparable but smaller than the baseline results. The estimates are again insignificant for the experienced politicians (column (5)-(6)). For out-of-state donors in Table 13, we find that the Twitter entry interacted with Twitter penetration is statistically insignificant for the overall sample and only marginally significant for new politicians, with significantly smaller coefficients in magnitude for the out-of-state compared to the in-state donations.

Overall, the results in Tables 10, 11, 12, and 13 are consistent with information rather than persuasion mechanisms, implying that Twitter helps politicians to make their candidacy and policy positions more visible, especially for the new politicians who were never elected before. In addition, Tables A21 and A22 in the online appendix are consistent with the theoretical prediction that using Twitter more informatively is associated with a greater increase in donations received following opening a Twitter account.

# 7 Conclusion

Electoral campaigns, from data collection and voter targeting to advertising and political communication, are among the most sophisticated and costly marketing efforts. A notable change in these efforts during the past decade in these efforts is the intensified use of social media platforms to reach out to and inform voters, partially eliminating dependence on traditional media outlets such as newspapers and television. The essential question is whether the adoption of social media technology alters any aspects of political competition or electoral politics. More broadly, can innovations in information technologies change the way markets operate? In this study, we document that entry on social media (Twitter, Facebook) can help to increase funds and attract new donors for new politicians. Overall, results imply that social media can intensify electoral competition by reducing the barriers for entrants to raise money.

Many avenues of future research lie at the intersection of adoption of new communication technologies and marketing. Future studies may expand the findings from our study to investigate whether the same trends follow for brands, specifically if the availability of new communication technologies such as Twitter can help new brands to ward off competition from incumbents. Similarly, studying the extent of substitution between new and traditional media channels is at the core of marketing and advertising and scholars may be interested in documenting whether traditional and social media are complements or substitutes in delivering brand and product information to consumers. If they are, readers may gain from learning how these new technologies alter the way consumers perceive products or form their consideration sets.

13(1):19.

Xiang, Y. and Sarvary, M. (2007). News consumption and media bias. *Marketing Science*, 26(5):611–628.

Xiong, R., Athey, S., Bayati, M., and Imbens, G. (2019). Optimal experimental design for staggered rollouts. *arXiv preprint arXiv:1911.03764*.

Yildirim, P., Gal-Or, E., and Geylani, T. (2013). User-generated content and bias in news media. *Management Science*, 59(12):2655–2666.

Zhang, W., Johnson, T. J., Seltzer, T., and Bichard, S. L. (2010). The revolution will be networked the influence of social networking sites on political attitudes and behavior. *Social Science Computer Review*, 28(1):75–92.

Zhu, Y. and Dukes, A. (2015). Selective reporting of factual content by commercial media. *Journal of Marketing Research*, 52(1):56–76.
35

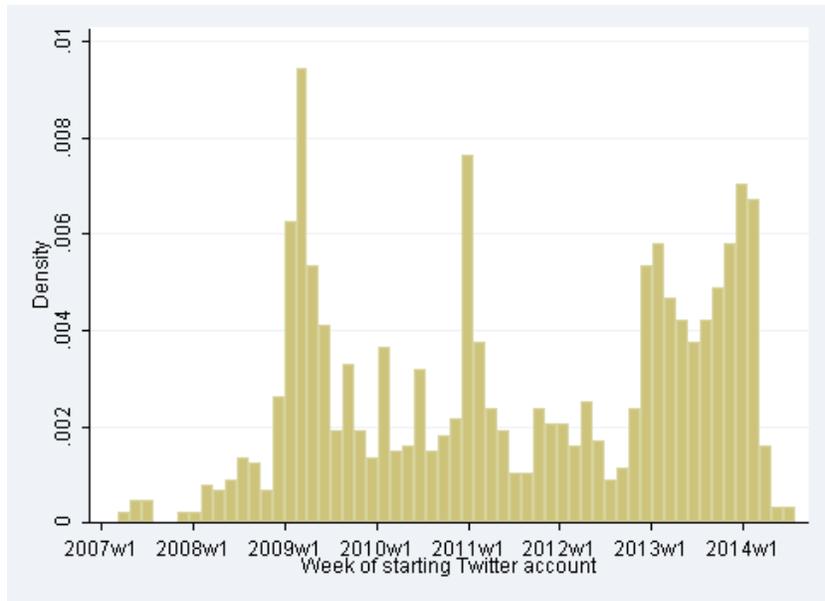

Figure 1: Dates (week) of Opening an Account on Twitter

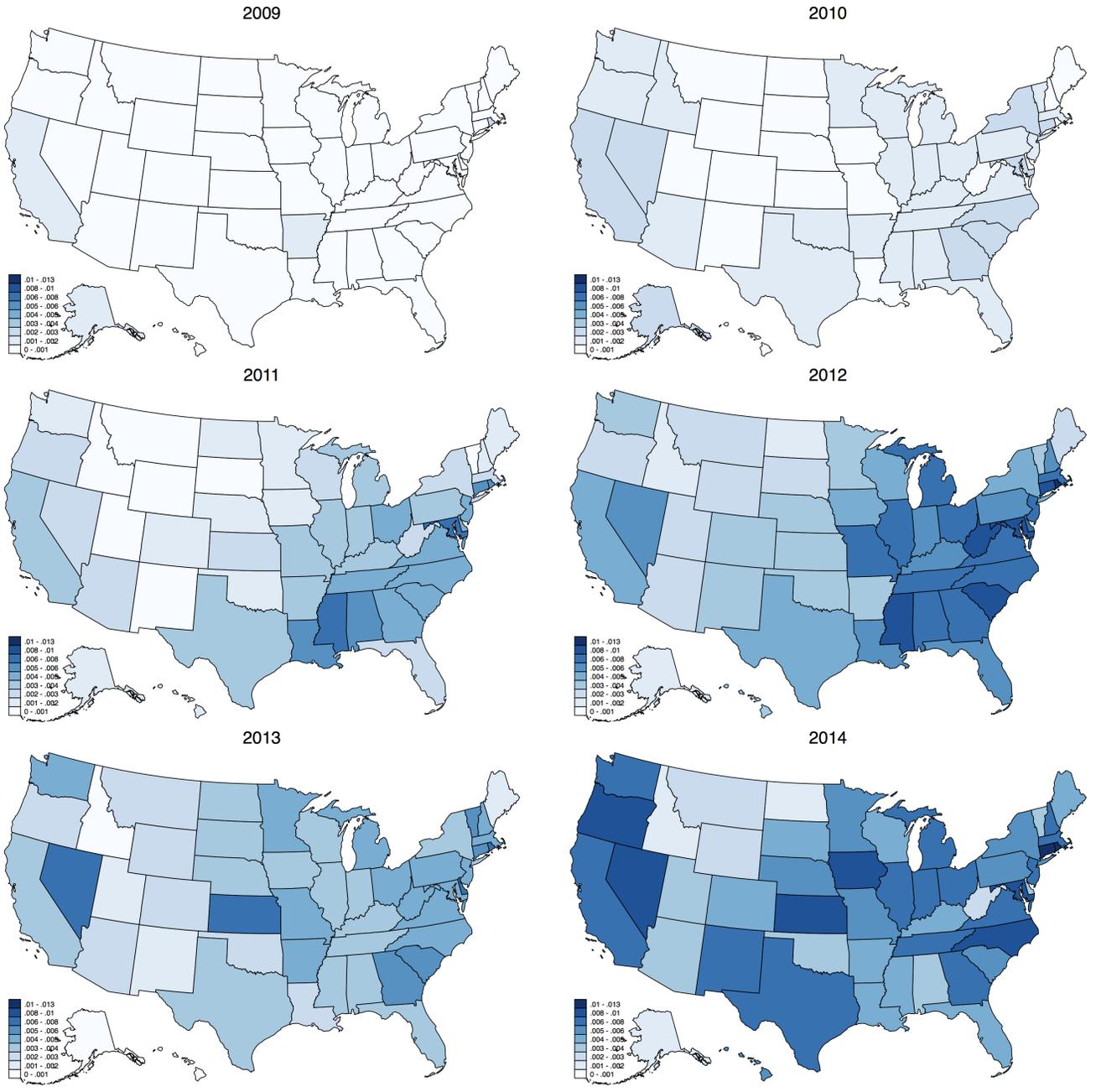

Figure 2: Twitter Penetration Over the years, In absolute numbers

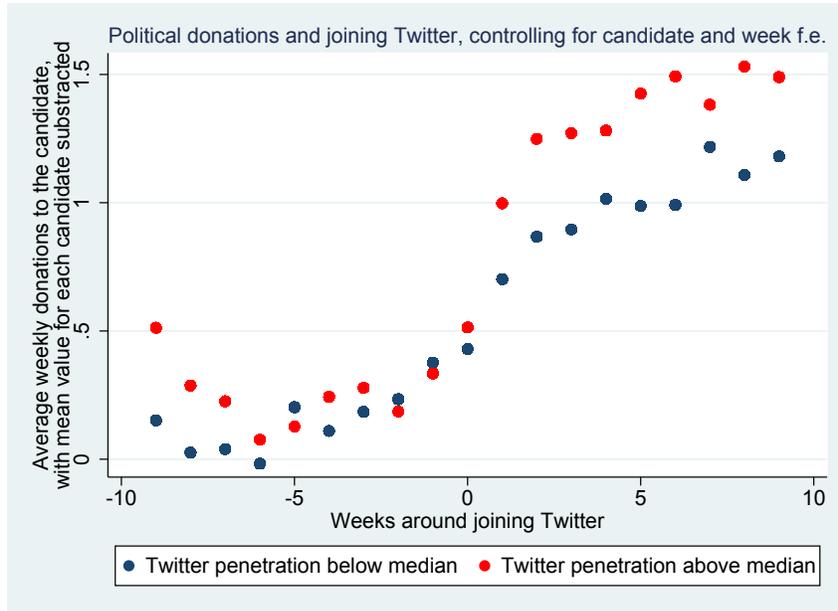

Figure 3: Donations and Twitter Penetration

r

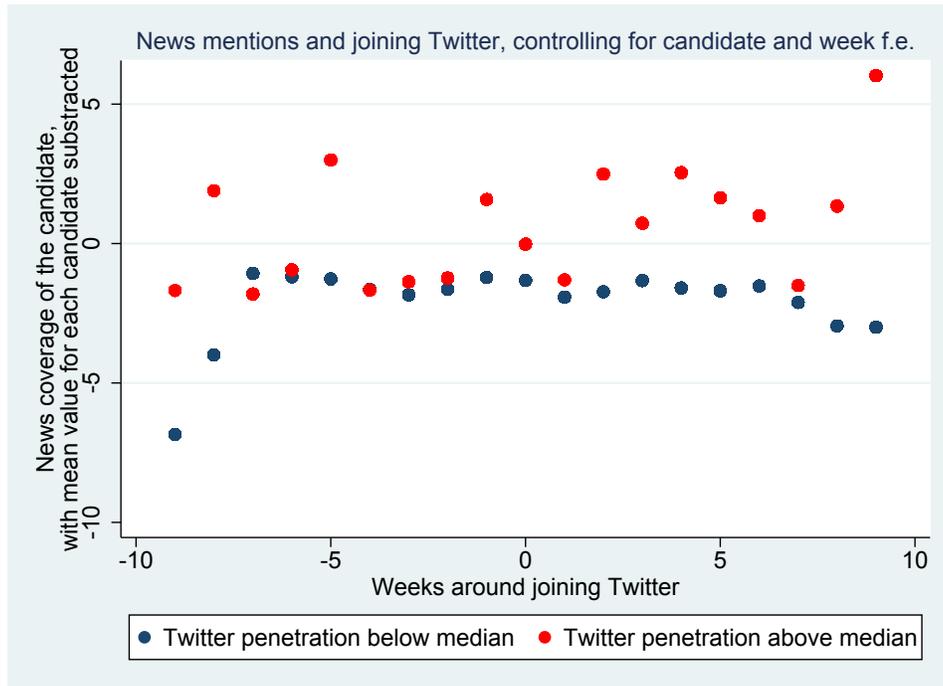

Figure 4: Number of News Mentions and Twitter Penetration

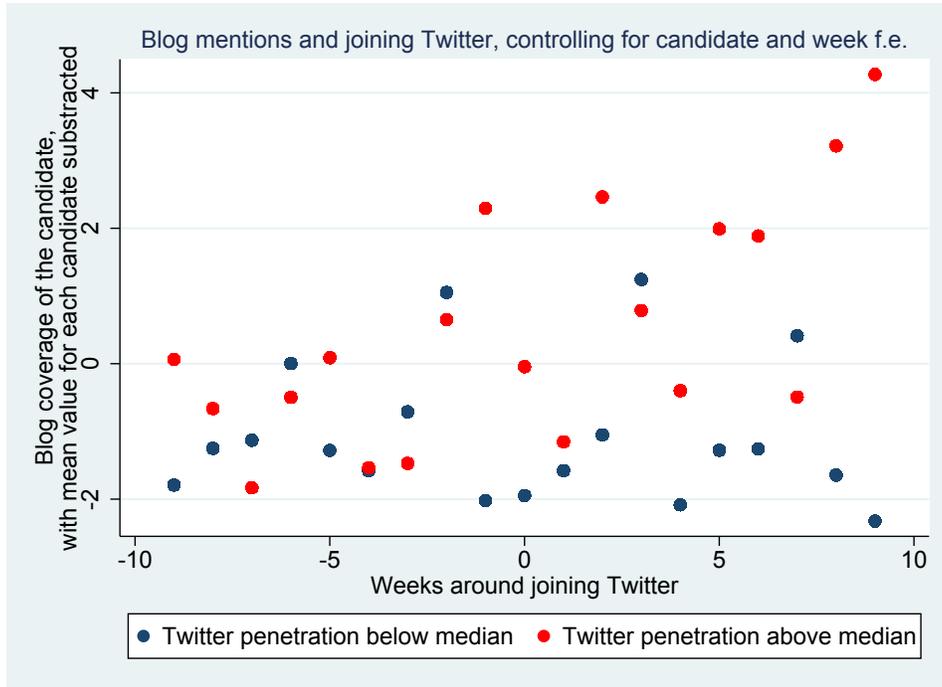

Figure 5: Number of Blog Mentions and Twitter Penetration

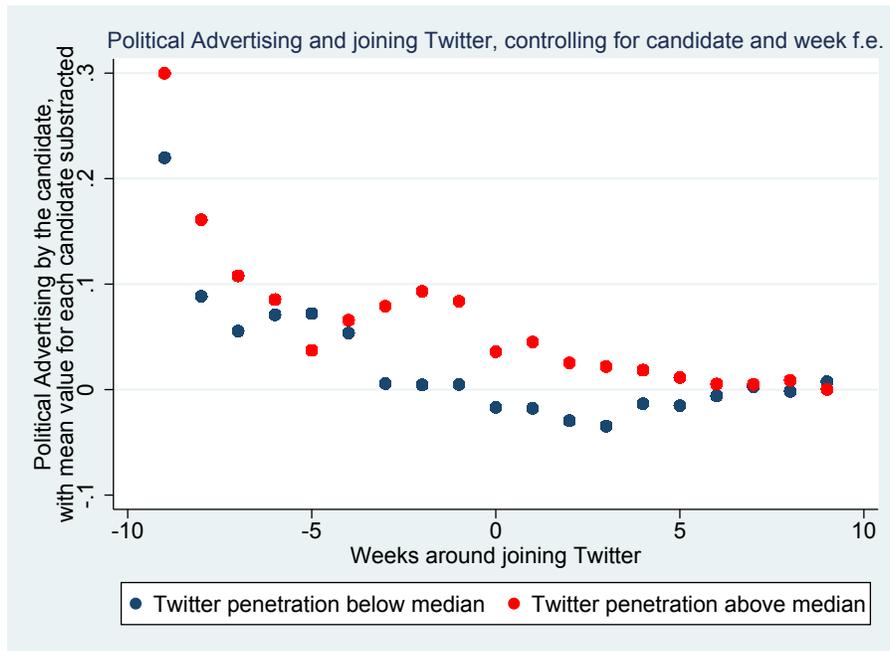

Figure 6: Political Advertising and Twitter Penetration

Table 1: Twitter Adoption and Log(Aggregate Donations)

| | Log(Aggregate Donations) | | | | | |
|---|---|---|---|---|---|---|
| | (1) | (2) | (3) | (4) | (5) | (6) |
| VARIABLES | All | All | New | New | Experienced | Experienced |
| onTwitter x Twit_Penet | 102.6499** | 106.1455** | 186.0268*** | 192.2501*** | -45.0641 | -47.0084 |
| | (45.5244) | (45.5292) | (44.9294) | (45.7696) | (93.7545) | (91.0756) |
| onTwitter | 0.1879 | 1.3162 | 0.1268 | 2.3393 | 0.3961* | -0.0791 |
| | (0.1126) | (1.0235) | (0.1140) | (1.5472) | (0.2248) | (2.0678) |
| Observations | 565,968 | 565,968 | 236,740 | 236,740 | 329,228 | 329,228 |
| R-squared | 0.8215 | 0.8215 | 0.8828 | 0.8828 | 0.7856 | 0.7856 |
| Politician-Month FE | Y | Y | Y | Y | Y | Y |
| Week of Month FE | Y | Y | Y | Y | Y | Y |
| Baseline controls x on Twitter | N | Y | N | Y | N | Y |
| Implied Twitter effect for 2009 | .029** | .03** | .047*** | .048*** | -.013 | -.013 |
| Implied Twitter effect for 2014 | .228** | .236** | .414*** | .428*** | -.1 | -.105 |

Note: Robust standard errors clustered at the level of the state and week in parenthesis. *** p<0.01, ** p<0.05, * p<0.1. The dependent variable is Log (Aggregate Donations). Columns (1)-(2) include all politicians while columns (3)-(4) include only new and columns (5)-(6) have the experienced politicians. State level baseline controls interacted with the politician being on Twitter include the median household income and population size. Average of pageviews relative to all pageviews in state-year is used as Twitter penetration measure. All specifications include Politician-month and week of month fixed effects.

Table 2: Twitter Adoption and Probability of Receiving at least One Donation

| | Prob of One Donation | | | | | |
|---|---|---|---|---|---|---|
| | (1) | (2) | (3) | (4) | (5) | (6) |
| VARIABLES | All | All | New | New | Experienced | Experienced |
| onTwitter x Twit_Penet | 13.7776** | 14.3168** | 22.5307*** | 23.4135*** | -1.9080 | -2.0584 |
| | (5.8924) | (5.9123) | (6.3244) | (6.4678) | (12.0430) | (11.7725) |
| onTwitter | 0.0243 | 0.1972 | 0.0218 | 0.3403 | 0.0407 | 0.0035 |
| | (0.0149) | (0.1503) | (0.0168) | (0.2057) | (0.0301) | (0.2909) |
| Observations | 565,968 | 565,968 | 236,740 | 236,740 | 329,228 | 329,228 |
| R-squared | 0.7865 | 0.7865 | 0.8448 | 0.8449 | 0.7508 | 0.7508 |
| Politician-Month FE | Y | Y | Y | Y | Y | Y |
| Week of Month FE | Y | Y | Y | Y | Y | Y |
| Baseline controls x on Twitter | N | Y | N | Y | N | Y |
| Implied Twitter effect for 2009 | .004** | .004** | .006*** | .006*** | -.001 | -.001 |
| Implied Twitter effect for 2014 | .031** | .032** | .05*** | .052*** | -.004 | -.005 |

Note: Robust standard errors clustered at the level of the state and week in parenthesis. *** p<0.01, ** p<0.05, * p<0.1. The dependent variable is Probability of Receiving at least One Donation. Columns (1)-(2) include all politicians while columns (3)-(4) include only new and columns (5)-(6) have the experienced politicians. State level baseline controls interacted with the politician being on Twitter include the median household income and population size. Average of pageviews relative to all pageviews in state-year is used as Twitter penetration measure. All specifications include Politician-month and week of month fixed effects.

Table 3: Twitter Adoption and Log (Number of Donations)

| | Log (Number of Donations) | | | | | |
|---|---|---|---|---|---|---|
| | (1) | (2) | (3) | (4) | (5) | (6) |
| VARIABLES | All | All | New | New | Experienced | Experienced |
| onTwitter x Twit_Penet | 16.9626 | 17.4442 | 41.2259*** | 42.5467*** | -25.3309 | -26.4579 |
| | (12.3434) | (12.3865) | (11.5946) | (11.5548) | (24.0077) | (23.0095) |
| onTwitter | 0.0533 | 0.2116 | 0.0226 | 0.4715 | 0.1328** | -0.1396 |
| | (0.0323) | (0.1734) | (0.0294) | (0.3578) | (0.0567) | (0.4622) |
| Observations | 565,968 | 565,968 | 236,740 | 236,740 | 329,228 | 329,228 |
| R-squared | 0.8384 | 0.8384 | 0.9007 | 0.9007 | 0.8006 | 0.8007 |
| Politician-Month FE | Y | Y | Y | Y | Y | Y |
| Week of Month FE | Y | Y | Y | Y | Y | Y |
| Baseline controls x on Twitter | N | Y | N | Y | N | Y |
| Implied Twitter effect for 2009 | .005 | .005 | .01*** | .011*** | -.007 | -.007 |
| Implied Twitter effect for 2014 | .038 | .039 | .092*** | .095*** | -.056 | -.059 |

Note: Robust standard errors clustered at the level of the state and week in parenthesis. *** p<0.01, ** p<0.05, * p<0.1. The dependent variable is Log (Number of Donations). Columns (1)-(2) include all politicians while columns (3)-(4) include only new and columns (5)-(6) have the experienced politicians. State level baseline controls interacted with the politician being on Twitter include the median household income and population size. Average of pageviews relative to all pageviews in state-year is used as Twitter penetration measure. All specifications include Politician-month and week of month fixed effects.

Table 4: Baseline Results: Eight Week Window with Politician Fixed Effects

| | Log (Aggregated Donations) | | | Prob of One Donation | | | Log (Number of Donations) | | |
|---|---|---|---|---|---|---|---|---|---|
| | (1) | (2) | (3) | (4) | (5) | (6) | (7) | (8) | (9) |
| VARIABLES | All | New | Old | All | New | Old | All | New | Old |
| onTwitter x Twit_Penet | 93.9519** | 126.7358** | -13.2064 | 12.5811** | 15.0845** | -0.6215 | 16.3467 | 25.8044** | -5.4879 |
| | (38.9716) | (50.8035) | (44.1856) | (5.0375) | (6.8434) | (5.8806) | (10.8844) | (12.3834) | (12.0133) |
| onTwitter | 0.0303 | 0.1225 | 0.0862 | 0.0071 | 0.0212 | 0.0133 | 0.0056 | 0.0304 | 0.0081 |
| | (0.1119) | (0.1288) | (0.1679) | (0.0148) | (0.0183) | (0.0232) | (0.0326) | (0.0319) | (0.0454) |
| Observations | 24,094 | 14,160 | 9,929 | 24,094 | 14,160 | 9,929 | 24,094 | 14,160 | 9,929 |
| R-squared | 0.5869 | 0.6077 | 0.5720 | 0.5460 | 0.5667 | 0.5237 | 0.6016 | 0.6354 | 0.6005 |
| Politician FE | Y | Y | Y | Y | Y | Y | Y | Y | Y |
| Week FE | Y | Y | Y | Y | Y | Y | Y | Y | Y |
| Implied Twitter effect for 2009 | .024** | .035** | -.003 | .003** | .004** | 0 | .004 | .007** | -.001 |
| Implied Twitter effect for 2014 | .209** | .287** | -.027 | .028** | .034** | -.001 | .036 | .058** | -.011 |

Note: Robust standard errors clustered at the level of the state and week in parenthesis. *** p<0.01, ** p<0.05, * p<0.1. The dependent variables are Log (Number of Donations), Probability of Receiving Donations and Log(Number of Donations). Columns (1), (4) and (7) include all politicians while columns (2), (5) and (8) include only new and columns (3),(6) and (9) have the experienced politicians. Average of pageviews relative to all pageviews in state-year is used as Twitter penetration measure. All specifications include Politician and week fixed effects.

Table 5: Amount of Donations: Different Windows with Politician Fixed Effects

| | All Politicians: Log (Aggregate Donations) | | | | | | |
|---|---|---|---|---|---|---|---|
| | 5 weeks | 6 weeks | 7 weeks | 8 weeks | 9 weeks | 10 weeks | 20 weeks |
| Variables | (1) | (2) | (3) | (4) | (5) | (6) | (7) |
| on Twitter x Twit_Penet | 100.1897** | 89.1422** | 92.7713** | 93.9519** | 85.6743** | 80.6435** | 50.1190 |
| | (40.6520) | (41.4398) | (39.0579) | (38.9716) | (36.8228) | (37.8201) | (37.0327) |
| onTwitter | 0.0081 | 0.0323 | 0.0348 | 0.0303 | 0.0705 | 0.1042 | 0.5155*** |
| | (0.1219) | (0.1194) | (0.1153) | (0.1119) | (0.1064) | (0.1105) | (0.1107) |
| Observations | 19,221 | 20,846 | 22,470 | 24,094 | 25,716 | 27,336 | 43,508 |
| R-squared | 0.6018 | 0.5932 | 0.5894 | 0.5869 | 0.5832 | 0.5807 | 0.5635 |
| Politician FE | Y | Y | Y | Y | Y | Y | Y |
| Week FE | Y | Y | Y | Y | Y | Y | Y |
| Implied Twitter effect for 2009 | .025** | .022** | .023** | .024** | .021** | .02** | .013 |
| Implied Twitter effect for 2014 | .223** | .198** | .206** | .209** | .191** | .179** | .111 |

Note: Robust standard errors clustered at the level of the state and week in parenthesis. *** p<0.01, ** p<0.05, * p<0.1. The dependent variable is Log(Aggregate Donations). Columns (1)-(7) include All politicians. Average of pageviews relative to all pageviews in state-year is used as Twitter penetration measure. All specifications include politician and week fixed effects.

Table 6: Amount of Donations: Different Windows with Politician Fixed Effects

| | New Politicians: Log (Aggregate Donations) | | | | | | |
|---|---|---|---|---|---|---|---|
| | 5 weeks | 6 weeks | 7 weeks | 8 weeks | 9 weeks | 10 weeks | 20 weeks |
| Variables | (1) | (2) | (3) | (4) | (5) | (6) | (7) |
| on Twitter x Twit_Penet | 124.1443** | 115.2081** | 121.8793** | 126.7358** | 124.9643** | 122.2513** | 133.0385** |
| | (49.4201) | (50.2453) | (49.4881) | (50.8035) | (51.3200) | (50.8119) | (52.2574) |
| onTwitter | 0.0976 | 0.1113 | 0.1293 | 0.1225 | 0.1720 | 0.2266 | 0.7307*** |
| | (0.1287) | (0.1262) | (0.1254) | (0.1288) | (0.1323) | (0.1393) | (0.1559) |
| Observations | 11,328 | 12,274 | 13,218 | 14,160 | 15,098 | 16,034 | 25,257 |
| R-squared | 0.6162 | 0.6106 | 0.6080 | 0.6077 | 0.6053 | 0.6048 | 0.5890 |
| Politician FE | Y | Y | Y | Y | Y | Y | Y |
| Week FE | Y | Y | Y | Y | Y | Y | Y |
| Implied Twitter effect for 2009 | .035** | .032** | .034** | .035** | .035** | .034** | .037** |
| Implied Twitter effect for 2014 | .281** | .26** | .276** | .287** | .283** | .276** | .301** |

Note: Robust standard errors clustered at the level of the state and week in parenthesis. *** p<0.01, ** p<0.05, * p<0.1. The dependent variable is Log(Aggregate Donations). Columns (1)-(7) include New politicians. Average of pageviews relative to all pageviews in state-year is used as Twitter penetration measure. All specifications include politician and week fixed effects.

Table 7: Amount of Donations: Different Windows with Politician Fixed Effects

| | Experienced Politicians: Log (Aggregate Donations) | | | | | | |
|---|---|---|---|---|---|---|---|
| | 5 weeks | 6 weeks | 7 weeks | 8 weeks | 9 weeks | 10 weeks | 20 weeks |
| Variables | (1) | (2) | (3) | (4) | (5) | (6) | (7) |
| on Twitter x Twit_Penet | 17.0589 | -2.7135 | -4.7562 | -13.2064 | -29.6278 | -30.0255 | -26.3213 |
| | (42.0674) | (46.2546) | (43.7108) | (44.1856) | (40.4304) | (40.4767) | (42.1172) |
| onTwitter | 0.0278 | 0.0856 | 0.0720 | 0.0862 | 0.1305 | 0.1337 | 0.2238 |
| | (0.1796) | (0.1884) | (0.1793) | (0.1679) | (0.1618) | (0.1674) | (0.1528) |
| Observations | 7,886 | 8,567 | 9,247 | 9,929 | 10,610 | 11,296 | 18,244 |
| R-squared | 0.5907 | 0.5792 | 0.5756 | 0.5720 | 0.5674 | 0.5637 | 0.5468 |
| Politician FE | Y | Y | Y | Y | Y | Y | Y |
| Week FE | Y | Y | Y | Y | Y | Y | Y |
| Implied Twitter effect for 2009 | .004 | -.001 | -.001 | -.003 | -.007 | -.007 | -.006 |
| Implied Twitter effect for 2014 | .034 | -.005 | -.01 | -.027 | -.059 | -.06 | -.043 |

Note: Robust standard errors clustered at the level of the state and week in parenthesis. *** p<0.01, ** p<0.05, * p<0.1. The dependent variable is Log(Aggregate Donations). Columns (1)-(7) include Experienced politicians. Average of pageviews relative to all pageviews in state-year is used as Twitter penetration measure. All specifications include politician and week fixed effects.

Table 8: Campaign Expenditure Placebo Analysis

| | Panel A: Campaign Expenditure Placebo I | | | | | | |
|---|---|---|---|---|---|---|---|
| | (1) | (2) | (3) | (4) | (5) | (6) | (7) |
| VARIABLES | Log(Total Expenditure) | Log(Contributions) | Log(Polling Expenses) | Log(Refund to Donors) | Log(Fundraising) | Log(Transfers to Committees) | Log(Travel Expenses) |
| onTwitter x Twit_Penet | 24.5295 | -12.5698 | -0.6529 | 1.0771 | 2.6789 | -2.8861 | -5.9670 |
| | (43.9563) | (9.9327) | (8.7970) | (1.0577) | (13.9863) | (2.8872) | (15.4036) |
| onTwitter | -0.0002 | -0.2113 | -0.0234 | 0.0226 | 0.6984 | 0.0214 | -0.2215 |
| | (1.5214) | (0.2611) | (0.2009) | (0.0221) | (0.6493) | (0.0278) | (0.4379) |
| Observations | 561,339 | 561,598 | 561,600 | 561,594 | 561,576 | 561,600 | 561,548 |
| R-squared | 0.8710 | 0.3316 | 0.3251 | 0.2718 | 0.6456 | 0.2717 | 0.5939 |
| Politician-Month FE | Y | Y | Y | Y | Y | Y | Y |
| Week of Month FE | Y | Y | Y | Y | Y | Y | Y |
| Baseline controls x on Twitter | Y | Y | Y | Y | Y | Y | Y |
| Implied Twitter effect for 2009 | .007 | -.003 | 0 | 0 | .001 | -.001 | -.002 |
| Implied Twitter effect for 2014 | .055 | -.028 | -.001 | .002 | .006 | -.006 | -.013 |

| | Panel B: Campaign Expenditure Placebo II | | | | | |
|---|---|---|---|---|---|---|
| | (8) | (9) | (10) | (11) | (12) | (13) |
| VARIABLES | Log(Administrative Expenses) | Log(Advertising Expenditure) | Log(Events) | Log(Materials) | Log(Donations) | Log(Loan Repayments) |
| onTwitter x Twit_Penet | 13.4955 | 17.5680 | 5.4400 | 8.1173 | -2.1121 | 0.3731 |
| | (22.9729) | (15.9738) | (12.5658) | (13.2697) | (3.3032) | (0.9381) |
| onTwitter | -0.8601 | -0.1807 | 0.5236 | 0.1613 | 0.1971 | -0.0880 |
| | (0.6406) | (0.4189) | (0.3131) | (0.3412) | (0.1686) | (0.0715) |
| Observations | 561,538 | 561,579 | 561,590 | 561,591 | 561,595 | 561,600 |
| R-squared | 0.8369 | 0.6210 | 0.5227 | 0.4963 | 0.3928 | 0.2637 |
| Politician-Month FE | Y | Y | Y | Y | Y | Y |
| Week of Month FE | Y | Y | Y | Y | Y | Y |
| Baseline controls x on Twitter | Y | Y | Y | Y | Y | Y |
| Implied Twitter effect for 2009 | .004 | .005 | .002 | .002 | -.001 | 0 |
| Implied Twitter effect for 2014 | .03 | .039 | .012 | .018 | -.005 | .001 |

Note: Robust standard errors clustered at the level of the state and week in parenthesis. *** p<0.01, ** p<0.05, * p<0.1. Columns (1)-(13) include all politicians. State level baseline controls interacted with the politician being on Twitter include the median household income and population size. Average of pageviews relative to all pageviews in state-year is used as Twitter penetration measure. All specifications include Politician-month and week of month fixed effects.

Table 9: Political Ads, News and Blogs Placebo Analysis

|  | Log(Political Advertising Expenditure) | | | Log(Number of News Mentions) | | | Log(Number of Blog Mentions) | | |
| --- | --- | --- | --- | --- | --- | --- | --- | --- | --- |
|  | (1) | (2) | (3) | (4) | (5) | (6) | (7) | (8) | (9) |
| VARIABLES | All | New | Experienced | All | New | Experienced | All | New | Experienced |
| onTwitter x Twit_Penet | -13.9824 | 11.1326 | -58.8757 | 0.4565 | -2.4421 | 5.9242 | -5.7416 | -7.2800* | -2.9491 |
|  | (18.0394) | (14.4600) | (52.2719) | (3.9450) | (3.3989) | (9.0824) | (4.1593) | (3.6545) | (8.2101) |
| onTwitter | 0.3212 | -0.0515 | 0.7542* | 0.0341 | -0.0358 | 0.1325 | -0.0359 | -0.0177 | -0.0533 |
|  | (0.1937) | (0.1642) | (0.4118) | (0.1092) | (0.0727) | (0.2323) | (0.1115) | (0.1038) | (0.2043) |
| Observations | 565,968 | 236,740 | 329,228 | 46,379 | 28,320 | 18,059 | 46,379 | 28,320 | 18,059 |
| R-squared | 0.8136 | 0.8301 | 0.8020 | 0.9207 | 0.9275 | 0.9106 | 0.8584 | 0.8588 | 0.8488 |
| Politician-Month FE | Y | Y | Y | Y | Y | Y | Y | Y | Y |
| Week of Month FE | Y | Y | Y | Y | Y | Y | Y | Y | Y |
| Baseline controls x on Twitter | Y | Y | Y | Y | Y | Y | Y | Y | Y |
| Implied Twitter effect for 2009 | -.004 | .003 | -.016 | 0 | -.001 | .001 | -.001 | -.002* | -.001 |
| Implied Twitter effect for 2014 | -.031 | .025 | -.131 | .001 | -.006 | .01 | -.013 | -.016* | -.005 |

Note: Robust standard errors clustered at the level of the state and week in parenthesis. *** p<0.01, ** p<0.05, * p<0.1. The dependent variable are Log(Political Advertising Expenditure), Log(Number of News Mentions), and Log(Number of Blog Mentions) respectively. Columns (1),(4) and (7) include all politicians while columns (2), (5) and (8) include only new and columns (3), (6) and (9) have the experienced politicians. State level baseline controls interacted with the politician being on Twitter include the median household income and population size. Average of pageviews relative to all pageviews in state-year is used as Twitter penetration measure. All specifications include Politician-month and week of month fixed effects.

Table 10: Amount of Donations and New vs. Repeat Donors

| Panel A: Log (Aggregate Donations) | | | | | | |
| --- | --- | --- | --- | --- | --- | --- |
|  | New Donors | | | Repeat Donors | | |
|  | (1) | (2) | (3) | (4) | (5) | (6) |
| VARIABLES | All | New | Experienced | All | New | Experienced |
| onTwitter x Twit_Penet | 99.6026** | 201.2764*** | -82.2807 | -23.4208 | 13.5236 | -80.3688 |
|  | (39.5139) | (42.8484) | (74.3038) | (37.0208) | (19.4423) | (84.5136) |
| onTwitter | 0.6378 | 2.2815 | -1.5874 | -0.2534 | -0.5769 | 0.2076 |
|  | (0.7134) | (1.5024) | (1.5364) | (0.8247) | (0.8508) | (1.5842) |
| Observations | 565,968 | 236,740 | 329,228 | 565,968 | 236,740 | 329,228 |
| R-squared | 0.7850 | 0.8725 | 0.7332 | 0.7626 | 0.8080 | 0.7359 |
| Politician-Month FE | Y | Y | Y | Y | Y | Y |
| Week of Month FE | Y | Y | Y | Y | Y | Y |
| Baseline controls x on Twitter | Y | Y | Y | Y | Y | Y |
| Implied Twitter effect for 2009 | .028** | .05*** | -.023 | -.007 | .003 | -.022 |
| Implied Twitter effect for 2014 | .222** | .448*** | -.183 | -.052 | .03 | -.179 |

| Panel B: Log (Number of Donations) | | | | | | |
| --- | --- | --- | --- | --- | --- | --- |
| onTwitter x Twit_Penet | 19.0058* | 43.1319*** | -24.0642 | -6.4079 | 3.1616 | -20.8431 |
|  | (10.0928) | (11.0250) | (16.4680) | (8.3034) | (4.4554) | (18.5322) |
| onTwitter | 0.2554* | 0.4975 | -0.0834 | -0.1068 | -0.0713 | -0.1357 |
|  | (0.1360) | (0.3617) | (0.4017) | (0.1962) | (0.1299) | (0.3680) |
| Observations | 565,968 | 236,740 | 329,228 | 565,968 | 236,740 | 329,228 |
| R-squared | 0.8169 | 0.8892 | 0.7642 | 0.7891 | 0.8484 | 0.7615 |
| Politician-Month FE | Y | Y | Y | Y | Y | Y |
| Week of Month FE | Y | Y | Y | Y | Y | Y |
| Baseline controls x on Twitter | Y | Y | Y | Y | Y | Y |
| Implied Twitter effect for 2009 | .005* | .011*** | -.007 | -.002 | .001 | -.006 |
| Implied Twitter effect for 2014 | .042* | .096*** | -.054 | -.014 | .007 | -.046 |

Notes: 1. Robust standard errors clustered at the level of the state and week in parenthesis. *** p<0.01, ** p<0.05, * p<0.1. The dependent variables are Log (Aggregate Donations) and Log (Number of Donations). Columns (1) and (4) include all politicians while columns (2) and (5) include only new and columns (3) and (6) have the experienced politicians. State level baseline controls interacted with the politician being on Twitter include the median household income and population size. Average of pageviews relative to all pageviews in state-year is used as Twitter penetration measure. All specifications include Politician-month and week of month fixed effects.
2. A statistical test comparing the weekly aggregate donations from new vs. repeat donors show the following p-values: 0.000 (all politicians), 0.000 (new politicians), 0.971 (experienced politicians). We do not see a significant difference in the number of donations.

Table 11: Joining Twitter and Amount of Donations: House vs. Senate Candidates

| | Panel A: Log (Aggregate Donations) | | | | | |
|---|---|---|---|---|---|---|
| | Senate | | | House | | |
| | (1) | (2) | (3) | (4) | (5) | (6) |
| VARIABLES | All | New | Experienced | All | New | Experienced |
| onTwitter x Twit_Penet | 34.7339 | 48.6398 | 43.2330 | 119.0656** | 214.1968*** | -57.8678 |
| | (73.9788) | (106.8262) | (107.1972) | (50.7529) | (48.7895) | (106.4910) |
| onTwitter | -3.1263 | -1.3372 | -4.6257 | 2.4824* | 2.7287 | 1.7364 |
| | (2.0686) | (2.5507) | (4.1722) | (1.3541) | (1.8950) | (2.2820) |
| Observations | 114,816 | 50,408 | 64,408 | 451,152 | 186,332 | 264,820 |
| R-squared | 0.8665 | 0.9249 | 0.8259 | 0.8077 | 0.8643 | 0.7748 |
| Politician-Month FE | Y | Y | Y | Y | Y | Y |
| Week of Month FE | Y | Y | Y | Y | Y | Y |
| Baseline controls x on Twitter | Y | Y | Y | Y | Y | Y |
| Implied Twitter effect for 2009 | .01 | .014 | .016 | .03** | .054*** | -.014 |
| Implied Twitter effect for 2014 | .09 | .126 | .114 | .248** | .484*** | -.083 |

| | Panel B: Log (Number of Donations) | | | | | |
|---|---|---|---|---|---|---|
| onTwitter x Twit_Penet | -6.5753 | -16.0041 | 5.7861 | 22.3269 | 51.2158*** | -30.8124 |
| | (20.5819) | (20.4530) | (35.6365) | (13.7621) | (12.0619) | (25.7055) |
| onTwitter | -0.6366 | -0.7959 | -0.3724 | 0.3957 | 0.6293 | -0.0380 |
| | (0.6569) | (0.7602) | (1.1739) | (0.2404) | (0.4758) | (0.4227) |
| Observations | 114,816 | 50,408 | 64,408 | 451,152 | 186,332 | 264,820 |
| R-squared | 0.8935 | 0.9339 | 0.8607 | 0.8131 | 0.8773 | 0.7785 |
| Politician-Month FE | Y | Y | Y | Y | Y | Y |
| Week of Month FE | Y | Y | Y | Y | Y | Y |
| Baseline controls x on Twitter | Y | Y | Y | Y | Y | Y |
| Implied Twitter effect for 2009 | -.002 | -.005 | .002 | .006 | .013*** | -.008 |
| Implied Twitter effect for 2014 | -.017 | -.041 | .015 | .047 | .116*** | -.044 |

Notes: 1. Robust standard errors clustered at the level of the state and week in parenthesis. *** p<0.01, ** p<0.05, * p<0.1. The dependent variables are Log (Aggregate Donations) and Log (Number of Donations). Columns (1) and (4) include all politicians while columns (2) and (5) include only new and columns (3) and (6) have the experienced politicians. State level baseline controls interacted with the politician being on Twitter include the median household income and population size. Average of pageviews relative to all pageviews in state-year is used as Twitter penetration measure. All specifications include Politician-month and week of month fixed effects.
2. Adding a triple interaction term to compare the aggregate weekly donations to the House vs. Senate candidate finds the following p-values: 0.249 (all politicians), 0.020 (new politicians), and 0.423 (experienced politicians). For the number of donations comparison, corresponding p-values are 0.162 (all), 0.001 (new) and 0.304 (experienced).

Table 12: Politicians With and Without Facebook Before

| | Log(Aggregate Donations) | | | | Prob of Donation | | | |
|---|---|---|---|---|---|---|---|---|
| | No FB before | No FB before | Had FB before | Had FB before | No FB before | No FB before | Had FB before | Had FB before |
| | (1) | (2) | (3) | (4) | (5) | (6) | (7) | (8) |
| VARIABLES | All | All | All | All | All | All | All | All |
| onTwitter x Twit_Penet | 107.1144*** | 110.6932*** | -208.8425 | -112.7912 | 14.4419*** | 14.9910*** | -40.1547 | -32.9786 |
| | (36.5950) | (36.5206) | (270.5396) | (443.9461) | (4.8938) | (4.9012) | (37.2622) | (56.4349) |
| onTwitter | 0.1537* | 1.2504 | 1.4660 | 6.8097 | 0.0182 | 0.1862 | 0.2243* | 1.3319* |
| | (0.0859) | (0.9892) | (1.0423) | (5.5263) | (0.0123) | (0.1447) | (0.1289) | (0.7843) |
| Observations | 550,239 | 550,239 | 15,689 | 15,689 | 550,239 | 550,239 | 15,689 | 15,689 |
| R-squared | 0.8240 | 0.8240 | 0.7674 | 0.7674 | 0.7885 | 0.7885 | 0.7129 | 0.7129 |
| Politician-Month FE | Y | Y | Y | Y | Y | Y | Y | Y |
| Week FE | Y | Y | Y | Y | Y | Y | Y | Y |
| Baseline controls x on Twitter | N | Y | N | Y | N | Y | N | Y |
| Implied Twitter effect for 2009 | .03*** | .031*** | -.061 | -.033 | .004*** | .004*** | -.012 | -.01 |
| Implied Twitter effect for 2014 | .238*** | .246*** | -.388 | -.209 | .032*** | .033*** | -.075 | -.061 |

Note: Robust standard errors clustered at the level of the state. *** p<0.01, ** p<0.05, * p<0.1. State level baseline controls interacted with the politician being on Twitter include the median household income and population size. Average of pageviews relative to all pageviews in state-year is used as Twitter penetration measure. All specifications include Politician-month and week fixed effects. We cluster only at state level because in the restricted sample we do not have enough clusters to compute standard errors. We control for week fixed effects instead of week of month fixed effects to ensure computation of standard errors.

Table 13: Joining Twitter and Within the State Donations

| | (1) | (2) | (3) | (4) | (5) | (6) |
|---|---|---|---|---|---|---|
| | \multicolumn{6}{c}{Within State} | | | | | |
| | \multicolumn{6}{c}{Log(Aggregate Donations)} | | | | | |
| VARIABLES | All | All | New | New | Experienced | Experienced |
| on Twitter x Twit_Penet | 99.5491** | 100.9357** | 164.8527*** | 167.4275*** | -14.4344 | -16.8130 |
| | (43.4447) | (43.5136) | (38.3601) | (38.8466) | (97.8290) | (95.5988) |
| on Twitter | 0.1757 | 0.5550 | 0.1347 | 1.2038 | 0.3262 | -0.3620 |
| | (0.1171) | (1.0078) | (0.1037) | (1.4707) | (0.2351) | (1.8667) |
| Observations | 543,504 | 543,504 | 225,904 | 225,904 | 317,600 | 317,600 |
| R-squared | 0.7961 | 0.7961 | 0.8640 | 0.8640 | 0.7573 | 0.7573 |
| Politician-Month FE | Y | Y | Y | Y | Y | Y |
| Week of Month FE | Y | Y | Y | Y | Y | Y |
| Baseline controls x on Twitter | N | Y | N | Y | N | Y |
| Implied Twitter effect for 2009 | .025** | .025** | .041*** | .042*** | -.004 | -.005 |
| Implied Twitter effect for 2014 | .221** | .225** | .373*** | .379*** | -.032 | -.037 |
| | \multicolumn{6}{c}{Outside State} | | | | | |
| on Twitter x Twit_Penet | 19.9318 | 22.6562 | 87.8919* | 93.0614* | -98.7879* | -101.0485* |
| | (42.1108) | (41.7695) | (46.1417) | (46.6828) | (58.7697) | (55.5853) |
| on Twitter | 0.1158 | 0.9958 | 0.0628 | 1.9295** | 0.3106** | -0.2368 |
| | (0.1001) | (0.6411) | (0.0990) | (0.9470) | (0.1541) | (1.3118) |
| Observations | 565,888 | 565,888 | 236,712 | 236,712 | 329,176 | 329,176 |
| R-squared | 0.7203 | 0.7203 | 0.8036 | 0.8036 | 0.6792 | 0.6792 |
| Politician-Month FE | Y | Y | Y | Y | Y | Y |
| Week of Month FE | Y | Y | Y | Y | Y | Y |
| Baseline controls x on Twitter | N | Y | N | Y | N | Y |
| Implied Twitter effect for 2009 | .006 | .006 | .022* | .023* | -.027* | -.028* |
| Implied Twitter effect for 2014 | .044 | .05 | .196* | .207* | -.22* | -.225* |

Note: Robust standard errors clustered at the level of the state and week in parenthesis. *** p<0.01, ** p<0.05, * p<0.1. The dependent variable is Log (Aggregate Donations). Columns (1)-(2) include all politicians while columns (3)-(4) include only new and columns (5)-(6) have the experienced politicians. State level baseline controls interacted with the politician being on Twitter include the median household income and population size. Average of pageviews relative to all pageviews in state-year is used as Twitter penetration measure. All specifications include Politician-month and week of month fixed effects.

# Online Appendix

## A.1 Calculating Persuasion Rate for Twitter

To be able to compare the magnitudes that we uncover with other studies in the literature, we compute persuasion rates (DellaVigna and Kaplan, 2007; DellaVigna and Gentzkow, 2010). We cannot compute persuasion rates for all the followers of all politicians because we do not observe how the number of followers evolve for every politician from the time they open an account on Twitter, and we can only estimate the impact of joining Twitter within a month of Twitter entry. However, using information on the number of followers gained in the first 3-4 months after opening an account for some politicians, we can compute persuasion rates under the assumption that early number of followers are similar for politicians who just joined Twitter.[40] We observe the number of the followers for a subset of the politicians within 3 months of their account opening for two points in time: at the time of data collection by Halberstam and Knight (2016) and at the time of our own data collection. For these politicians (21% of the politicians who opened their accounts in 2012), the average number of followers gained is 104 within the first 3 months after opening a Twitter account and is 151 within the first four months. The persuasion rate calculated based on these early group of followers is likely not generalizable to the remaining parts of the population, since the characteristics of these individuals may significantly vary in terms of their interest in the politician, politics, policy issues, or engagement in technology. Further, in contrast to other studies reporting on the persuasive effects of media, we employ temporal rather than spatial variation.

To estimate the persuasion rate associated with opening a Twitter channel, we use the formula

$$f = \frac{y_t - y_c}{e_t - e_c} \times \frac{1}{1 - y_0} \times 100$$

used for reporting the persuasive effects of various media by DellaVigna and Gentzkow (2010). Here the treatment (control) group is represented by $T$ ($C$), $e_j$ is the share of group $j \in \{T, C\}$ receiving the message, $y_j$ is the share of group $j$ adopting the behavior of interest (donate), and $y_0$ is the share that would adopt the behavior if there were no messages. DellaVigna and Gentzkow (2010) assume that where $y_0$ is not observed, it can be approximated by $y_C$. To approximate this number, using Table 3, we estimate the expected number of donations. The persuasion rate captures the effect of the persuasion treatment on the relevant behavior $(y_T - y_C)$, adjusting for exposure to the message $(e_T - e_C)$ and for the size of the population left to be convinced $(1 - y_0)$ (DellaVigna and Gentzkow,

---
[40]Unfortunately, we were not lucky to have politicians who join Twitter within a month of our data collection. But we believe that the average number of followers in these early months is unlikely to change significantly. If anything, if we assume that the average number of followers is smaller than the ones we find for 3 months, our persuasion rates should be larger.

2010, pg. 645).

In this study, our treatment is the entry of politician to Twitter. Similar to DellaVigna and Gentzkow (2010), we assume that $e_t - e_c$ is 100%, that is, all followers of a politician observe the treatment (entry on Twitter and the subsequent tweets within the first month). The $y_t - y_c$ is computed by using the implied effect of Twitter for 2014 from column (4) of Table 3 with the benchmark estimations, multiplied by the average number of donations (2.95), divided by the number of followers (151 or 104, depending on the number of months after account opening). We then multiply this with the remaining 2.79 weeks in the month after Twitter entry since our estimates are credibly identified for this period due to politician-month fixed effects.

Finally, $y_0$ is a counterfactual estimate for the share of donations in the absence of Twitter entry. To compute $y_0$, we multiply 0.374 (a counterfactual estimate for donations in the absence of Twitter entry computed by predicting donations assuming that entry to Twitter variable is equal to zero) by the average length of exposure (2.79 weeks), and divide it by the number of followers. The number of donations from all the followers in the absence of Twitter entry could be computed as the (logarithm of) average realized number of donations (0.469) minus the implied effect of being on Twitter when the dependent variable is the number of donations (0.095).

As before, 2.79 weeks is the average number of weeks in the month after Twitter entry. Using the numbers above, we obtain that the persuasion rate (for 104 and 151 followers) is equal to:

$$f_{104} = (2.95 \times 0.095 \times 2.79/104)/(1 - .374 \times 2.79/104) = .0074 = 0.75\%$$

$$f_{151} = (2.95 \times 0.095 \times 2.79/151)/(1 - .374 \times 2.79/151) = 0.005 = 0.51\%$$

These persuasion rates associated with opening a Twitter channel are rather at the lower end of the estimates found in the literature. It is lower than the reported persuasion rates of news media (which range from 2 p.p. to 20 p.p. for media in the United States), but it is comparable with the 1.0 p.p. persuasion rate of direct mailing (Gerber and Green, 2000) and the 0.1-1.0 p.p. persuasion rate of political advertising (Spenkuch and Toniatti, 2016). The similarity to the persuasion rate of direct mailing and advertising is not surprising, the studies of traditional media that we cite compute persuasion rate for voting, not donating, and the fraction of the public who votes is greater than the fraction of those who donate to politicians. Moreover, for donations to be reported in the FEC database, the donations from an individual must exceed $200 (either as the sum of smaller installments or in one single donation). Therefore it is likely that we are underestimating the actual persuasive power of communication via Twitter.

As another exercise, we can compute persuasion rate for implied donations by assuming that for every donation we observe in our dataset, a fixed multiple of donations under $200 exist. For instance, if there were 2 donations below $200 for every donation above $200, our persuasion rate becomes three times the estimates given above (1.5-2.2 p.p.) and approaches the persuasion rates reported for newspaper endorsements (6 p.p. for unexpected endorsements, or 2 p.p. for expected endorsement for Chiang and Knight (2011) and the influence of TV adoption on turnout (4.4 p.p. for Gentzkow (2006)). Moreover, we would get similar persuasion rates (1.5 p.p.) if we were to assume that the effect persists upto eight weeks as implied by our window specifications (Table 4).

## A.2 Bias from Unobservables

Our identification strategy is built around a difference-in-differences strategy which controls for fine-grained fixed effects and heterogeneity with respect to the observable state characteristics. Though there is no particular reason for why unobservable determinants of donations could affect high- and low- Twitter penetration states differentially,[41] and placebo tests are not consistent with some obvious violations of our identifying assumptions, we cannot rule out with certainty that some unobservable factors explain our results. Below, we follow the strategy proposed by Altonji et al. (2005) (AET), which was implemented by Nunn and Wantchekon (2011), Galor and Özak (2016), DellaVigna et al. (2014), Adena et al. (2014), among others. This approach exploits the idea that we can use the selection on observables to assess the potential bias due to unobservables. Such a strategy allows us to determine how much stronger the selection on unobservables would have to be compared to the selection on observables in order to fully explain our results.

To perform this test we need two entities: (1) $\beta_f$, which is the coefficient of interest from the regression including the full set of observables and (2) $\beta_r$, which is the coefficient of interest from the regression which includes a restricted set of controls, or no controls at all. When the unobservables are positively correlated with the observables, to assess the potential bias coming from unobservables, we compute the ratio

$$\frac{\beta_f}{\beta_r - \beta_f}.$$

The intuition behind the formula is simple. First, consider why the ratio is decreasing in $\beta_r - \beta_f$. The smaller the difference between $\beta_r$ and $\beta_f$ is, the less is the estimate affected by selection on observables, and the stronger the selection on unobservables needs to be (relative to observables) to explain away the entire effect. Next, consider the intuition behind $\beta_f$ in the numerator. The larger $\beta_f$ is, the greater is the effect that needs to be explained away by selection on unobservables, and therefore the higher is the ratio.

---

[41]One potential reason is that Twitter penetration merely serves as a proxy for income, education, internet access, or other socioeconomic characteristics of a state, and what we observe is a higher responsiveness to the shock (joining Twitter) in richer, better connected, or more liberal places. To see if this is the case, we test whether donations received can be explained by differential effects of entry on Twitter with different controls (logarithm of population in the state, median household income in a state, the share of people who earn over $250,000 annually, share of people with a college education, share who voted for Bush in 2004 and share of African Americans). We report these results in Table A19. For comparison, the coefficient for Twitter entry interacted with Twitter penetration (from our baseline specification) is reproduced in column (1). Results suggest that the interactions of being on Twitter with each of the mentioned controls are small in magnitude and statistically insignificant (columns (2)-(6)). We also find that states with higher Twitter penetration do not see significantly more Twitter account openings. (Please see Table A20).

In our analysis, we consider the restricted regression which only has time and politician level fixed effects while the regressions which are 'unrestricted' have the full set of controls of 'onTwitter' interacted with demographic controls. We also contrast the restricted regression with fixed effects relative to one which controls for campaign expenditures. In discussing these estimates, we include results with subsamples referring to a set of analyses we use in the rest of the paper (new vs. repeat donors, donations to Democrats vs. Republicans, results excluding the year 2009, House vs. Senate candidates).

The coefficient ratio tests given in Table A1 measures the statistic for both the amount of donations and the number of donations as dependent variables in the regressions. Of the ratios reported, none of them is less than one. The minimum value is 10.4, in absolute terms. The ratios range from 10.4 to 158.1 in absolute terms with the mean of 47.7. Therefore, to attribute our entire difference-in-differences estimates to selection effects, selection on unobservables would have to be at least 10.4 times greater than the selection on observables and, on average, over forty-seven times greater. In our view, these results make it less likely that the estimated effect reported for adopting Twitter is fully driven by unobservables. To add some context, a similar exercise carried out in Nunn and Wantchekon (2011) reports ratios which range from three to eleven, with an average of four. Galor and Özak (2016) report AET coefficient ratio statistics ranging from 3 to 21.5, with a mean of 10.

It is also important that most of the coefficient ratio tests in the cases in which we include interactions with demographics yield negative statistics since, with the introduction of controls, our estimate of interest increases in a majority of cases. Hence, to rule out our results, we would need to construct an argument based on negative selection on unobservables.

Finally, if we control for campaign expenditures, our estimates have marginally smaller magnitudes, implying that campaign expenditures could indeed be controlling for some unobserved campaign activity. Even if we focus only on just the coefficient ratio tests from this set, it is unlikely that our results will be explained away by such unobservables since the minimum statistic here is 15.95 in absolute value.

Overall, testing for the bias from unobservables increases the confidence in our estimates as the tests we report provide bounds for the potential bias.

Table A1: Using Selection on Observables to Analyze Bias from Unobservables

| Regression Specification | Controls in Restricted Set (R) | Controls in Full (F) Set | Coeff. Ratio Test Amount of Donations | Coeff. Ratio Test Count of Donation |
|---|---|---|---|---|
| Baseline | Only Fixed Effects | Interacted Controls | -30.37 | -36.22 |
| New Donors | Only Fixed Effects | Interacted Controls | -63.28 | -26.86 |
| Democrat | Only Fixed Effects | Interacted Controls | 66.71 | -65.24 |
| Without 2009 | Only Fixed Effects | Interacted Controls | -11.2615 | -10.4647 |
| House | Only Fixed Effects | Interacted Controls | -17.23 | -17.30 |
| Baseline | Only Fixed Effects | Only Campaign Exp. | 50.3 | 26.35 |
| New Donors | Only Fixed Effects | Only Campaign Exp. | 64.33 | -158.1 |
| Democrat | Only Fixed Effects | Only Campaign Exp. | 45.39 | 46.37 |
| Without 2009 | Only Fixed Effects | Only Campaign Exp. | 41 | 88.17 |
| House | Only Fixed Effects | Only Campaign Exp. | 73.15 | 15.95 |

Column (1) in this Table describes which regressions we analyze in each case. Columns (2) and (3) describe the controls used in the 'restricted' regression ($\beta_r$) and the regression which has the full set of controls ($\beta_f$). The coefficient ratio test is computed as $\frac{\beta_f}{\beta_r - \beta_f}$.

## A.3 Theoretical Model

We sketch out a simple partial equilibrium framework of donation decisions by potential political donors. We analyze donation decisions in situations where politicians do and do not use Twitter. In this framework, we abstract away from explicitly modeling the strategic decision of politicians to join Twitter. We use the model to derive some testable predictions on donation decisions which we then take to data.

Consider a setting where politicians can be either new or experienced, indexed by $i \in \{e, n\}$. A politician $i$ has a 'type' or quality, $\theta_i$. The politician knows her $\theta_i$. There is a unit mass of potential donors. We assume that all potential donors want a higher 'quality' politician which, in this context, can be interpreted as competence, honesty, or experience of a politician.[42] We adopt a separable utility framework for donors similar to Chiang and Knight (2011) and Matějka and Tabellini (2016).

---
[42] Analyzing quality instead of ideology is more pertinent in our context, since we analyze donations within states, where ideological differentiation within a party would be limited. This modeling choice is in line with Durante and Knight (2012) as well as Knight and Chiang (2011).

A5

An individual donor $d$ has the following utility from donating to politician $i$:[43]

$$U_{di} = \theta_i - c_d$$

The term $c_d$ captures the cost of donating. We normalize the outside option of the donors to 0. The donors do not observe $\theta_i$ but hold (unbiased) prior beliefs such that

$$\theta_i \sim N(\bar{\theta}_i, \sigma_{i0}^2)$$

We assume that $\bar{\theta}_e > \bar{\theta}_n$ which will imply that ex-ante, without Twitter, experienced politicians have an advantage in receiving higher donations relative to newer politicians. We will focus on the case where also $\sigma_{n0}^2 > \sigma_{e0}^2$. A higher variance for new politicians implies that ex-ante, the donors place less confidence in their estimate of $\theta_n$ relative to $\theta_e$. This structure is in line with the evidence that experienced politicians hold an informational advantage over newer candidates as documented by Oliver and Ha (2007).

If a politician joins Twitter then she can provide information to the donors or could send persuasive messages. The politician can send a message $m$ to voters such that:

$$m_i = \bar{\theta}_i + \epsilon_i$$

with $\epsilon_i \sim N(\mu, \sigma_{i\epsilon}^2)$ with $\mu > 0$.

To highlight how joining Twitter affects donations differently for new and experienced politicians, we analyze the donations received by each type of politician with and without Twitter. If the politician does not join Twitter, donor $d$ will donate if

$$E(\theta_i) \geq c_d$$

If a politician joins Twitter then she will send a message $m_i$ which will be used by the donors to update their beliefs about $\theta_i$. The posterior belief after seeing $m_i$ is:

$$E(\theta_i|m_i) = V_{i0}m_i + V_{i\epsilon}\bar{\theta}_i$$

where $V_{i0} = \left(\frac{\sigma_{i0}^2}{\sigma_{i0}^2 + \sigma_{i\epsilon}^2}\right)$ and $V_{i\epsilon} = \left(\frac{\sigma_{i\epsilon}^2}{\sigma_{i0}^2 + \sigma_{i\epsilon}^2}\right)$. If a politician joins Twitter, donor $d$ will donate if

$$E(\theta_i|m_i) \geq c_d$$

---

[43]The linear utility framework is in line with (Chiang and Knight, 2011) and Durante and Knight (2012). Matějka and Tabellini (2016) adopt a more general framework where $u(\theta_i)$ is concave and differentiable.

We define $\Delta_i \equiv E(\theta_i|m_i) - E(\theta_i)$. If $\epsilon_i > 0$, then we can establish the following proposition.

**Proposition 1.** *A new politician is more likely to gain from joining Twitter relative to an experienced one if her messages are more informative than those of an experienced politician (i.e., $\sigma_{e\epsilon}^2 > \sigma_{n\epsilon}^2$).*

**Proof.** The proof follows straight from writing out the expressions for $\Delta_i$. $E(\theta_i|m_i) - E(\theta_i)$ is simply $\left(\frac{\sigma_{i0}^2}{\sigma_{i0}^2 + \sigma_{i\epsilon}^2}\right)\epsilon_i$. This implies that $\Delta_n - \Delta_e = V_{n0}\epsilon_n - V_{e0}\epsilon_e$. The results in the proposition follow directly.

**Donations and Twitter Penetration:** We listed the results when there is universal access to Twitter and all donors observe how informative the use of Twitter is for all politicians. As in our empirical model, we assume that there are different geographical regions (states), $s \in \{1, 2, ..., S\}$ with different Twitter usage. Each state has a unit mass of potential donors. Moreover, we assume that only a (random) fraction $\phi_s$ uses Twitter. This assumption is in line with Butters (1977). This penetration coefficient varies across states with $\phi_1 \geq \phi_2 \geq .... \geq \phi_S$. Assuming that Twitter penetration is the only dimension which varies across regions, we can easily see that politicians in regions with a higher $\phi_s$ will receive a bigger increase in donations by joining Twitter:

$$\phi_s \Delta_i \leq \phi_{s-1} \Delta_i$$

This also shows that if $\phi_s = 0$ for some $s$ then in that region there will be an insignificant increase in donations for both experienced and new politicians.

Based on the simple theoretical model we presented, we develop four key testable hypotheses:

1. The total donations received by a politician increase upon adopting Twitter and sending messages.

2. The gain from Twitter adoption is higher for new politicians compared to experienced politicians when they send more informative messages.

3. The gain from adopting Twitter is higher for a politician who sends more informative tweets.

4. The states with high Twitter penetration donate more to politicians adopting Twitter compared to the states with low Twitter penetration.

## A.4 Robustness Checks and Heterogeneous Effects

### A.4.1 Heterogeneous Effects for Democrats and Republicans

Republican and Democratic voters differ in demographic characteristics. Democratic voters are generally ethnically more diverse, have higher education, are religiously unaffiliated, and have lower income. One or more of these characteristics may correlate with internet and social media use, implying that candidates registered with the Democratic Party may have higher returns from adopting Twitter because the medium appeals to their constituents. We test whether Twitter has an asymmetric effect on candidates from the two parties.

The results in Table A2 suggest that the impact of Twitter is more pronounced for Democratic candidates, with the corresponding coefficients for Republicans being smaller and not statistically significant. Moreover, these differences in donation amount and number of donations is statistically significant across Democrats and Republicans, but only for new politicians and not for experienced ones. These results suggest that Twitter adoption has heterogeneous effects across the two party candidates, possibly because of the demographic differences between the target audiences of the two parties.

### A.4.2 Twitter Followers and Donations

To test for the overlap between the users on Twitter and donors, using the first and last names along with the information on geographical locations (state of the Twitter user), we match the names of the donors to a politician to the names of her followers on Twitter. We are able to achieve a match rate of 3.65%. Note that this match rate is low because the proportion of users who share their location information is very small, reported to be less than 1% of Twitter users (Sloan and Morgan, 2015). However, this seemingly low rate is comparable to the match rates in other studies of Twitter users (e.g., 10% in Barberá et al. (2013)).

We acknowledge that matching is not perfect, but we expect it to only introduce attenuation bias to our results. To assess the impact of joining Twitter and whether the donations came from followers of politicians on Twitter, we focus on candidate-weeks which have donations from both followers and non followers. Moreover, due to power issues, we focus on the full set of politicians and do not split the sample into new and experienced politicians.

Table A3 shows that a significantly higher amount of total donations come from followers after a politician adopts Twitter (columns (1) and (2)). We find a similar significant difference when we look at the donation count (columns (3)-(4)). The direction of the results is in line with our expectations.

Table A2: Joining Twitter and Amount of Donations: Democrat vs. Republican Candidates

| | Log (Aggregate Donations) | | | | | |
|---|---|---|---|---|---|---|
| | Democrat | | | Republican | | |
| | (1) | (2) | (3) | (4) | (5) | (6) |
| VARIABLES | All | New | Experienced | All | New | Experienced |
| onTwitter x Twit_Penet | 191.3717*** | 294.0677*** | -17.1925 | 42.2747 | 105.1078 | -66.6532 |
| | (62.6423) | (69.1821) | (113.6123) | (62.1364) | (63.2500) | (121.6479) |
| onTwitter | 0.6599 | 4.4521** | -3.7025 | 2.2577 | 1.1090 | 3.7109 |
| | (1.4409) | (1.8825) | (2.7531) | (1.5641) | (1.8858) | (2.4564) |
| Observations | 234,936 | 92,236 | 142,700 | 331,032 | 144,504 | 186,528 |
| R-squared | 0.8263 | 0.9029 | 0.7855 | 0.8167 | 0.8686 | 0.7840 |
| Politician-Month FE | Y | Y | Y | Y | Y | Y |
| Week of Month FE | Y | Y | Y | Y | Y | Y |
| Baseline controls x on Twitter | Y | Y | Y | Y | Y | Y |
| Implied Twitter effect for 2009 | .054*** | .083*** | -.005 | .011 | .026 | -.019 |
| Implied Twitter effect for 2014 | .391*** | .623*** | -.024 | .094 | .234 | -.148 |
| | Log (Number of Donations) | | | | | |
| onTwitter x Twit_Penet | 42.0835** | 71.8004*** | -14.1937 | -0.7878 | 16.9751 | -32.5209 |
| | (19.2961) | (18.9703) | (37.6522) | (13.8848) | (12.8675) | (28.4836) |
| onTwitter | 0.3556 | 1.0184* | -0.3770 | 0.2663 | 0.2111 | 0.1784 |
| | (0.4055) | (0.5364) | (0.8295) | (0.3077) | (0.4372) | (0.4381) |
| Observations | 234,936 | 92,236 | 142,700 | 331,032 | 144,504 | 186,528 |
| R-squared | 0.8474 | 0.9175 | 0.8081 | 0.8298 | 0.8873 | 0.7926 |
| Politician-Month FE | Y | Y | Y | Y | Y | Y |
| Week of Month FE | Y | Y | Y | Y | Y | Y |
| Baseline controls x on Twitter | Y | Y | Y | Y | Y | Y |
| Implied Twitter effect for 2009 | .012** | .02*** | -.004 | 0 | .004 | -.009 |
| Implied Twitter effect for 2014 | .086** | .152*** | -.02 | -.002 | .038 | -.072 |

Note: Robust standard errors clustered at the level of the state and week in parenthesis. *** p<0.01, ** p<0.05, * p<0.1. The dependent variables are Log (Aggregate Donations) and Log (Number of Donations). Columns (1) and (4) include all politicians while columns (2) and (5) include only new and columns (3) and (6) have the experienced politicians. State level baseline controls interacted with the politician being on Twitter include the median household income and population size. Average of pageviews relative to all pageviews in state-year is used as Twitter penetration measure. All specifications include Politician-month and week of month fixed effects.
2. A triple interaction specification comparing the aggregate donations to Democrats and Republicans finds the p-values of 0.003 (all), 0.000 (new) and 0.619 (experienced). For the number of donations, corresponding p-values are 0.003 (all), 0.000 (new) and 0.512 (experienced).

Table A3: Joining Twitter and Donations from Followers

| | Log(Aggregate Donations) from Followers | | Log(Number of Donations) from Followers | |
|---|---|---|---|---|
| | (1) | (2) | (3) | (4) |
| VARIABLES | All | All | All | All |
| onTwitter x Twit_Penet | 714.7294 | 546.8622*** | 133.7473* | 187.2788*** |
| | (582.2572) | (11.7478) | (77.7619) | (3.1360) |
| onTwitter | -4.1190 | 80.5549*** | -0.5922* | 6.6755*** |
| | (2.7593) | (0.9960) | (0.3182) | (0.2901) |
| | | | | |
| Observations | 6,616 | 6,616 | 6,616 | 6,616 |
| R-squared | 0.4859 | 0.4863 | 0.6879 | 0.6880 |
| Politician-Month FE | Y | Y | Y | Y |
| Week of Month FE | Y | Y | Y | Y |
| Baseline controls x on Twitter | N | Y | N | Y |
| Implied Twitter effect for 2009 | .147 | .113*** | .028* | .039*** |
| Implied Twitter effect for 2014 | 1.767 | 1.352*** | .331* | .463*** |

Note: Robust standard errors clustered at the level of the state in parenthesis. *** p<0.01, ** p<0.05, * p<0.1. The dependent variables are Log (Aggregate Donations) and Log (Number of Donations). Columns (1)-(4) include all politicians. State level baseline controls interacted with the politician being on Twitter include the median household income and population size. Average of pageviews relative to all pageviews in state-year is used as Twitter penetration measure. All specifications include Politician-month and week of month fixed effects.

### A.4.3 Excluding Campaign Periods

While the placebo checks we reported rule out the possibility of simultaneous campaign events driving our results, as another test we run our specification using only data from off-election years and the months over which a politician is less likely to be actively involved in campaign efforts. Elections take place in even-numbered years (2010, 2012 and 2014 in our sample), so it is likely that in the first half of each of the odd-numbered years (2009, 2011, 2013) campaign activities would be limited. We re-estimate our diff-in-diff specifications using data only from the first six months of 2009, 2011 and 2013. In Table A4, we find that even focusing only on this disconnected 18 month period, the effect of using Twitter persists for new politicians looking at the interaction term (column (2)) for aggregate donations, and remains insignificant for the experienced politicians (columns (3)). Our qualitative results hold outside of campaign periods as well, suggesting that they are less likely to be driven by unobserved contemporaneous campaign events. Our data highlights that a disproportionate number of Twitter accounts were opened in 2009. While including the politician-month fixed effects and week of month fixed effect (or week fixed effects) account for any idiosyncrasies of a particular time period, we also test if our estimates are driven by only this year's data. Columns (4)-(6) of Table A4 show that our baseline results replicate after excluding politicians who joined Twitter in 2009.

### A.5 Donations above $1,000

Our results up to here are based on donations below $1,000. We focused on these small donations as they are more likely to be coming from individuals and are likely to be affected by the information disseminated via Twitter. Larger donations are more likely to be driven by instrumental motivation

Table A4: Joining Twitter Outside Campaign Periods and Excluding 2009

| | Log(Aggregate Donations) | | | | | |
|---|---|---|---|---|---|---|
| | Outside Campaign Periods | | | Excluding the Year 2009 | | |
| | (1) | (2) | (3) | (4) | (5) | (6) |
| VARIABLES | All | New | Experienced | All | New | Experienced |
| onTwitter x Twit_Penet | 140.7951 | 345.8642*** | -54.2303 | 187.9085*** | 250.5351*** | 61.6658 |
| | (104.2754) | (125.3045) | (123.9820) | (56.8522) | (60.4253) | (114.0301) |
| onTwitter | -0.2632 | 1.1923 | -2.0445 | 3.0789** | 4.8293** | 0.8521 |
| | (1.7494) | (1.6251) | (3.2406) | (1.3851) | (2.2532) | (2.4786) |
| Observations | 141,492 | 61,984 | 79,508 | 471,640 | 173,028 | 298,612 |
| R-squared | 0.7923 | 0.8759 | 0.7556 | 0.8254 | 0.8838 | 0.7959 |
| Politician-Month FE | Y | Y | Y | Y | Y | Y |
| Week of Month FE | Y | Y | Y | Y | Y | Y |
| Baseline controls x on Twitter | Y | Y | Y | Y | Y | Y |
| Implied Twitter effect for 2009 | .039 | .087*** | -.015 | - | - | - |
| Implied Twitter effect for 2014 | - | - | - | .418*** | .557*** | .137 |

Note: Robust standard errors clustered at the level of the state and week in parenthesis. *** p<0.01, ** p<0.05, * p<0.1. The dependent variables are Log (Aggregate Donations). All politicians are included in columns (1) and (4), while it is the new politicians in columns (2) and (5) and experienced politicians in (3) and (6). State level baseline controls interacted with the politician being on Twitter include the median household income and population size. Average of pageviews relative to all pageviews in state-year is used as Twitter penetration measure. All specifications include Politician-month and week of month fixed effects.

(for example, lobbying, or gaining access to a candidate). In this section, we investigate what happens if we look at larger donations between the values of $1,000 and $3,000. Table A5 summarizes these results. In the tables, the interaction term is not statistically significant for the sample of all politicians (columns (1)-(2)). However, we still see some increase in the donations for new politicians (columns (3)-(4)), perhaps compensated by a decrease in the donations for more experienced politicians (columns (5)-(6)). For donations in the range of $1000-$3,000, the average weekly amount received by a candidate in the month of opening a Twitter account is $2,031. For a new politician, the same number is $1,799.[44] A calculation based on Table A5 (column (3)) suggests that joining Twitter increases donations received by the new candidates by $1,799×0.024 × 2.79 = $120 for 2009 and by $1,799×0.214 × 2.79 = $1,074 for 2014. This corresponds to an additional implied gain of 0.05%-0.4% of the total campaign funds for a new House politician. The estimates in Table A6 are suggestive of no effects beyond a 4-week window. We were also not able to find any impact of Twitter for political donations larger than $3,000 (see Table A7), consistent with the likelihood that these donations are driven by different motivations and major donors are less impacted by the information originating from Twitter. Thus for larger donations Twitter shows an effect in the same direction as donations under $1,000, but this effect is much less precisely estimated.

---

[44]See Tables A8 and A9 in the online appendix.

Table A5: Amount of Donations between $1,000-$3,000

|  | Log(Aggregate Donations) | | | | | |
|---|---|---|---|---|---|---|
|  | (1) | (2) | (3) | (4) | (5) | (6) |
| VARIABLES | All | All | New | New | Experienced | Experienced |
| onTwitter x Twit_Penet | -5.3026 | -2.2504 | 96.1866** | 96.0919** | -181.9087** | -172.9096** |
|  | (32.1576) | (31.7351) | (46.6266) | (45.6727) | (70.1591) | (69.6449) |
| onTwitter | 0.3893*** | 1.4379 | 0.2169** | 0.2460 | 0.7615*** | 2.9007* |
|  | (0.0895) | (0.8925) | (0.0929) | (1.3505) | (0.1740) | (1.6458) |
| Observations | 565,968 | 565,968 | 236,740 | 236,740 | 329,228 | 329,228 |
| R-squared | 0.6856 | 0.6856 | 0.7640 | 0.7640 | 0.6440 | 0.6440 |
| Politician-Month FE | Y | Y | Y | Y | Y | Y |
| Week of Month FE | Y | Y | Y | Y | Y | Y |
| Baseline controls x on Twitter | N | Y | N | Y | N | Y |
| Implied Twitter effect for 2009 | -.001 | -.001 | .024** | .024** | -.051** | -.048** |
| Implied Twitter effect for 2014 | -.012 | -.005 | .214** | .214** | -.405** | -.385** |

Note: Robust standard errors clustered at the level of the state and week in parenthesis. *** p<0.01, ** p<0.05, * p<0.1. The dependent variable is Log (Aggregate Donations). Columns (1)-(2) include all politicians while columns (3)-(4) include only new and columns (5)-(6) have the experienced politicians. State level baseline controls interacted with the politician being on Twitter include the median household income and population size. Average of pageviews relative to all pageviews in state-year is used as Twitter penetration measure. All specifications include Politician-month and week of month fixed effects.

Table A6: Amount of Donations between $1,000-$3,000: Window Specification

|  | Log(Aggregate Donations) | | | | | | | | |
|---|---|---|---|---|---|---|---|---|---|
|  | (1) | (2) | (3) | (4) | (5) | (6) | (7) | (8) | (9) |
| Window | 4 weeks | 5 weeks | 6 weeks | 4 weeks | 5 weeks | 6 weeks | 4 weeks | 5 weeks | 6 weeks |
| VARIABLES | All | All | All | New | New | New | Experienced | Experienced | Experienced |
| onTwitter x Twit_Penet | 41.3812 | 41.9163 | 30.0603 | 69.6228 | 58.1174 | 49.8498 | -31.6855 | -18.5865 | -34.8913 |
|  | (35.6075) | (35.0243) | (34.9839) | (55.7036) | (55.9863) | (55.8109) | (57.0875) | (56.2695) | (53.0211) |
| onTwitter | 0.1680* | 0.1615 | 0.1870* | 0.2531* | 0.2797** | 0.2857** | 0.1460 | 0.1046 | 0.1622 |
|  | (0.0980) | (0.0989) | (0.0981) | (0.1281) | (0.1251) | (0.1203) | (0.1791) | (0.1860) | (0.1863) |
| Observations | 17,596 | 19,221 | 20,846 | 10,382 | 11,328 | 12,274 | 7,205 | 7,886 | 8,567 |
| R-squared | 0.4884 | 0.4774 | 0.4671 | 0.5251 | 0.5170 | 0.5131 | 0.4817 | 0.4662 | 0.4510 |
| Politician FE | Y | Y | Y | Y | Y | Y | Y | Y | Y |
| Week FE | Y | Y | Y | Y | Y | Y | Y | Y | Y |
| Implied Twitter effect for 2009 | .01 | .011 | .008 | .019 | .016 | .014 | -.008 | -.005 | -.009 |
| Implied Twitter effect for 2014 | .092 | .093 | .067 | .157 | .131 | .113 | -.06 | -.037 | -.07 |

Note: Robust standard errors clustered at the level of the state and week in parenthesis. *** p<0.01, ** p<0.05, * p<0.1. Columns (1)-(3) include all politicians while columns (4)-(6) include only new and columns (7)-(9) have the experienced politicians. Average of pageviews relative to all pageviews in state-year is used as Twitter penetration measure. All specifications include Politician and week fixed effects.

Table A7: Amount and Number of Donations (Over $3,000)

|  | Log (Aggregate Donations) | | | Log (Number of Donations) | | |
|---|---|---|---|---|---|---|
|  | (1) | (2) | (3) | (4) | (5) | (6) |
| VARIABLES | All | New | Experienced | All | New | Experienced |
| on Twitter x Twit_Penet | 1.0314 | -0.4855 | 3.4377 | 0.0742 | -0.5837 | 1.1508 |
|  | (13.3907) | (18.9927) | (25.9391) | (1.7804) | (2.5629) | (2.9272) |
| onTwitter | 0.4457 | 0.6035 | 0.2451 | 0.0641 | 0.0955 | 0.0251 |
|  | (0.3459) | (0.4955) | (0.4688) | (0.0464) | (0.0697) | (0.0521) |
| Observations | 565,968 | 236,740 | 329,228 | 565,968 | 236,740 | 329,228 |
| R-squared | 0.5885 | 0.6082 | 0.5780 | 0.6625 | 0.6948 | 0.6451 |
| Politician-Month FE | Y | Y | Y | Y | Y | Y |
| Week of Month FE | Y | Y | Y | Y | Y | Y |
| Baseline controls x on Twitter | Y | Y | Y | Y | Y | Y |
| Implied Twitter effect for 2009 | 0 | 0 | .001 | 0 | 0 | 0 |
| Implied Twitter effect for 2014 | .002 | -.001 | .008 | 0 | -.001 | .003 |

Note: Robust standard errors clustered at the level of the state and week in parenthesis. *** p<0.01, ** p<0.05, * p<0.1. The dependent variables are Log (Aggregate Donations) and Log (Number of Donations). Columns (1)-(4) include all politicians while columns (2)-(5) include only new and columns (3)-(6) have the experienced politicians. State level baseline controls interacted with the politician being on Twitter include the median household income and population size. Average of pageviews relative to all pageviews in state-year is used as Twitter penetration measure. All specifications include Politician-month and week of month fixed effects.

## A.6  Additional Tables

In this section, we provide a set of tables to support the exercises provided in the main document. First set of tables we provide are aimed at describing the data we work with. Tables A8 – A9 provide the summary statistics for politicians and their donations. In Table A10, we provide a summary of our main Twitter penetration measure for each year and each state.

The next set of tables provided in this section are aimed at demonstrating the robustness of our key findings. Table A11 replicates our benchmark specification, including the campaign accounts. Table A12 carries out the benchmark specification excluding the politicians who had an account on Facebook prior to opening an account on Twitter. Table A13 tests if the timing of opening an account on Twitter is correlated with the timing of opening a Twitter account, and demonstrates that it is not. In Tables A14–A15, we replicate our benchmark specification with a number of alternate penetration measures and show that the same qualitative relationship follows.

Table A16 displays the coefficients for the controls added to the benchmark specifications for transparency. In the table, population and income largely do not show a correlation. Table A17 demonstrates that most of the expense categories are correlated with aggregate donations. Since expenses are largely paid from funds coming from the donations, this is expected. In Table A18, we show that the effect of Twitter is robust to controlling for the contemporaneous and lagged campaign expenditures. Table A19 tests if opening a Twitter penetration is a proxy for another characteristic of a state by including the onTwitter dummy interacted with a set of demographics rather than Twitter penetration. We find that none of the other characteristics predict aggregate donations. Table A20 shows that states with higher Twitter penetration do not see significantly more Twitter account openings.

Tables A21 and A22 show that gain of new politicians is correlated with their tweeting activity and specific content. Donations go up for politicians who post more original tweets, rather than retweets, use more hyperlinks, use more anti-establishment related words or appear to be more "plugged in." Table A23 repeats the analysis we show for Twitter for Facebook, and finds a similar result. Finally, our baseline results are robust to adding week fixed effects (Table A24) as well as higher order (cubic) time trends (Table A25).

Table A8: Summary Statistics

| All Politicians | | | | | |
|---|---|---|---|---|---|
| | Observations | Mean | Std. Dev. | Min | Max |
| Aggregate Donations | 1,834 | 2,562.3 | 19,954.6 | 0 | 707,487.9 |
| Probability of Donations | 1,834 | 0.25 | 0.25 | 0 | 0.990 |
| Campaign Expenditures | 1,834 | 10,216.7 | 80157.4 | 0 | 2,753,732 |
| No. of News Mentions | 1,834 | 8.95 | 252.4 | 0 | 10,768.6 |
| No. of Blog Mentions | 1,834 | 6.07 | 150.17 | 0 | 6340 |
| Facebook Account Before | 1,834 | 0.028 | 0.141 | 0 | 1 |
| No. of Tweets | 1,834 | 0.50 | 1.42 | 0 | 10.29 |
| No. of Retweets | 1,834 | 24.27 | 285.55 | 0 | 10872.410 |
| No. of URLs | 1,834 | 0.33 | 0.99 | 0 | 8.70 |
| Anti-establishment Tweets | 1,834 | 0.01 | 0.01 | 0 | 0.32 |
| No. of Tweets with 'We' | 1,834 | 0.04 | 0.10 | 0 | 1.37 |
| Worried Score | 1,834 | 34.97 | 21.41 | 0 | 100 |

| Politician-Weeks at the Month of Joining Twitter | | | | | |
|---|---|---|---|---|---|
| | Observations | Mean | Std. Dev. | Min | Max |
| Aggregate Donations | 7,295 | 1,533.7 | 6,429.8 | 0 | 158269 |
| Probability of Donations | 7,295 | 0.29 | 0.46 | 0 | 1 |
| Campaign Expenditures | 7,223 | 6,718.50 | 53,410.2 | -798.02 | 2,662,027 |
| No. of News Mentions | 7,295 | 9.62 | 299.32 | 0 | 12400 |
| No. of Blog Mentions | 7,295 | 6.95 | 181.78 | 0 | 7,480 |
| Facebook Account Before | 7,295 | 0.01 | 0.09 | 0 | 1 |
| No. of Tweets | 7,295 | 0.41 | 2.87 | 0 | 85 |
| No. of Retweets | 7,295 | 2.58 | 71.49 | 0 | 4,700 |
| No. of URLs | 7,295 | 0.21 | 1.85 | 0 | 60 |
| Anti-establishment Tweets | 7,295 | 0.003 | 0.06 | 0 | 2 |
| Aggregate Donations $1K-$3K | 7,295 | 2,030.57 | 11,043.7 | 0 | 311,700 |

Table A9: Summary Statistics: New and Experienced

| | Politician-Weeks (Full Sample) | | | | | | | | | | |
|---|---|---|---|---|---|---|---|---|---|---|---|
| | New Politicians | | | | | Experienced Politicians | | | | Comparison | |
| | N | Mean | Std. Dev. | Min | Max | N | Mean | Std. Dev. | Min | Max | Difference | Std. Error |
| Aggregate Donations | 237548 | 1680.232 | 11893.629 | 0 | 2317786 | 334660 | 3188.400 | 83007.713 | 0 | 1.36e+07 | 1508.168*** | (145.549) |
| Probability of Donations | 237548 | 0.169 | 0.374 | 0 | 1 | 334660 | 0.309 | 0.462 | 0 | 1 | 0.140*** | (0.001) |
| Campaign Expenditures | 235036 | 6613.936 | 60315.829 | -19002.91 | 6613983 | 331556 | 12943.745 | 331925.131 | -1346930 | 5.93e+07 | 6329.809*** | (589.723) |
| News Mentions | 29011 | 3.278 | 118.069 | 0 | 19800 | 18665 | 18.295 | 407.248 | 0 | 19800 | 15.017*** | (3.060) |
| Blog Mentions | 29011 | 2.185 | 36.286 | 0 | 2440 | 18665 | 11.689 | 227.141 | 0 | 8220 | 9.504*** | (1.676) |
| Facebook Account Before | 237548 | 0.017 | 0.131 | 0 | 1 | 334660 | 0.035 | 0.183 | 0 | 1 | 0.0173*** | (0.000) |
| Tweets | 237548 | 0.285 | 2.656 | 0 | 278 | 334660 | 0.657 | 4.164 | 0 | 432 | 0.371*** | (0.009) |
| Retweets | 237548 | 0.976 | 97.235 | 0 | 37069 | 334660 | 40.810 | 6092.835 | 0 | 3338067 | 39.83*** | (10.534) |
| No. of URLs | 237548 | 0.175 | 1.748 | 0 | 113 | 334660 | 0.439 | 2.943 | 0 | 386 | 0.264*** | (0.006) |
| Tweets with 'We' | 237548 | 0.021 | 0.303 | 0 | 41 | 334660 | 0.044 | 0.419 | 0 | 53 | 0.023*** | (0.001) |
| Anti-establishment Tweets | 237548 | 0.004 | 0.091 | 0 | 8 | 334660 | 0.005 | 0.102 | 0 | 23 | 0.001*** | (0.000) |
| Worried Score | 210364 | 38.941 | 18.962 | 19 | 100 | 302876 | 39.023 | 18.718 | 19 | 100 | 0.0817 | (0.054) |
| | Politician-Weeks (At the Month of Joining Twitter) | | | | | | | | | | |
| | New Politicians | | | | | Experienced Politicians | | | | Comparison | |
| | N | Mean | Std. Dev. | Min | Max | N | Mean | Std. Dev. | Min | Max | Difference | Std. Error |
| Aggregate Donations | 4181 | 1073.584 | 5915.194 | 0 | 158269 | 3114 | 2151.353 | 7015.686 | 0 | 114623 | 1077.769*** | (155.482) |
| Probability of Donations | 4181 | 0.235 | 0.424 | 0 | 1 | 3114 | 0.378 | 0.485 | 0 | 1 | 0.143*** | (0.011) |
| Campaign Expenditures | 4132 | 4773.146 | 50369.890 | 0 | 1915575 | 3091 | 9319.025 | 57127.687 | -798.02 | 2662027 | 4545.878*** | (1292.227) |
| News Mentions | 4181 | 1.063 | 11.647 | 0 | 259 | 3114 | 21.118 | 457.723 | 0 | 12400 | 20.055* | (8.204) |
| Blog Mentions | 4181 | 1.357 | 38.936 | 0 | 2270 | 3114 | 14.459 | 274.386 | 0 | 7480 | 13.102** | (4.954) |
| Facebook Account Before | 4181 | 0.004 | 0.060 | 0 | 1 | 3114 | 0.015 | 0.123 | 0 | 1 | 0.012*** | (0.002) |
| Tweets | 4181 | 0.432 | 3.213 | 0 | 85 | 3114 | 0.384 | 2.331 | 0 | 60 | -0.048 | (0.065) |
| Retweets | 4181 | 0.361 | 14.962 | 0 | 956 | 3114 | 5.546 | 107.989 | 0 | 4700 | 5.184** | (1.949) |
| Tweets with URLs | 4181 | 0.220 | 2.127 | 0 | 60 | 3114 | 0.196 | 1.392 | 0 | 41 | -0.024 | (0.041) |
| Tweets with 'We' | 4181 | 0.027 | 0.280 | 0 | 6 | 3114 | 0.019 | 0.168 | 0 | 4 | -0.008 | (0.005) |
| Anti-establishment Tweets | 4181 | 0.002 | 0.044 | 0 | 1 | 3114 | 0.004 | 0.069 | 0 | 2 | -0.777 | (0.001) |
| Worried Score | 3766 | 38.966 | 18.919 | 19 | 100 | 2801 | 38.189 | 18.285 | 19 | 100 | 0.0817 | (0.463) |
| Aggregate Donation $1000-$3000 | 4181 | 1798.56 | 11560.62 | 0 | 279732 | 3114 | 2342.069 | 10302.69 | 0 | 311700 | -543.5086** | (261.3548) |

Table A10: Benchmark Twitter Penetration Measure (Avg. Pageviews) by state-year

|       | 2009    | 2010    | 2011    | 2012    | 2013    | 2014    |
|-------|---------|---------|---------|---------|---------|---------|
| AK    | 0.00120 | 0.00244 | 0.00127 | 0.00149 | 0.00047 | 0.00102 |
| AL    | 0.00066 | 0.00123 | 0.00515 | 0.00748 | 0.00391 | 0.00321 |
| AR    | 0.00106 | 0.00110 | 0.00393 | 0.00377 | 0.00427 | 0.00416 |
| AZ    | 0.00069 | 0.00148 | 0.00236 | 0.00264 | 0.00299 | 0.00390 |
| CA    | 0.00125 | 0.00206 | 0.00344 | 0.00431 | 0.00382 | 0.00644 |
| CO    | 0.00064 | 0.00074 | 0.00143 | 0.00385 | 0.00214 | 0.00453 |
| CT    | 0.00075 | 0.00268 | 0.00514 | 0.00848 | 0.00581 | 0.01181 |
| DC    | 0.00345 | 0.00143 | 0.00881 | 0.00994 | 0.00906 | 0.00992 |
| DE    | 0.00050 | 0.00068 | 0.00474 | 0.00695 | 0.00661 | 0.00379 |
| FL    | 0.00064 | 0.00162 | 0.00240 | 0.00500 | 0.00330 | 0.00490 |
| GA    | 0.00089 | 0.00215 | 0.00471 | 0.00748 | 0.00507 | 0.00712 |
| HI    | 0.00025 | 0.00088 | 0.00138 | 0.00357 | 0.00076 | 0.00504 |
| IA    | 0.00043 | 0.00056 | 0.00188 | 0.00464 | 0.00392 | 0.00907 |
| ID    | 0.00019 | 0.00143 | 0.00036 | 0.00105 | 0.00072 | 0.00143 |
| IL    | 0.00065 | 0.00121 | 0.00332 | 0.00631 | 0.00375 | 0.00616 |
| IN    | 0.00065 | 0.00124 | 0.00357 | 0.00514 | 0.00319 | 0.00666 |
| KS    | 0.00074 | 0.00093 | 0.00276 | 0.00328 | 0.00601 | 0.00846 |
| KY    | 0.00090 | 0.00107 | 0.00321 | 0.00519 | 0.00312 | 0.00445 |
| LA    | 0.00072 | 0.00079 | 0.00578 | 0.00527 | 0.00279 | 0.00452 |
| MA    | 0.00059 | 0.00132 | 0.00229 | 0.00698 | 0.00465 | 0.00624 |
| MD    | 0.00091 | 0.00241 | 0.00644 | 0.00957 | 0.00441 | 0.00806 |
| ME    | 0.00062 | 0.00061 | 0.00108 | 0.00214 | 0.00118 | 0.00467 |
| MI    | 0.00081 | 0.00141 | 0.00393 | 0.00658 | 0.00454 | 0.00650 |
| MN    | 0.00049 | 0.00133 | 0.00172 | 0.00396 | 0.00465 | 0.00591 |
| MO    | 0.00049 | 0.00093 | 0.00367 | 0.00713 | 0.00426 | 0.00582 |
| MS    | 0.00072 | 0.00173 | 0.00606 | 0.00872 | 0.00305 | 0.00481 |
| MT    | 0.00043 | 0.00018 | 0.00043 | 0.00209 | 0.00250 | 0.00246 |
| NC    | 0.00080 | 0.00214 | 0.00448 | 0.00697 | 0.00466 | 0.00913 |
| ND    | 0.00058 | 0.00038 | 0.00192 | 0.00106 | 0.00370 | 0.00145 |
| NE    | 0.00036 | 0.00050 | 0.00197 | 0.00378 | 0.00317 | 0.00551 |
| NH    | 0.00049 | 0.00031 | 0.00114 | 0.00575 | 0.00429 | 0.00770 |
| NJ    | 0.00074 | 0.00162 | 0.00410 | 0.00628 | 0.00469 | 0.00708 |
| NM    | 0.00086 | 0.00058 | 0.00082 | 0.00341 | 0.00166 | 0.00665 |
| NV    | 0.00080 | 0.00250 | 0.00233 | 0.00562 | 0.00671 | 0.00872 |
| NY    | 0.00078 | 0.00245 | 0.00284 | 0.00425 | 0.00394 | 0.00598 |
| OH    | 0.00088 | 0.00123 | 0.00492 | 0.00609 | 0.00430 | 0.00654 |
| OK    | 0.00091 | 0.00156 | 0.00101 | 0.00387 | 0.00232 | 0.00346 |
| OR    | 0.00082 | 0.00149 | 0.00213 | 0.00265 | 0.00205 | 0.00863 |
| PA    | 0.00061 | 0.00123 | 0.00310 | 0.00543 | 0.00402 | 0.00597 |
| RI    | 0.00230 | 0.00070 | 0.00572 | 0.01286 | 0.00743 | 0.01045 |
| SC    | 0.00089 | 0.00225 | 0.00403 | 0.00813 | 0.00509 | 0.00569 |
| SD    | 0.00024 | 0.00084 | 0.00068 | 0.00236 | 0.00313 | 0.00455 |
| TN    | 0.00057 | 0.00146 | 0.00407 | 0.00689 | 0.00367 | 0.00625 |
| TX    | 0.00089 | 0.00149 | 0.00302 | 0.00453 | 0.00383 | 0.00777 |
| UT    | 0.00060 | 0.00063 | 0.00093 | 0.00275 | 0.00171 | 0.00347 |
| VA    | 0.00090 | 0.00183 | 0.00493 | 0.00742 | 0.00413 | 0.00716 |
| VT    | 0.00048 | 0.00113 | 0.00088 | 0.00350 | 0.00519 | 0.00341 |
| WA    | 0.00060 | 0.00113 | 0.00182 | 0.00312 | 0.00453 | 0.00737 |
| WI    | 0.00072 | 0.00104 | 0.00235 | 0.00468 | 0.00366 | 0.00469 |
| WV    | 0.00045 | 0.00069 | 0.00242 | 0.00816 | 0.00490 | 0.00272 |
| WY    | 0.00018 | 0.00018 | 0.00081 | 0.00246 | 0.00208 | 0.00245 |
| Total | 0.00076 | 0.00127 | 0.00301 | 0.00520 | 0.00384 | 0.00576 |

Table A11: Joining Twitter and Donations with Campaign Accounts Included

|  | Log(Aggregate Donations) | | | Prob. of a Donation | | | Log(Number of Donations) | | |
|---|---|---|---|---|---|---|---|---|---|
|  | (1) | (2) | (3) | (4) | (5) | (6) | (7) | (8) | (9) |
| VARIABLES | All | New | Experienced | All | New | Experienced | All | New | Experienced |
| on Twitter x Twit_Penet | 129.7702*** | 217.1976*** | -50.4068 | 15.9449*** | 24.5988*** | -2.2660 | 24.3760* | 52.4972*** | -31.8362 |
|  | (46.9017) | (48.9254) | (87.1296) | (5.6369) | (6.1483) | (10.3824) | (13.4746) | (12.9017) | (24.9576) |
| onTwitter | 1.2803 | 2.1692 | 0.2374 | 0.1518 | 0.2436 | 0.0458 | 0.3459* | 0.7141* | -0.1098 |
|  | (1.0032) | (1.4021) | (2.1598) | (0.1371) | (0.1704) | (0.2788) | (0.1949) | (0.3711) | (0.5257) |
| Observations | 667,368 | 290,596 | 376,772 | 667,368 | 290,596 | 376,772 | 667,368 | 290,596 | 376,772 |
| R-squared | 0.8210 | 0.8732 | 0.7888 | 0.7876 | 0.8350 | 0.7565 | 0.8405 | 0.8978 | 0.8052 |
| Politician-Month FE | Y | Y | Y | Y | Y | Y | Y | Y | Y |
| Week of Month FE | Y | Y | Y | Y | Y | Y | Y | Y | Y |
| Baseline controls x on Twitter | Y | Y | Y | Y | Y | Y | Y | Y | Y |
| Implied Twitter effect for 2009 | .033*** | .054*** | -.014 | .004*** | .006*** | -.001 | .006* | .013*** | -.009 |
| Implied Twitter effect for 2014 | .289*** | .491*** | -.112 | .035*** | .056*** | -.005 | .054* | .119*** | -.071 |

Note: Robust standard errors clustered at the level of the state and week in parenthesis. *** p<0.01, ** p<0.05, * p<0.1. The dependent variables are Log (Aggregate Donations), Probability of receiving at least one donations, and Log (Number of Donations). All politicians are included in columns (1), (4) and (7), while it is the new politicians in columns (2), (5) and (8) and experienced politicians in (3), (6) and (9). State level baseline controls interacted with the politician being on Twitter include the median household income and population size. Average of pageviews relative to all pageviews in state-year is used as Twitter penetration measure. All specifications include Politician-month and week of month fixed effects.

Table A12: Baseline Excluding Politicians Who had Facebook Before

|  | Log(Aggregate Donations) | | | Log(Number of Donations) | | |
|---|---|---|---|---|---|---|
|  | (1) | (2) | (3) | (4) | (5) | (6) |
| VARIABLES | All | New | Experienced | All | New | Experienced |
| onTwitter x Twit_Penet | 107.5880** | 191.7927*** | -42.8838 | 16.8552 | 42.4916*** | -28.2385 |
|  | (45.3533) | (45.9474) | (90.4304) | (12.3295) | (11.6057) | (22.6353) |
| onTwitter | 1.3406 | 2.2891 | 0.0576 | 0.2185 | 0.4643 | -0.1147 |
|  | (1.0281) | (1.5473) | (2.0682) | (0.1765) | (0.3592) | (0.4692) |
| Observations | 550,239 | 232,574 | 317,665 | 550,239 | 232,574 | 317,665 |
| R-squared | 0.8218 | 0.8833 | 0.7855 | 0.8387 | 0.9008 | 0.8002 |
| Politician-Month FE | Y | Y | Y | Y | Y | Y |
| Week of Month FE | Y | Y | Y | Y | Y | Y |
| Baseline controls x on Twitter | Y | Y | Y | Y | Y | Y |
| Implied Twitter effect for 2009 | .03** | .048*** | -.012 | .005 | .011*** | -.008 |
| Implied Twitter effect for 2014 | .239** | .427*** | -.095 | .037 | .095*** | -.063 |

Note: 1. Robust standard errors clustered at the level of the state and week in parenthesis. *** p<0.01, ** p<0.05, * p<0.1. The dependent variables are Log (Aggregate Donations) and Log (Number of Donations). Columns (1) and (4) include all politicians while columns (2) and (5) include only new and columns (3) and (6) have the experienced politicians. State level baseline controls interacted with the politician being on Twitter include the median household income and population size. Average of pageviews relative to all pageviews in state-year is used as Twitter penetration measure. All specifications include Politician-month and week of month fixed effects.
2. A triple interaction specification comparing the aggregate donations to politicians with and without a Facebook account before finds the p-values of 0.044 (all), 0.009 (new) and 0.590 (experienced). For the number of donations, corresponding p-values are 0.011(all), 0.021 (new) and 0.108 (experienced).

Table A13: Correlation of Timing in Politician's Twitter Entry and Opening a Facebook Account

|  | Whether Politicians Had a Facebook Account | | | | | |
|---|---|---|---|---|---|---|
|  | (1) | (2) | (3) | (4) | (5) | (6) |
| VARIABLES | All | All | New | New | Experienced | Experienced |
| onTwitter x Twit_Penet | -0.4597 | -0.5020 | -0.0209 | -0.0088 | -1.1860 | -1.3389 |
|  | (0.2880) | (0.3199) | (0.0261) | (0.0225) | (0.7649) | (0.8568) |
| onTwitter | 0.0021 | -0.0107 | 0.0000 | 0.0047 | 0.0053 | -0.0312 |
|  | (0.0014) | (0.0114) | (0.0004) | (0.0051) | (0.0035) | (0.0258) |
| Observations | 565,968 | 565,968 | 236,740 | 236,740 | 329,228 | 329,228 |
| R-squared | 0.9961 | 0.9961 | 0.9926 | 0.9926 | 0.9974 | 0.9974 |
| Politician-Month FE | Y | Y | Y | Y | Y | Y |
| Week of Month FE | Y | Y | Y | Y | Y | Y |
| Baseline controls x on Twitter | N | Y | N | Y | N | Y |
| Implied Twitter effect for 2009 | 0 | 0 | 0 | 0 | 0 | 0 |
| Implied Twitter effect for 2014 | -.001 | -.001 | 0 | 0 | -.003 | -.003 |

Note: Robust standard errors clustered at the level of the state and week in parenthesis. *** p<0.01, ** p<0.05, * p<0.1. The dependent variable is whether the politician had a Facebook account before or not. Columns (1)-(2) include all politicians while columns (3)-(4) include only new and columns (5)-(6) have the experienced politicians. State level baseline controls interacted with the politician being on Twitter include the median household income and population size. Average of pageviews relative to all pageviews in state-year is used as Twitter penetration measure. All specifications include Politician-month and week of month fixed effects.

Table A14: Benchmark Specifications with Alternative Penetration Measures – I

| | (1) | (2) | (3) | (4) | (5) | (6) | (7) | (8) | (9) | (10) | (11) | (12) |
|---|---|---|---|---|---|---|---|---|---|---|---|---|
| | | | | | | Log(Aggregate Donations) | | | | | | |
| VARIABLES | New | Experienced | New | Experienced | New | Experienced | New | Experienced | New | Experienced | New | Experienced |
| onTwitter x 7day household | 7.2629* | 9.5706 | | | | | | | | | | |
| | (4.0512) | (6.5756) | | | | | | | | | | |
| onTwitter x Total_duration | | | 0.000039*** | 0.000035 | | | | | | | | |
| | | | (0.0000) | (0.0000) | | | | | | | | |
| onTwitter x Total_Households | | | | | 0.0018*** | 0.0011 | | | | | | |
| | | | | | (0.0007) | (0.0018) | | | | | | |
| onTwitter x Twit_Penet | | | | | | | 190.9910*** | -52.7947 | | | | |
| | | | | | | | (44.2721) | (90.7344) | | | | |
| onTwitter x Avg_Household_Share | | | | | | | | | 6.3244** | 2.0112 | | |
| | | | | | | | | | (2.7242) | (4.2373) | | |
| onTwitter x Duration Avg | | | | | | | | | | | 108.5391*** | -62.9260 |
| | | | | | | | | | | | (34.5968) | (70.3464) |
| onTwitter | 2.4675 | 0.8685 | 2.9673* | 0.7164 | 3.1687* | 0.6020 | 2.4187 | -0.0324 | 1.9289 | 0.0754 | 2.221 | -0.2188 |
| | (1.5684) | (1.9641) | (1.6305) | (2.0960) | (1.6998) | (2.2499) | (1.5299) | (2.0024) | (1.4758) | (1.9939) | (1.544) | (2.0776) |
| Observations | 232,204 | 324,716 | 236,740 | 329,228 | 236,740 | 329,228 | 236,740 | 329,228 | 236,740 | 329,228 | 236,740 | 329,228 |
| R-squared | 0.8821 | 0.7854 | 0.8827 | 0.7856 | 0.8827 | 0.7856 | 0.8818 | 0.7853 | 0.8827 | 0.7856 | 0.8827 | 0.7856 |
| Politician-Month FE | Y | Y | Y | Y | Y | Y | Y | Y | Y | Y | Y | Y |
| Week of Month FE | Y | Y | Y | Y | Y | Y | Y | Y | Y | Y | Y | Y |
| Baseline controls x on Twitter | Y | Y | Y | Y | Y | Y | Y | Y | Y | Y | Y | Y |
| Implied Twitter effect for 2009 | .138* | .182 | .001*** | .001 | .005*** | .003 | .048*** | -.015 | .013** | .003 | .027*** | -.018 |
| Implied Twitter effect for 2014 | .269* | .303 | .138*** | .15 | .177*** | .107 | .425*** | -.117 | .119** | .037 | .241*** | -.14 |

Note: Robust standard errors clustered at the level of the state and week in parenthesis. *** p<0.01, ** p<0.05, * p<0.1. The dependent variable is Log (Aggregate Donations). State level baseline controls interacted with the politician being on Twitter include the median household income and population size. All specifications include Politician-month and week of month fixed effects. Estimates in columns (7) and (8) come from a population weighted regression with the baseline penetration measure used.

A20

| VARIABLES | (1) New | (2) Experienced | (3) New | (4) Experienced | (5) New | (6) Experienced | (7) New | (8) Experienced | (9) New | (10) Experienced | (11) New | (12) Experienced |
|---|---|---|---|---|---|---|---|---|---|---|---|---|
| onTwitter x Sessions Median | 155.933*** (42.027) | -78.098 (82.324) | | | | | | | | | | |
| onTwitter x Median_Household_Share | | | 8.3843*** (2.6432) | -2.8516 (5.9464) | | | | | | | | |
| onTwitter x Sessions Avg | | | | | 155.9186*** (41.9372) | -81.4213 (80.9901) | | | | | | |
| onTwitter x Median Duration | | | | | | | 112.0688*** (40.1836) | -69.2784 (72.7620) | | | | |
| onTwitter x 30day household | | | | | | | | | 5.776* (3.1612) | 8.452* (4.348) | | |
| onTwitter x Median Views | | | | | | | | | | | 134.9873 (126.3772) | -113.2331 (129.9525) |
| onTwitter | 2.2000 (1.5445) | -0.1446 (2.0619) | 2.3417 (1.5594) | -0.1623 (2.0593) | 2.0761 (1.5051) | -0.1435 (2.0793) | 2.5873 (1.6079) | -0.4089 (2.0840) | 2.432 (1.576) | 0.775 (1.950) | 2.5679 (1.6295) | -0.4554 (2.0896) |
| Observations | 236,740 | 329,228 | 236,740 | 329,228 | 236,740 | 329,228 | 236,740 | 329,228 | 232,204 | 324,716 | 236,740 | 329 |
| R-squared | 0.8827 | 0.7856 | 0.8827 | 0.7856 | 0.8827 | 0.7856 | 0.8827 | 0.7856 | 0.882 | 0.785 | 0.8827 | 0.7856 |
| Politician-Month FE | Y | Y | Y | Y | Y | Y | Y | Y | Y | Y | Y | Y |
| Week of Month FE | Y | Y | Y | Y | Y | Y | Y | Y | Y | Y | Y | Y |
| Baseline controls x on Twitter | Y | Y | Y | Y | Y | Y | Y | Y | Y | Y | Y | Y |
| Implied Twitter effect for 2009 | .002*** | -.001 | .002*** | -.001 | .06*** | -.03 | .028*** | -.019 | 0.001* | 0.002* | 0.034 | -0.031 |
| Implied Twitter effect for 2014 | .017*** | -.008 | .019*** | -.006 | .256*** | -.111 | .249*** | -.154 | 0.013* | 0.019* | .3 | -0.252 |

Table A15: Benchmark Specifications with Alternative Penetration Measures – II

Note: Robust standard errors clustered at the level of the state and week in parenthesis. *** p<0.01, ** p<0.05, * p<0.1. The dependent variable is Log (Aggregate Donations). State level baseline controls interacted with the politician being on Twitter include the median household income and population size. All specifications include Politician-month and week of month fixed effects.

A21

Table A16: Aggregate Donations with Control Coefficients Displayed

| | | | | | Log (Aggregate Donations) | | | | | |
|---|---|---|---|---|---|---|---|---|---|---|
| VARIABLES | (1) All | (2) All | (3) All | (4) All | (5) New | (6) New | (7) New | (8) Experienced | (9) Experienced | (10) Experienced |
| onTwitter x Twit_Penet | -43.7011* | 102.6499** | 105.1469** | 106.1455** | 186.0268*** | 192.5849*** | 192.2501*** | -45.0641 | -51.3934 | -47.0084 |
| | (22.7379) | (45.5244) | (46.0190) | (45.5292) | (44.9294) | (45.7327) | (45.7696) | (93.7545) | (94.6549) | (91.0756) |
| onTwitter | 1.1938*** | 0.1879 | 0.9505 | 1.3162 | 0.1268 | 2.5061 | 2.3393 | 0.3961* | -1.1174 | -0.0791 |
| | (0.1212) | (0.1126) | (1.0889) | (1.0235) | (0.1140) | (1.6336) | (1.5472) | (0.2248) | (2.1463) | (2.0678) |
| onTwitter × log(Population) | | | -0.0482 | -0.0253 | | -0.1501 | -0.1600 | | 0.0958 | 0.1659 |
| | | | (0.0678) | (0.0786) | | (0.1037) | (0.1076) | | (0.1301) | (0.1180) |
| onTwitter × Income | | | | -0.0172 | | | 0.0076 | | | -0.0506* |
| | | | | (0.0124) | | | (0.0128) | | | (0.0270) |
| Observations | 565,968 | 565,968 | 565,968 | 565,968 | 236,740 | 236,740 | 236,740 | 329,228 | 329,228 | 329,228 |
| R-squared | 0.0188 | 0.8215 | 0.8215 | 0.8215 | 0.8828 | 0.8828 | 0.8828 | 0.7856 | 0.7856 | 0.7856 |
| Politician-Month FE | N | Y | Y | Y | Y | Y | Y | Y | Y | Y |
| Week of Month FE | N | Y | Y | Y | Y | Y | Y | Y | Y | Y |
| Implied Twitter effect for 2009 | -.012* | .029** | .029** | .03** | .047*** | .048*** | .048*** | -.013 | -.014 | -.013 |
| Implied Twitter effect for 2014 | -.097* | .228** | .234** | .236** | .414*** | .428*** | .428*** | -.1 | -.114 | -.105 |

Note: Robust standard errors clustered at the level of the state and week in parenthesis. *** p<0.01, ** p<0.05, * p<0.1. The dependent variable is Log (Aggregate Donations). Columns (1)-(4) include all politicians while columns (5)-(7) include only new and columns (8)-(10) have the experienced politicians. Columns (3), (6) and (9) include state level baseline controls the politician being on Twitter interacted with population size. Columns (4), (7) and (10) include state level baseline controls the politician being on Twitter interacted with the median household income and population size. Average of pageviews relative to all pageviews in state-year is used as Twitter penetration measure. Columns (2)-(10) specifications include Politician-month and week of month fixed effects.

A22

Table A17: Correlation of Expenditure Categories with Amount of Donations

| VARIABLES | Total Expend. (1) All | Contributions (2) All | Poll (3) All | Refunds (4) All | Fundraise (5) All | Transfers (6) All | Travel (7) All | Admin (8) All | Ads (9) All | Events (10) All | Materials (11) All | Donations (12) All | Loan Repay (13) All |
|---|---|---|---|---|---|---|---|---|---|---|---|---|---|
| Log(Aggregate Donations) | 0.0819*** | 0.0007 | 0.0016** | 0.0002 | 0.0237*** | 0.0000 | 0.0097*** | 0.0257*** | 0.0122*** | 0.0070*** | 0.0055*** | 0.0008 | 0.0002* |
|  | (0.0049) | (0.0006) | (0.0008) | (0.0001) | (0.0030) | (0.0001) | (0.0015) | (0.0028) | (0.0020) | (0.0015) | (0.0010) | (0.0006) | (0.0001) |
| Observations | 562,579 | 562,846 | 562,848 | 562,842 | 562,824 | 562,848 | 562,796 | 562,786 | 562,826 | 562,838 | 562,839 | 562,843 | 562,848 |
| R-squared | 0.8715 | 0.3302 | 0.3252 | 0.2702 | 0.6456 | 0.2717 | 0.5933 | 0.8367 | 0.6212 | 0.5231 | 0.4965 | 0.3923 | 0.2619 |
| Politician-Month FE | Y | Y | Y | Y | Y | Y | Y | Y | Y | Y | Y | Y | Y |
| Week of Month FE | Y | Y | Y | Y | Y | Y | Y | Y | Y | Y | Y | Y | Y |

Note: Robust standard errors clustered at the level of the state and week in parenthesis. *** p<0.01, ** p<0.05, * p<0.1. All politicians included in columns (1)-(13). Average of pageviews relative to all pageviews in state-year is used as Twitter penetration measure. All specifications include Politician-month and week of month fixed effects.

Table A18: Donations and Contemporaneous, and Lagged Campaign Expenditure Controls

| | (1) | (2) | (3) | (4) | (5) | (6) |
|---|---|---|---|---|---|---|
| | \multicolumn{6}{c}{Log (Aggregate Donations)} | | | | | |
| VARIABLES | All | New | Experienced | All | New | Experienced |
| on Twitter x Twit_Penet | 100.6431** | 176.5487*** | -40.2461 | 117.9161** | 181.8973*** | -2.1690 |
| | (45.0527) | (41.9974) | (93.6875) | (45.8447) | (45.7522) | (96.5402) |
| onTwitter | 0.1695 | 0.0922 | 0.3840* | 0.1126 | 0.1220 | 0.2095 |
| | (0.1114) | (0.1074) | (0.2271) | (0.1141) | (0.1201) | (0.2345) |
| Log(Campaign Expenditure) | 0.0831*** | 0.1219*** | 0.0704*** | | | |
| | (0.0055) | (0.0076) | (0.0059) | | | |
| Log(Last month Campaign Expenditure) | | | | 0.0128*** | 0.0243*** | 0.0088 |
| | | | | (0.0041) | (0.0038) | (0.0056) |
| Observations | 561,339 | 234,393 | 326,946 | 551,580 | 228,214 | 323,366 |
| R-squared | 0.8225 | 0.8848 | 0.7862 | 0.8223 | 0.8828 | 0.7870 |
| Politician-Month FE | Y | Y | Y | Y | Y | Y |
| Week of Month FE | Y | Y | Y | Y | Y | Y |
| Baseline controls x on Twitter | N | N | N | N | N | N |
| Implied Twitter effect for 2009 | .028** | .044*** | -.011 | .033** | .046*** | -.001 |
| Implied Twitter effect for 2014 | .224** | .393*** | -.09 | .262** | .405*** | -.005 |
| | \multicolumn{6}{c}{Log (Number of Donations)} | | | | | |
| on Twitter x Twit_Penet | 16.3200 | 38.4066*** | -24.0397 | 19.0968 | 39.5161*** | -19.1702 |
| | (12.3205) | (10.8266) | (24.0987) | (12.4689) | (11.6067) | (24.2931) |
| onTwitter | 0.0479 | 0.0122 | 0.1296** | 0.0388 | 0.0154 | 0.1077* |
| | (0.0322) | (0.0278) | (0.0573) | (0.0344) | (0.0302) | (0.0636) |
| Log(Campaign Expenditure) | 0.0240*** | 0.0354*** | 0.0203*** | | | |
| | (0.0016) | (0.0026) | (0.0016) | | | |
| Log(Last month Campaign Expenditure) | | | | 0.0070*** | 0.0120*** | 0.0051*** |
| | | | | (0.0013) | (0.0014) | (0.0017) |
| Observations | 561,339 | 234,393 | 326,946 | 551,580 | 228,214 | 323,366 |
| R-squared | 0.8394 | 0.9024 | 0.8015 | 0.8389 | 0.9011 | 0.8014 |
| Politician-Month FE | Y | Y | Y | Y | Y | Y |
| Week of Month FE | Y | Y | Y | Y | Y | Y |
| Baseline controls x on Twitter | N | N | N | N | N | N |
| Implied Twitter effect for 2009 | .005 | .01*** | -.007 | .005 | .01*** | -.005 |
| Implied Twitter effect for 2014 | .036 | .085*** | -.053 | .042 | .088*** | -.043 |

Note: Robust standard errors clustered at the level of the state and week in parenthesis. *** p<0.01, ** p<0.05, * p<0.1. The dependent variables are Log (Aggregate Donations) and Log (Number of Donations). Columns (1) and (4) include all politicians while columns (2) and (5) include only new and columns (3) and (6) have the experienced politicians. State level baseline controls interacted with the politician being on Twitter include the median household income and population size. Average of pageviews relative to all pageviews in state-year is used as Twitter penetration measure. All specifications include Politician-month and week of month fixed effects.

Table A19: Demographics, Twitter Penetration and Donations

| VARIABLES | Log(Aggregate Donations) | | | | | | |
|---|---|---|---|---|---|---|---|
| | (1) All | (2) All | (3) All | (4) All | (5) All | (6) All | (7) All |
| onTwitter x Twit_Penet | 102.6499** | | | | | | |
| | (45.5243) | | | | | | |
| on Twitter x Education | | -2.2186 | | | | | |
| | | (1.6557) | | | | | |
| on Twitter x Population | | | -0.0238 | | | | |
| | | | (0.0693) | | | | |
| on Twitter x Bush Vote Share | | | | 0.1699 | | | |
| | | | | (0.7847) | | | |
| on Twitter x African American | | | | | 0.1986 | | |
| | | | | | (0.9545) | | |
| on Twitter x Household Income | | | | | | -0.0165 | |
| | | | | | | (0.0117) | |
| on Twitter x Share of Rich | | | | | | | -0.0304 |
| | | | | | | | (0.0655) |
| on Twitter | 0.1879 | 2.2290 | 0.8172 | 0.3533 | 0.4161*** | 1.1418** | 0.5106*** |
| | (0.1126) | (1.3366) | (1.1305) | (0.3816) | (0.1224) | (0.5074) | (0.1901) |
| Observations | 565,968 | 565,968 | 566,280 | 565,968 | 565,968 | 565,968 | 565,968 |
| R-squared | 0.8215 | 0.8215 | 0.8212 | 0.8215 | 0.8215 | 0.8215 | 0.8215 |
| Politician-Month FE | Y | Y | Y | Y | Y | Y | Y |
| Week of Month FE | Y | Y | Y | Y | Y | Y | Y |
| Baseline controls x on Twitter | N | N | N | N | N | N | N |
| Implied Twitter effect for 2009 | .029** | -.001 | 0 | 0 | 0 | 0 | 0 |
| Implied Twitter effect for 2014 | .228** | -.005 | 0 | 0 | 0 | 0 | 0 |

Note: Robust standard errors clustered at the level of the state and week in parenthesis. *** $p<0.01$, ** $p<0.05$, * $p<0.1$. The dependent variable is Log (Aggregate Donations). Average of pageviews relative to all pageviews in state-year is used as Twitter penetration measure. All specifications include Politician-month and week of month fixed effects.

Table A20: Survival Analysis –Adopting Twitter and Correlation with Penetration

| VARIABLES | (1) Join Twitter | (2) Join Twitter | (3) Join Twitter | (4) Join Twitter |
|---|---|---|---|---|
| Penetration | 41.925 | 41.376 | 16.222 | 20.058 |
| | (25.520) | (25.279) | (20.074) | (20.095) |
| Observations | 183,171 | 183,171 | 186,636 | 186,636 |
| Baseline controls | Y | Y | Y | Y |
| Model Assumption | Proportional Odds | Proportional Hazard | Proportional Hazard | Proportional Hazard |
| Baseline Hazard Assumption | N/A | N/A | Gompertz | Weibull |

Note: Robust standard errors in parenthesis. *** $p<0.01$, ** $p<0.05$, * $p<0.1$.

Table A21: Politicians' Tweets and Retweets

| | Log(Aggregate Donations) | | | | | |
|---|---|---|---|---|---|---|
| | Tweets | | | Re-Tweets | | |
| | (1) | (2) | (3) | (4) | (5) | (6) |
| VARIABLES | All | New | Experienced | All | New | Experienced |
| onTwitter x Twit_Penet x log(tweets) | 160.5434 | 1,292.0855*** | 100.2382 | | | |
| | (141.5208) | (376.8166) | (163.7363) | | | |
| onTwitter x Twit_Penet | 118.1129** | 209.1624*** | -42.1596 | 107.3255** | 194.4538*** | -45.6113 |
| | (45.3456) | (46.4943) | (90.0369) | (45.4012) | (45.5389) | (90.8583) |
| onTwitter x Twit_Penet x log(retweets) | | | | 1.0821 | 29.4087 | 16.4282 |
| | | | | (82.1314) | (191.3594) | (93.6309) |
| onTwitter | 1.2716 | 2.3430 | -0.2047 | 1.1762 | 2.1948 | -0.2402 |
| | (1.0127) | (1.5445) | (2.0509) | (1.0277) | (1.5537) | (2.0571) |
| | | | | | | |
| Observations | 77,433 | 46,562 | 30,871 | 77,433 | 46,562 | 30,871 |
| R-squared | 0.7931 | 0.8394 | 0.7233 | 0.7930 | 0.8394 | 0.7233 |
| Politician-Month FE | Y | Y | Y | Y | Y | Y |
| Week of Month FE | Y | Y | Y | Y | Y | Y |
| Baseline controls x on Twitter | Y | Y | Y | Y | Y | Y |
| Implied Twitter effect for 2009 | .04 | .359*** | .025 | 0 | .008 | .004 |
| Implied Twitter effect for 2014 | .357 | 2.921*** | .164 | .002 | .066 | .027 |

Note: Robust standard errors clustered at the level of the state and week in parenthesis. *** p<0.01, ** p<0.05, * p<0.1. The dependent variable is Log (Aggregate Donations). Columns (1)-(4) include all politicians while columns (2)-(5) include only new and columns (3)-(6) have the experienced politicians. State level baseline controls interacted with the politician being on Twitter include the median household income and population size. Average of pageviews relative to all pageviews in state-year is used as Twitter penetration measure. All specifications include Politician-month and week of month fixed effects.

Table A22: Sentiment Analysis

| | Log (Aggregate Donations) | | | | | | | | |
|---|---|---|---|---|---|---|---|---|---|
| | Hyperlinks | | | Plugged-In | | | Anti Establishment | | |
| | (1) | (2) | (3) | (4) | (5) | (6) | (7) | (8) | (9) |
| VARIABLES | All | New | Experienced | All | New | Experienced | All | New | Experienced |
| on Twitter x Twit_Penet x Links | -211.7399 | 2,334.8629*** | -202.6737 | | | | | | |
| | (264.4008) | (286.9852) | (299.6514) | | | | | | |
| onTwitter x Twit_Penet x log(Plugged-In) | | | | 195.0489** | 213.5259** | 135.4634 | | | |
| | | | | (73.2620) | (82.8572) | (124.5946) | | | |
| onTwitter x Twit_Penet x log(Anti Establishment) | | | | | | | 529.4737*** | 501.3115*** | 167.9429 |
| | | | | | | | (62.9517) | (92.0858) | (104.3727) |
| onTwitter x Twit_Penet | 112.8880** | 200.6463*** | -41.5839 | -666.9876** | -661.8708** | -555.8361 | 106.0181** | 191.9797*** | -45.8876 |
| | (45.4109) | (45.4234) | (90.9508) | (287.2844) | (315.9802) | (515.1979) | (45.3790) | (45.5413) | (90.5551) |
| onTwitter | 1.1995 | 2.2182 | -0.2566 | 1.7570 | 2.9232* | 0.0823 | 1.2143 | 2.2278 | -0.2306 |
| | (1.0169) | (1.5543) | (2.0532) | (1.6334) | (1.5150) | (3.0865) | (1.0250) | (1.5478) | (2.0603) |
| | | | | | | | | | |
| Observations | 77,433 | 46,562 | 30,871 | 69,658 | 41,644 | 28,014 | 77,433 | 46,562 | 30,871 |
| R-squared | 0.7930 | 0.8393 | 0.7233 | 0.7906 | 0.8410 | 0.7159 | 0.7930 | 0.8392 | 0.7233 |
| Politician-Month FE | Y | Y | Y | Y | Y | Y | Y | Y | Y |
| Week of Month FE | Y | Y | Y | Y | Y | Y | Y | Y | Y |
| Baseline controls x on Twitter | Y | Y | Y | Y | Y | Y | Y | Y | Y |
| Implied Twitter effect for 2009 | -.053 | .649*** | -.05 | .054** | .059** | .034 | .133*** | .139*** | .041 |
| Implied Twitter effect for 2014 | -.471 | 5.278*** | -.332 | .434** | .483** | .222 | 1.178*** | 1.133*** | .275 |

Note: Robust standard errors clustered at the level of the state and week in parenthesis. *** p<0.01, ** p<0.05, * p<0.1. The dependent variable is Log (Aggregate Donations). Columns (1), (4) and (7) include all politicians while columns (2), (5) and (8) include only new and columns (3), (6) and (9) have the experienced politicians. State level baseline controls interacted with the politician being on Twitter include the median household income and population size. Average of pageviews relative to all pageviews in state-year is used as Twitter penetration measure. All specifications include Politician-month and week of month fixed effects.

Table A23: Facebook Adoption, Amount of Donations and the Number of Donations

|  | Log (Aggregate Donations) | | | Log (Number of Donations) | | |
|---|---|---|---|---|---|---|
|  | (1) | (2) | (3) | (4) | (5) | (6) |
| VARIABLES | All | New | Experienced | All | New | Experienced |
| on Facebook x Facebook_Penet | 28.7949 | 69.1274*** | 4.8686 | 10.2927** | 22.2321*** | 4.0376 |
|  | (18.3529) | (23.8433) | (23.2144) | (4.6895) | (5.7106) | (6.6132) |
| on Facebook | 4.6736 | 11.5255*** | -6.4195 | 2.3640* | 2.4361** | 1.9086 |
|  | (6.8758) | (3.0804) | (14.8822) | (1.2795) | (0.9130) | (2.6534) |
| Observations | 565,968 | 236,740 | 329,228 | 565,968 | 236,740 | 329,228 |
| R-squared | 0.8215 | 0.8826 | 0.7856 | 0.8384 | 0.9006 | 0.8006 |
| Politician-Month FE | Y | Y | Y | Y | Y | Y |
| Week FE | Y | Y | Y | Y | Y | Y |
| Baseline controls x on Facebook | Y | Y | Y | Y | Y | Y |
| Implied Facebook effect for 2009 | .034 | .082*** | .006 | .012** | .026*** | .005 |
| Implied Facebook effect for 2014 | .33 | .921*** | .047 | .118** | .296*** | .039 |

Note: Robust standard errors clustered at the level of the state and week in parenthesis. *** p<0.01, ** p<0.05, * p<0.1. The dependent variables are Log (Aggregate Donations) and Log (Number of Donations). Columns (1)-(4) include all politicians while columns (2)-(5) include only new and columns (3)-(6) have the experienced politicians. State level baseline controls interacted with the politician being on Facebook include the median household income and population size. Average of pageviews relative to all pageviews in state-year is used as Facebook penetration measure. All specifications include Politician-month and week of month fixed effects.

Table A24: Baseline Results with Week Fixed Effects

|  | Log(Aggregate Donations) | | | Prob. of Donations | | | Log(Number of Donations) | | |
|---|---|---|---|---|---|---|---|---|---|
|  | (1) | (2) | (3) | (4) | (5) | (6) | (7) | (8) | (9) |
| VARIABLES | All | New | Experienced | All | New | Experienced | All | New | Experienced |
| on Twitter x Twit_Penet | 108.3569** | 144.6814*** | 64.2917 | 14.6596*** | 18.4853*** | 9.8438 | 17.6377 | 25.5938** | 9.2538 |
|  | (40.8647) | (44.3573) | (74.7063) | (5.3966) | (6.2203) | (9.9655) | (11.1982) | (11.4311) | (18.4330) |
| onTwitter | 1.2285 | 2.1541 | 0.0273 | 0.1853 | 0.3226 | 0.0040 | 0.1950 | 0.3945 | -0.0622 |
|  | (1.0401) | (1.5106) | (2.0119) | (0.1510) | (0.2008) | (0.2860) | (0.1900) | (0.3474) | (0.4655) |
| Observations | 565,968 | 236,740 | 329,228 | 565,968 | 236,740 | 329,228 | 565,968 | 236,740 | 329,228 |
| R-squared | 0.8238 | 0.8840 | 0.7900 | 0.7881 | 0.8457 | 0.7540 | 0.8422 | 0.9034 | 0.8074 |
| Politician-Month FE | Y | Y | Y | Y | Y | Y | Y | Y | Y |
| Week FE | Y | Y | Y | Y | Y | Y | Y | Y | Y |
| Baseline controls x on Twitter | Y | Y | Y | Y | Y | Y | Y | Y | Y |
| Implied Twitter effect for 2009 | .03** | .036*** | .018 | .004*** | .005*** | .003 | .005 | .006** | .003 |
| Implied Twitter effect for 2014 | .241** | .322*** | .143 | .033*** | .041*** | .022 | .039 | .057** | .021 |

Note: Robust standard errors clustered at the level of the state and week in parenthesis. *** p<0.01, ** p<0.05, * p<0.1. The dependent variables are Log (Aggregate Donations), Probability of at least One Donation and Log (Number of Donations). All politicians are included in columns (1), (4) and (7), while it is the new politicians in columns (2), (5) and (8) and experienced politicians in (3), (6) and (9). State level baseline controls interacted with the politician being on Twitter include the median household income and population size. Average of pageviews relative to all pageviews in state-year is used as Twitter penetration measure. All specifications include Politician-month and week fixed effects.

Table A25: Baseline Results with Cubic Week Time Trend

|  | Log(Aggregate Donations) | | | Prob. of Donations | | | Log(Number of Donations) | | |
|---|---|---|---|---|---|---|---|---|---|
|  | (1) | (2) | (3) | (4) | (5) | (6) | (7) | (8) | (9) |
| VARIABLES | All | New | Experienced | All | New | Experienced | All | New | Experienced |
| on Twitter x Twit_Penet | 108.3569** | 144.6814*** | 64.2917 | 14.6596*** | 18.4853*** | 9.8438 | 17.6377 | 25.5938** | 9.2538 |
|  | (40.8647) | (44.3573) | (74.7063) | (5.3966) | (6.2203) | (9.9655) | (11.1982) | (11.4311) | (18.4330) |
| onTwitter | 1.2285 | 2.1541 | 0.0273 | 0.1853 | 0.3226 | 0.0040 | 0.1950 | 0.3945 | -0.0622 |
|  | (1.0401) | (1.5106) | (2.0119) | (0.1510) | (0.2008) | (0.2860) | (0.1900) | (0.3474) | (0.4655) |
| Observations | 565,968 | 236,740 | 329,228 | 565,968 | 236,740 | 329,228 | 565,968 | 236,740 | 329,228 |
| R-squared | 0.8238 | 0.8840 | 0.7900 | 0.7881 | 0.8457 | 0.7540 | 0.8422 | 0.9034 | 0.8074 |
| Politician-Month FE | Y | Y | Y | Y | Y | Y | Y | Y | Y |
| Cubic Week Trend | Y | Y | Y | Y | Y | Y | Y | Y | Y |
| Baseline controls x on Twitter | Y | Y | Y | Y | Y | Y | Y | Y | Y |
| Implied Twitter effect for 2009 | .03** | .036*** | .018 | .004*** | .005*** | .003 | .005 | .006** | .003 |
| Implied Twitter effect for 2014 | .241** | .322*** | .143 | .033*** | .041*** | .022 | .039 | .057** | .021 |

Note: Robust standard errors clustered at the level of the state and week in parenthesis. *** p<0.01, ** p<0.05, * p<0.1. The dependent variables are Log (Aggregate Donations), Probability of Atleast One Donation and Log (Number of Donations). All politicians are included in columns (1), (4) and (7), while it is the new politicians in columns (2), (5) and (8) and experienced politicians in (3), (6) and (9). State level baseline controls interacted with the politician being on Twitter include the median household income and population size. Average of pageviews relative to all pageviews in state-year is used as Twitter penetration measure. All specifications include Politician-month and cubic week time trend fixed effects.

A27

# Online Appendix B: Data Appendix

## B.1 Data Collection from Twitter

We provide guidelines for Twitter data collection here. Twitter allows researchers and developers to pull data from API in two different forms.

1. **REST API.** The API allows researchers to look up any user or tweet from the past conditional on a unique identifier (i.e. a user's Twitter handle, a tweet's ID, etc). However, Twitter places pretty tight constraints on the amount of data one can get in a given window of time. Due to the limitations in data gathering, we use the REST API to collect information about the politicians and their tweets.

2. **Streaming API.** This API is the most commonly used tool for gathering Twitter data in academic research. The Streaming API allows researchers to tap into 1% of all incoming tweets in a random fashion and without the data extraction limits of the REST API. Via the Streaming API, we are unable to obtain every tweet posted on Twitter, but we obtain a consistent random sample of them. We use this API when we need massive amounts of data: the followers' profile information and their tweeting activity data.

**Verification of Politician Twitter Accounts.** After data collection, a research assistant who is blind to the research question manually verified the politician accounts. The verification of the politician accounts could also partially be handled via the Twitter API field "verified," which shows whether or not an account is verified. However, some congressman hold unverified accounts, although from the posted information on the profiles, it is plausible to assume the accounts are authentic.

**Searching for a Candidate's Account**. The search for a candidate account on Twitter is initiated by searching for each candidate's name via the Twitter API, and deduced which handle was his or hers algorithmically and subsequently checked manually by a research assistant.

## B.2 Alternate Measures of Twitter Penetration

Below are the alternate measures we use for Twitter penetration in Equation (1), based on data from comScore's online browsing panel and an alternate social panel, Simmons OneView. As described in the main text, comScore records all online browsing activity of nearly 50,000 households in the United States. From comScore browsing data, we create the following various penetration measures which indicate the amount of information users receive from Twitter.

1. **Median Share of Pageviews on Twitter:** Median share of weekly pageviews (or page visits) on Twitter, relative to all pageviews in the state-year.

2. **Average Duration Spent on Twitter:** Average amount of weekly time spent on Twitter relative to the total time spent on Twitter in the same state-year.

3. **Median Duration Spent on Twitter:** Median of share of weekly time spent on Twitter relative to the total time spent on Twitter in the same state-year.

4. **Total Duration Spent on Twitter (Absolute Measure):** Total amount of weekly time spent on Twitter in the same state-year.

5. **Total Number of Households on Twitter (Absolute Measure):** Total number of weekly households visiting Twitter in a week in the same state-year.

6. **Average Share of Households on Twitter:** Average of weekly share of households visiting Twitter relative to all households in same state-year.

7. **Median Proportion of Households on Twitter:** Median of weekly share of households visiting Twitter relative to all households in same state-year.

8. **Average Share of Site Sessions on Twitter:** Average of weekly share of browsing sessions that include Twitter relative to all browsing sessions in the state-year. A session is a combination of page views starting from the time a household opens a browser until they close it.

9. **Median Share of Site Sessions on Twitter:** Median of weekly share of browsing sessions that include Twitter relative to all browsing sessions in the state-year.

Simmons Oneview is a database of surveys of households from the 48 contagious states and Washington DC. In the survey, there are questions about media habits of individuals. We use two questions from these surveys which report individuals' Twitter usage, "Have you visited Twitter in the last 30 days?" and "Have you visited Twitter in the last 7 days?" Individuals' responses are coded as yes or no and we look at the proportion of households which answered yes. The two measures we generate from this database are:

1. **7day Share of Households:** Share of households who stated that they visited Twitter in the last 7 days in each state-year.

2. **30day Share of Households:** Share of households who stated that they visited Twitter in the last 30 days in each state-year.

The correlation between the benchmark measure used in the paper (given in the first row/column, average pageviews on Twitter relative to all pageviews) and these alternate measures ranges between 0.24 and 0.89 (see Table B1). Correlation between any other two measure ranges from 0.18 to 0.98. As expected, the correlation between measures obtained from the two different data sources (comScore and Simmons) show lower correlation, whereas measures from the same source have higher correlation among them. Altogether, the correlation values demonstrate that the penetration patterns captured by the benchmark is informative.[45]

Table B1: Correlation Between the Penetration Measures

| Variables | Avg Pagev. | 7day Share HH | Total Durat. | Total HH | Avg Sess. | Avg Share HH | Med Share Durat. | Med Share Sess. | Med Share HH | 30d Share HH | Med Med Pagev. | Avg Share Durat. |
|---|---|---|---|---|---|---|---|---|---|---|---|---|
| Avg Pageviews | 1 | | | | | | | | | | | |
| 7d Share-Household | 0.405 | 1 | | | | | | | | | | |
| Total duration | 0.244 | 0.189 | 1 | | | | | | | | | |
| Total Households | 0.283 | 0.200 | 0.932 | 1 | | | | | | | | |
| Sessions_Avg | 0.791 | 0.323 | 0.369 | 0.343 | 1 | | | | | | | |
| Household_Share | 0.704 | 0.330 | 0.460 | 0.448 | 0.833 | 1 | | | | | | |
| Med Share-Duration | 0.735 | 0.295 | 0.241 | 0.232 | 0.716 | 0.577 | 1 | | | | | |
| Med Share-Sessions | 0.885 | 0.363 | 0.272 | 0.291 | 0.817 | 0.686 | 0.877 | 1 | | | | |
| Med Share-Household | 0.846 | 0.394 | 0.270 | 0.325 | 0.782 | 0.701 | 0.721 | 0.899 | 1 | | | |
| 30d Share-Household | 0.422 | 0.792 | 0.182 | 0.191 | 0.351 | 0.373 | 0.310 | 0.385 | 0.434 | 1 | | |
| Median Pageviews | 0.671 | 0.271 | 0.237 | 0.226 | 0.662 | 0.504 | 0.933 | 0.834 | 0.726 | 0.286 | 1 | |
| Avg Share-Duration | 0.739 | 0.303 | 0.200 | 0.191 | 0.724 | 0.564 | 0.988 | 0.885 | 0.730 | 0.324 | 0.924 | 1 |

Note: "Avg." stands for average, "Med." stands for median. Benchmark penetration variable is in the first column, average pageviews.

The distributions of the alternate measures are given in Table B2.

---

[45] Almost all penetration measures are statistically significantly correlated to each other at the 1% level.

Table B2: Summary Statistics (Politician-year level)

|  | mean | sd | min | max | p10 | p25 | p50 | p75 | p90 |
|---|---|---|---|---|---|---|---|---|---|
| Avg Pageviews | 0.00348 | 0.00231 | 0.00018 | 0.01286 | 0.00074 | 0.00127 | 0.00344 | 0.00492 | 0.00661 |
| Med Share-Pageviews | 0.00197 | 0.00122 | 0.00000 | 0.01083 | 0.00060 | 0.00115 | 0.00184 | 0.00236 | 0.00350 |
| Total (Weekly) Duration | 2605.589 | 4145.918 | 0.0000 | 19650 | 5 | 96 | 964 | 2779 | 8287 |
| Med Share-Duration | 0.00370 | 0.00270 | 0.00000 | 0.02796 | 0.00068 | 0.00136 | 0.00361 | 0.00508 | 0.00717 |
| Avg Share-Households | 0.07002 | 0.03108 | 0.00760 | 0.21554 | 0.02829 | 0.04552 | 0.07097 | 0.09110 | 0.10992 |
| Number of (Weekly) Households | 67.600 | 88.963 | 0.00000 | 456 | 2 | 7 | 34 | 88 | 172 |
| Avg Share-Household on Twitter | 0.0538 | 0.0402 | 0.00000 | 0.2131 | 0.0022 | 0.02707 | 0.0446 | 0.08 | 0.11 |
| Med Share-Households | 0.07142 | 0.03134 | 0.00000 | 0.22222 | 0.03175 | 0.04480 | 0.07244 | 0.09213 | 0.11069 |
| Avg Share-Sessions | 0.00372 | 0.00216 | 0.00017 | 0.01272 | 0.00093 | 0.00177 | 0.00401 | 0.00499 | 0.00656 |
| Med Share-Sessions | 0.00372 | 0.00219 | 0.00000 | 0.01434 | 0.00093 | 0.00171 | 0.00388 | 0.00505 | 0.00691 |
| 7d Share-Households | 0.04244 | 0.02464 | 0.00000 | 0.19145 | 0.01538 | 0.02431 | 0.03881 | 0.05668 | 0.07372 |
| 30d Share-Households | 0.06272 | 0.02898 | 0.00059 | 0.29931 | 0.02661 | 0.04391 | 0.06051 | 0.08107 | 0.09963 |

Note: Benchmark penetration measure is Average Twitter pageviews. Avg stands for average

Measures in first 10 rows are from comScore's browsing panel and the last two rows are from Simmons Oneview. N=10884.

## B.3 FEC Campaign Expense Categories

All registered politicians are required by law to report how they spend their funds, with date, purpose, amount and payee information. FEC classifies expenses into 12 spending categories, as given in Table B3 and leaves the uncommon items unclassified. Notice that expenditures can take negative values due to refunds.

Table B3: FEC Disbursement Categories

| Disbursement category | Description |
|---|---|
| Administrative | Salary, overhead, rent, postage, office supplies, equipment, furniture, ballot access fees, petition drives, party fees and legal and accounting expenses |
| Travel expenses | Commercial carrier tickets; reimbursements for use of private vehicles, advance payments for use of corporate aircraft; lodging and meal expenses incurred during travel |
| Solicitation & fundraising | costs for direct mail solicitations & fundraising events, printing, mailing lists, consultant fees, call lists, invitations, catering costs and room rental |
| Advertising expenses | General public advertising, purchases of radio/television broadcast/cable time, print advertisements and related production costs |
| Polling expenses | (No definition provided) |
| Campaign materials | Buttons, bumper stickers, brochures, mass mailings, pens, posters, balloons |
| Campaign event expenses | Costs associated with candidate appearances, campaign rallies, town meetings, phone banks, including catering costs, door to door get-out-the-vote efforts and driving voters to the polls |
| Transfers | Funds to other authorized committees of the same candidate |
| Loan repayments | Repayments of loans made or guaranteed by the candidate or other person |
| Refunds of contributions | refunds to individuals/ persons, political party committees or other political committees |
| Political contributions | Contributions to other federal candidates and committees, donations to nonfederal candidates and committees |
| Donations | Donations to charitable or civic organizations |